# Does generative AI narrow education-based productivity gaps? Evidence from a randomized experiment*

Guillermo Cruces† Diego Fernández Meijide‡ Sebastian Galiani§
Ramiro H. Gálvez¶ María Lombardi¶

May 2026

**Abstract**

Does generative artificial intelligence (AI) reinforce or reduce productivity differences across workers? Existing evidence largely studies AI within firms and occupations, where organizational selection compresses educational heterogeneity, leaving unclear whether AI narrows productivity gaps across individuals with different levels of education. A related question is whether the productivity gains from AI reflect productive use of the tool or mere delegation, fading once AI is unavailable, and whether the answer differs across education groups. We address both questions using a randomized online experiment outside firms, in which 1,174 adults aged 25–45 complete an incentivized, workplace-style business problem-solving task with or without a generative-AI assistant, followed by a non-AI-assisted follow-up module. AI increases performance for all participants, with substantially larger gains for lower-education individuals. In the control group without AI, higher-education participants outperform lower-education participants by 0.548 standard deviations; with AI, this gap falls to 0.139 standard deviations, closing about three-quarters of the initial gap. Chat-log measures explain both why this gap narrows and why it does not disappear: lower-education participants successfully obtain substantial assistance from AI, while higher-education participants use the tool somewhat more effectively across several margins. The follow-up results show that these gains are not purely driven by delegation; treated participants do not perform worse than controls once AI is removed, and lower-education participants retain part of their gain, although a sizable education gap re-emerges. Consistent with this interpretation, intensive use of the assistant produces strong task performance regardless of how much time and effort participants devote themselves, but follow-up performance is substantially higher only when intensive AI use is combined with sustained effort on the task. We interpret these findings as evidence that generative AI narrows effective productivity differences in task execution while underlying human-capital differences continue to shape unassisted performance and effective use of the tool.

**JEL Codes**: J24, O33
**Keywords**: Productivity, artificial intelligence, education, human capital, inequality.

*This work was conducted with the support of Sur Futuro as part of IDRC's FutureWORKS initiative and the Latin America and the Caribbean Research Network of the Inter-American Development Bank. The opinions expressed in this publication are those of the authors and do not necessarily reflect the views of the Inter-American Development Bank, its Board of Directors, or the countries they represent. We thank Julián Blanco for his assistance in the design and implementation of the AI-based assistant, Santiago Afonso for developing alternative procedures for AI-assisted grading, and Genaro Damiani and Valentín Pérez Gaitán for providing manual grading. We also thank Santiago Afonso and the SurFuturo and IADB teams for valuable feedback and suggestions, and José Guinot-Saporta for the management of the UoN CEDEX Laboratory. We thank David Autor, Alex Imas, and Craig McIntosh for comments on our first draft. We also thank seminar and workshop participants at the IADB Network Meeting, the LACEA Labor/World Bank Workshop on Technological Change and Labor Markets, the University of Liverpool, the University of Oklahoma, Tulane University, UdeSA, UTDT, UNLP and the University of Nottingham AI Reading Group. This study was reviewed and approved by the Nottingham School of Economics Research Ethics Committee and the CEDLAS-UNLP Research Ethics Committee (CEDLAS-REC). AEA-RCT Registry: 0016607.

†Universidad de San Andres and University of Nottingham; ‡Universidad de San Andres; §University of Maryland and NBER; ¶Universidad Torcuato Di Tella

# 1 Introduction

Technological change has long been a driver of inequality in labor markets. Historically, innovations have often disproportionately benefited highly educated workers, contributing to widening gaps in employment and earnings (Acemoglu and Autor, 2011; Goldin and Katz, 2008). Generative artificial intelligence (AI) now presents a new paradigm: more directly than many previous technologies, these systems can perform complex cognitive tasks. This raises the possibility that generative AI could be "skill-democratizing" rather than skill-biased, as suggested by Autor (2024), enabling workers with limited formal training to perform tasks previously requiring extensive education and thus empowering workers across the education distribution. Acemoglu (2024), by contrast, cautions that AI's effects on labor markets may be more modest and unequally distributed than these early discussions suggest. Whether AI weakens the link between formal education and productivity or instead reproduces existing skill advantages in a new form remains an open empirical question. Direct experimental evidence on how AI affects productivity differences across education groups, and through what mechanisms, remains scarce.

This paper asks a simple but consequential question: does generative AI reduce or reinforce productivity gaps across individuals with different levels of formal education, and how? In answering this question, we tackle an additional related question: do the productivity gains from AI reflect productive use of the tool or merely delegation to AI, fading once AI is unavailable? Existing evidence largely comes from workplace settings and specific occupations, where hiring and task assignment compress educational heterogeneity. These studies provide valuable evidence on how AI affects variation within relatively homogeneous groups (e.g., higher- versus lower-performing workers within a firm or occupation), but by design are less well suited to address whether AI narrows or widens the education-based productivity gaps that feature prominently in the inequality debate (Goldin and Katz, 2008). To address these questions we designed a randomized online experiment outside firms in which high school-educated and postsecondary-educated adults complete the same incentivized business problem-solving task either with or without access to a GPT-based assistant. By holding the task fixed and randomizing access across education-defined groups, we identify how AI changes the productivity advantage associated with formal education—that is, the between-group performance gap. We pair this task with an immediate non-AI-assisted follow-up module to ask whether the AI-induced performance gains are solely contemporaneous or partly carry over once the tool is removed. Our experimental design deliberately trades off organizational realism for identification of educational heterogeneity that is typically compressed within firms.[1] This choice allows us to

[1]Experimental designs differ from naturally-occurring settings along multiple dimensions (Harrison

isolate a dimension of technological change that existing workplace evidence cannot capture.

In our preregistered online experiment, 1,174 individuals aged 25 to 45 from Argentina were recruited to complete a workplace-relevant task with randomized access to an AI assistant. Crucially, we drew our sample and implemented our experiment with two distinct groups. We define low-education individuals as those who only have a high school diploma or have completed less than half of a postsecondary degree (university or non-university tertiary programs), and high-education individuals as those who have completed more than half of such a program.[2] This education classification was preregistered, and our results are robust to alternative definitions of the threshold for high education. The task involves solving a realistic hypothetical business scenario: participants receive an email from their boss, and to answer it they must analyze several sources of information. They must write a reply to the email that diagnoses the problem and proposes a solution. Importantly, our task is self-contained and does not rely on domain or firm-specific knowledge, allowing us to study AI's effects on performance in a general business problem-solving setting. Furthermore, our task was designed to tap into multiple cognitive dimensions: reading and data comprehension, reasoning, creative problem-solving, and writing. The AI assistant is based on OpenAI's GPT-4.1 model and, for the treatment group, is embedded in the experiment interface. Immediately after submitting the task, all participants completed an incentivized follow-up module without access to the assistant, including an open-ended question asking them to articulate the root cause of the business problem and some recall questions about the task content.

Our findings indicate that AI assistance substantially raises performance for both education groups, but with larger gains among low-education participants. Measured in standard deviations (SD) relative to the low-education control group, AI access increases the overall task score (our main outcome of interest) by 1.242 SD for low-education individuals and by 0.834 SD for high-education individuals. Among treated participants, performance remains dispersed, indicating that the treatment effect is not mechanically driven by generalized convergence to a perfect score. In the absence of AI access, high-education individuals outperform low-education individuals by 0.548 SD. Importantly, this gap between low- and high-education participants falls to 0.139 SD with AI access, closing 75% of the baseline difference. Finally, through an analysis and List, 2004); ours relaxes several of these in exchange for educational variation that is compressed within firms.

[2]In Argentina, open admission and free public tertiary education generate substantial heterogeneity within this intermediate group because many individuals enroll but complete only a limited portion of their studies. We therefore distinguish between incomplete postsecondary education below and above half completion, assigning them to the low- and high-education groups accordingly, which yields a more informative stratification of formal skills for our experimental design.

of the chat logs, we explain this distributional result. Lower-education participants obtain substantial assistance from AI, which helps explain why the gap narrows so much. Higher-education participants, however, use the tool somewhat more effectively across several margins, helping explain why the gap does not disappear.

Our results establish that AI raises performance for both groups and disproportionately benefits lower-education participants. However, these gains may partly reflect delegation to the tool rather than improved productivity. Consistent with this concern, experimental literature shows that AI assistance can reduce subsequent unassisted performance when users rely on the tool as a substitute for effortful problem-solving (Bastani et al., 2025; Shen and Tamkin, 2026; Liu et al., 2026). We use the non-AI-assisted follow-up module to address this concern in our setting, treating it as a diagnostic for the AI-induced gains rather than as a general test of skill formation. The results do not support a pure delegation interpretation: treated participants do not perform worse than controls once AI is removed. If anything, AI access modestly improves follow-up performance among lower-education participants by about 0.171 SD, while higher-education participants also show a small, non-significant gain. Despite these gains, lower-education treated participants still lag behind their higher-education counterparts in the follow-up by 0.200 SD, indicating that underlying human capital continues to shape unassisted performance. Consistent with this interpretation, when we divide treated participants into four mutually exclusive groups based on levels of AI assistance and engagement with the task, we find that intensive AI assistance predicts strong task performance even when engagement is low, but follow-up performance is substantially higher only when intensive AI assistance is combined with high engagement with the task.

Through what mechanism does generative AI disproportionately increase productivity among lower-education participants? A key distinction is between underlying skills and effective productivity in task execution. Formal education reflects accumulated human capital, such as abstract reasoning and experience with complex written communication. For low-education participants, the absence of these inputs acts as a binding constraint, as reflected in their lower baseline performance. By assisting participants in formulating their responses and producing the final written output, AI substitutes for these scarcer inputs and leads to larger productivity gains for this group, even though higher-education participants also benefit. In a sense, AI reduces the effective expertise required to complete the task, consistent with the task simplification mechanism emphasized in recent work (Autor and Thompson, 2025; Althoff and Reichardt, 2026; Hosseini and Lichtinger, 2026) and with the potential of AI as an "expertise-leveling" technology that may also change which workers are qualified for which jobs (Acemoglu et al., 2026; Agrawal et al., 2024; Freund and Mann, 2025). However, even with AI assistance, the task remains cognitively demanding, and par-ticipants must still evaluate, select, and integrate AI-generated content. Accordingly, despite the compression in the productivity gap, higher-education participants continue to outperform lower-education individuals on average. The chat-log analysis is consistent with this interpretation: both groups obtain substantial assistance from the tool, but higher-education participants use it somewhat more effectively across margins such as prompt detail and workflow structure. Ultimately, the labor market and distributional consequences of this task-level equalization depend on a series of factors, such as organizational context, task allocation, access to and adoption of AI by different groups (Bursztyn et al., 2025; Almog, 2025; Imas and Shukla, 2026), policy environment and equilibrium adjustments, which are beyond the scope of this study.

Our study builds on a rapidly growing literature on generative AI in labor-market settings, which generally finds larger productivity gains among workers with low baseline performance or experience. In an online experiment with 444 college-educated professionals, Noy and Zhang (2023) find that access to ChatGPT increases output quality by about 0.45 SD and reduces completion time by about 0.83 SD, with larger gains for lower baseline performers. In a lab-in-the-field experiment with 758 Boston Consulting Group consultants, Dell'Acqua et al. (2026) show that AI access raises productivity and output quality for tasks within the "jagged technological frontier" (e.g., completing 12.2% more tasks and doing so 25.1% faster, with quality gains exceeding 40%), and that improvements are larger for initially lower-performing consultants (about 43% versus 17% for top-half performers). In a field rollout of an AI assistant to 5,172 customer-support agents, Brynjolfsson et al. (2025) estimate an average 15% increase in issues resolved per hour, driven by gains of roughly 30–34% among novice or lower-skilled agents and small effects for the most experienced agents. Finally, Cui et al. (Forthcoming) study three field experiments with 4,867 software developers and report that access to an AI coding assistant increases completed tasks by 26% on average, with larger effects among less experienced developers.[3]

A smaller set of recent studies point to settings in which AI adoption instead benefits higher-ability or more experienced users. Using data from more than 160,000 software developers, Daniotti et al. (2026) find productivity gains concentrated among experienced developers, with little evidence of benefits for early-career developers. Relatedly, Kreitmeir and Raschky (2024) study the Italian ban of ChatGPT using a difference-in-differences estimation, and document experience-based heterogeneity in GitHub activity, with productivity differences widening in favor of more experienced users following the ban. In a field experiment providing AI-based business assistance to Kenyan entrepreneurs, Otis et al. (Forthcoming) find no significant average effect on revenues and profits, with gains for high-baseline performers and losses for low-baseline performers. Finally, in a university debating competition, Roldan (2024) finds larger performance gains for higher-ability participants, consistent with differences in users' ability to extract value from AI assistance.

[3] Additional studies examining the effects of AI within-occupations/firms also find larger gains among workers with low baseline performance or experience (Chen and Chan, 2024; Choi et al., 2024; Kanazawa et al., 2025; Marsal and Perkowski, 2025). For a detailed and regularly updated review of the evidence on AI and productivity, see Imas (2026).

Taken together, the emerging evidence suggests that generative AI often raises performance and compresses the productivity distribution within relatively homogeneous groups of workers, though heterogeneous effects can vary across settings and tasks. Benzell and Myers (2026) provide a model that rationalizes these mixed findings and highlights the role played by the interaction between the correlation of workers' skills across tasks and technological capabilities, allowing for the possibility of non-monotonic inequality effects as AI capabilities improve.

We make several contributions to the emerging literature on AI and labor markets. First, and most importantly, our experimental setup is explicitly designed to measure whether AI compresses performance gaps across groups with different education levels. Rather than focusing on within-occupation/firm heterogeneity in baseline performance, our design targets the between-education-group performance gap as the primary outcome. To the best of our knowledge, this is the first study to directly compare AI-assisted performance on the same task between education-based groups,[4] thereby shifting the focus from performance compression within relatively homogeneous worker populations to the effect of AI on productivity gaps across the education distribution.

Second, we contribute to the literature by distinguishing between contemporaneous AI-enabled performance gains and gains that persist once assistance is removed. To do so, we pair the main AI-assisted task with a non-AI-assisted follow-up module, allowing us to assess whether observed gains reflect delegation to the tool or whether they carry over to unassisted performance.

Third, we use chat logs to characterize the types of AI use associated with the distributional pattern behind our main result. This allows us to move beyond estimating heterogeneous treatment effects and examine how participants interact with the assistant, whether lower-education participants obtain substantial support from AI, and how differences in effective tool use help explain why the education gap narrows but does not disappear.

In summary, the empirical analysis yields four main findings. First, on whether AI narrows the education-based performance gap in the task, we find that AI raises performance for both groups, with larger gains for lower-education participants closing about three-quarters of the baseline gap. Second, on the nature of these gains, the results do not support a pure temporary-delegation interpretation: treated participants do not perform worse than controls once AI is removed, and lower-education participants retain part of their gain. Third, on when these gains carry over, intensive AI assistance predicts strong task performance even when task engagement is low, but follow-up performance is substantially higher when AI assistance is combined with sustained engagement with the task. Fourth, on why the gap narrows but does not disappear, we find that lower-education participants obtain substantial assistance from AI, while higher-education participants use the tool somewhat more effectively across several margins.

[4]Related unpublished work by Haslberger et al. (2025) also studies the effect of AI in an intrinsically heterogeneous sample—a representative sample of the UK working-age population. However, in their setting, performance without AI assistance does not differ by educational attainment. Our design focuses on tasks that generate baseline differentiation across education levels, allowing us to test whether AI narrows productivity gaps associated with formal human capital.

The rest of the paper is organized as follows. Section 2 provides details on the experimental design, and Section 3 describes the sample and estimation strategy. Section 4 reports the main results on task performance and estimates how AI changes the education-based productivity gap. Section 5 examines whether the AI-induced gains carry over once the tool is removed, including an engagement analysis distinguishing between the intensity of AI assistance and engagement with the task. Section 6 turns to the chat logs to explain why the education gap narrows substantially but does not disappear. Section 7 presents robustness checks, and Section 8 concludes.

# 2 Experimental design

## 2.1 Recruitment strategy

We conducted an online experiment in Argentina between September and November of 2025, in which participants were required to complete an incentivized task on a personal computer. We recruited participants using three reputable panel survey companies (see Appendix F.1 for more details). In addition to each panel's standard participation arrangements, participants could earn up to AR$ 25,000 (approximately US$ 18) in supermarket gift cards based on their performance in the task and some follow-up questions related to the task.[5] The experiment was designed to be completed in 20 minutes, with a maximum time limit of one hour. Participants took 21 minutes on average to complete the experiment.

[5]Participants were informed that their compensation could range from AR$ 0 to AR$ 25,000 depending on their performance in the task and follow-up questions, evaluated as in a professional setting based on content and format, with top performance receiving the full amount and poor performance receiving AR$ 0 (see Appendix F.2 for the exact wording). For simplicity and for fairness reasons, we assigned participants to one of three equally sized payment tiers (AR$ 0, AR$ 12,500, or AR$ 25,000) based on their relative performance compared to other participants within the same treatment and education group (low/high). For context, the hourly minimum wage at this time was AR$ 1,610.

We recruited low- and high-education individuals aged 25 to 45, and aimed to obtain a broadly balanced sample by age and gender within each education group. As specified in our preregistration,[6] we define low-education individuals as those with a high school diploma who have no postsecondary education or less than half of a postsecondary degree (university or non-university tertiary program), and high-education individuals as those who have completed more than half of such a program. Standard education groupings in the technology and inequality literature often treat individuals with postsecondary education short of a bachelor's degree as an intermediate "some college" category (Goldin and Katz, 2008; Acemoglu and Autor, 2011). In Argentina, non-university tertiary programs comprise higher-education tracks mostly outside universities (e.g., technical and professional institutes). Since they usually involve three years of studies or more, we include them in the high-education group. Another feature of the Argentine higher-education system is that public tertiary education is tuition-free and open admission, generating substantial heterogeneity among individuals with incomplete postsecondary education, as many enroll but complete only a small portion of their studies. We therefore distinguish between those below and above half completion, assigning them to the low- and high-education groups respectively, to obtain a more informative stratification of formal skills for our experimental design. Our results are robust to alternative classification choices, such as excluding individuals with incomplete postsecondary education altogether, or assigning all of them to either the low- or the high-education group. We obtained a sample of 1,174 individuals, 520 low-education and 654 high-education.[7]

## 2.2 Experiment structure

The experimental platform was based in a web browser. Upon entering the platform, and after consenting to participate and completing a short questionnaire on sociodemographic characteristics,[8] participants were randomly assigned by the experimental platform into either a treatment or control group. All participants were presented with the incentivized task. The main difference is that those in the treatment group had access to an AI assistant to complete the task, while those in the control group completed the same task without access to the assistant.[9] Participants were instructed not to leave the experimental platform while completing the task, and the platform recorded whether participants navigated away from the page. Treated participants interacted with an OpenAI's GPT-4.1-based AI assistant through a chat window embedded into the experimental platform, and all exchanges were recorded for subsequent analysis. Before proceeding to the task, individuals in the treatment group were presented with a brief tutorial on how to use this assistant. The tutorial was designed to familiarize treated participants with the mechanics of interacting with the assistant and with the need to provide it with the relevant instructions or context to respond appropriately, using a brief unrelated practice exercise. Importantly, the practice exercise was deliberately simple and did not provide guidance on how to analyze or solve the substantially more complex experimental task (see Appendix F.3 for details). Overall, 94% of treated participants interacted with the virtual assistant during the tutorial, and 77% completed the tutorial module (76% low- and 77% high-education).

[6]We preregistered the design of the experiment, including the definition of our education groups and detailed definitions of our outcome variables, among other specific characteristics of the experiment and the analysis. Further details available at https://www.socialscienceregistry.org/trials/16607.

[7]Appendix B.2 provides further details on recruitment and sample composition from initial contact to the final sample. Our sample is moderately unbalanced in terms of educational attainment (44% low- and 56% high-education) because the requirement of completing the task using a personal computer proved to be slightly more limiting for participation among low-education subjects. Importantly, this does not affect our results because our main coefficient of interest is the difference in treatment effects between the two education groups.

[8]Appendix B.1 depicts the different stages of the experiment, and Appendix F.2.1 presents the exact wording of the initial instructions and consent form translated from Spanish.

The experimental task presented a realistic hypothetical business scenario which consisted of an email from the participant's boss describing a problem that required the analysis of several sources of information (including text, a figure, and a table in an attached report). Participants were requested to reply to the email providing an answer to three questions to diagnose the problem and proposing a solution.[10] The task was designed to be self-contained, providing all the information needed to respond, and not requiring any specific industry or contextual knowledge. We designed three variations of the task with the same structure but set in different types of businesses: a cafeteria, a food delivery service, and a home appliances store.[11] Participants were randomly assigned to one of three task variations. The three variations of the task were carefully designed and thoroughly piloted to ensure that they were challenging for low-education participants, while still being understandable and completable without the support of the AI assistant. This design ensures that in the absence of AI access, performance reflects differences in human capital between low- and high-education participants while avoiding both ceiling and floor effects. Consistent with this, low-education participants in the control group reported higher perceived difficulty than high-education participants, but very few described the task as very difficult (see

[9]At the time of fieldwork, Qualtrics (the platform used to implement the browser-based experiment) did not natively support embedding a conversational LLM interface (such as ChatGPT) with the level of programmatic control and interaction logging required by our design. We therefore developed a custom web module to integrate the GPT-based assistant with the required experimental functionality and to log all participant-assistant interactions, as well as to monitor whether subjects left the platform page.

[10]For treated participants, the AI assistant was preloaded with the same informational report attached to the email (namely, the text, figure, and table that participants had to analyze) and initialized with a basic system prompt (see Appendix F.7 for details).

[11]The full wording of these tasks (translated from Spanish) is provided in Appendices F.4 to F.6.

Panel C of Appendix Figure A.1). At the same time, the majority of low-education participants in the control group reported some level of familiarity with the task (only 25% said it was not at all similar to their job or to other tasks they had faced before), as shown in Appendix Figure A.2.

After completing the task, all participants were asked a series of incentivized follow-up questions without access to the AI assistant, regardless of their treatment status. These questions were designed to measure short-run unassisted performance on related task content, allowing us to examine whether prior AI access affected participants' ability to articulate the root cause and recall key information once the assistant was unavailable. First, participants were asked to summarize the root cause of the problem, framed as a scenario in which their boss, who had not yet read their response email, asked about it directly. Next, they were presented with two multiple-choice questions: one assessing their interpretation of the figure from the task, and another probing factual recall (see Appendices F.4 to F.6 for details).

In the final section of the experiment, participants answered questions about their experience completing the task, their previous experience using AI, and, for those in the treatment group, their experience using the AI assistant. Participants in the control group were then given the option to access the tutorial on how to use the AI assistant and to view the task page with the assistant enabled. A large fraction (62%) of individuals in the control group expressed interest in this option, with near-identical rates across education groups (61% low education and 63% high education).

## 2.3 Outcomes of interest

Our main outcome of interest is performance in the task. We assess it on a preregistered 0–10 scale along two dimensions, content and writing. The main performance measure is an overall score, calculated as a weighted average of the two components (two thirds content, one third writing). For content, our preregistration indicates that participants receive up to 6 points for diagnosing the problem (up to 2 points for each of the questions posed in the email), and 4 points for proposing a solution (1 point for addressing the root cause, up to 2 for offering a concrete proposal, and 1 for realism). If the solution does not address the root cause, the participant receives 0 points for the entire solution subsection. For writing, we preregistered a 10-point scale divided into spelling and grammar (2 points), clarity (3 points), organization (4 points), and tone (1 point). To avoid giving high writing scores to very short incorrect responses, those under 200 characters that receive 0 points in content are also assigned 0 points in writing. In addition to the overall score, our main outcome variable, we also analyze each dimension (content and writing) separately, and construct an indicator variable for whether the participant correctly identified the root cause of the problem. All scores are standardized relative to the average and standard deviation of the score of the low-education control group. We also examine effects on task completion time, which is measured in minutes and top-coded at 20.

For the follow-up questions, the main outcome is a weighted average of the score on an open-ended question (graded on a 0–2 scale) and two binary indicators for correct answers to multiple-choice questions. We construct an overall follow-up score that assigns half the weight to the performance in the open-ended question and one quarter to each multiple-choice question. As with the task, scores are standardized relative to the low-education control group, and we analyze each component separately. Although all these outcomes were preregistered as primary, the overall task score is the natural focal outcome, with the remaining measures either direct decompositions of it or outcomes that are contingent on participants' performance in the task (such as completion time or performance in the follow-up questions). This structure follows the logic of hierarchical outcome families discussed in Calónico and Galiani (2025). Finally, we preregistered as a secondary outcome participants' perceived difficulty of the task, measured by means of a five-point Likert scale question.

We grade each participant's response to the task and the open-ended follow-up question using an AI-assisted scoring procedure powered by OpenAI's GPT-5-mini model. Building on our preregistered grading criteria, we developed a detailed task-specific grading rubric for both the task and the open-ended follow-up question. Grading is performed one response at a time. For each participant, the model is provided with the task-specific rubric, and the model returns item-level subscores aligned with this rubric.[12] We then aggregate these subscores to construct the preregistered outcomes using the preregistered weights. Appendix D.1 provides the full system prompts, the task-specific rubrics, and the aggregation procedure. Because the AI-based scoring exhibits some variation across iterations, we run the grading procedure ten times and use the average score across all iterations. Importantly, we show that our results are robust across grading iterations.

We also validate our grading procedure using alternative LLMs (Anthropic's Claude Haiku 4.5 and Google's Gemini Flash 3.0) for grading. To further assess the reliability of our AI grading approach, we requested an independent researcher not linked to this project to produce three different AI grading sets based solely on our preregistered criteria without access to our more detailed correction guidelines. Our AI-generated task scores closely match the AI-generated scores produced by the independent researcher. Finally, we had two Economics students independently grade a 10% random sample of responses manually (i.e., without using AI), and find a strong correspondence between these scores with our AI-generated grades. Importantly, both the independent researcher and the human graders were blind to participants' treatment status and education group. Overall, our AI-based grading procedure is reliable and well aligned with human judgment, as shown in detail in Section 7 below.

[12]Using an LLM for rubric-based grading is well suited to our setting, which requires consistent application of detailed criteria across a large number of open-ended responses. Recent evidence shows that modern LLMs can achieve high reliability when applying structured scoring rubrics and, in some text-annotation and coding tasks, can match or even outperform human annotators (Pack et al., 2024; Gilardi et al., 2023; Bermejo et al., 2025).

# 3 Sample, takeup, and estimation strategy

## 3.1 Sample and balance in baseline characteristics

The 1,795 participants who fulfilled the demographic criteria were randomly assigned by the experimental platform to a treatment or control group. As expected, both groups are balanced in their baseline characteristics, as indicated by the non-significance at conventional levels of the joint test of differences in characteristics (Panel A of Appendix Table A.1).

Of the 1,795 participants who met the eligibility criteria, 1,174 (65%) completed the experiment.[13] Completion rates are similar across treatment arms (64% in the treatment group and 66% in the control group, p-value=0.373) and are also balanced between treatment and control within each education group (Appendix Figure A.3).[14] In terms of observables, the balance of the original randomization remains in the analysis sample: baseline characteristics are balanced between treatment and control among completers (Panel B of Appendix Table A.1). Furthermore, finishers and non-finishers look similar on observables within each education group (Appendix Table A.2), and among completers, treatment and control remain balanced within each education group (Appendix Table A.3), with no joint differences detectable at conventional significance levels.

Appendix Table A.4 presents descriptive statistics for our low- and high-education samples. Both groups have a similar average age (approximately 35 years), and are balanced in terms of gender. In the low-education group, 64% of participants have only a high school diploma, and the remaining 36% have less than half of a postsecondary degree. In the high-education group, 29% of participants completed more than half of a postsecondary degree without finishing it, 63% completed their post-

[13] Stantcheva (2023) reports that non-completion rates in online survey experiments "tend to range between 15% and somewhat above 30%". Our non-completion rate is therefore near the upper end of this range, which likely reflects the greater demands of our setting: unlike the short information-provision survey experiments reviewed by Stantcheva (2023), participants in our study must complete a demanding job-like problem-solving and writing task.

[14] In the low-education group, the difference in completion rates between treatment and control is 0.321 percentage points; in the high-education group it is 3.681 percentage points. The corresponding p-values are 0.925 and 0.221, respectively. We also cannot reject equality of completion rates across all four treatment-by-education cells (p-value=0.531).

secondary studies, and 8% have a graduate degree. In Section 7, we show that our results are robust to dropping participants with incomplete postsecondary education, defining high education as having any postsecondary education (regardless of completion), or restricting the high-education group to participants who completed their postsecondary studies. In terms of labor-market characteristics, 78% of low-education participants are currently employed (15% are employers or self-employed, and 63% are employees), whereas 89% of the high-education participants currently work (14% as employers/self-employed, and 75% as employees). High-education individuals also report, on average, one additional year of work experience (11.6 vs. 10.3 years).

In terms of previous AI use, most participants in both education groups report having used AI before (68% low education and 77% high education), with usage mainly concentrated in occasional and constant use categories, and slightly higher overall prior use among high-education participants (Appendix Figure A.4).

## 3.2 Treatment takeup and compliance

Although treated participants were encouraged to use the assistant, its use was entirely optional, and they were informed that their rewards depended solely on their responses, regardless of whether they used the assistant or not (see Appendix F.2.2). Takeup was high, with 84% of treated participants using the virtual assistant during the task (81% low education and 86% high education).

A potential concern is that participants in the control group used AI assistance. To prevent this, we instructed all participants not to consult external sources or seek help from others at the start of the experiment and immediately before the task. To monitor compliance, we tracked whether they exited the experimental module; those who did saw a popup reminder that leaving the page was not permitted. In the control group, 31% of participants left the page during the task (vs. 19% in the treatment group). Furthermore, we took snapshots of responses every minute to detect abrupt changes or content that could indicate AI use, following Noy and Zhang (2023).

We used two complementary approaches to detect AI use for control group participants who exited the experiment page during the task. First, we searched for direct evidence of AI use such as emojis, bullet-point formatting, or excerpts of the conversation with the AI in both the snapshots and the participants' final response. Second, to detect AI use that is less apparent, we computed the largest word-count increase between consecutive snapshots for each participant. We then classified control-group participants as suspected AI users if they exited the experiment page and have a maximum word-count change comparable to that of treated participants who used the virtual assistant (see Appendix B.3 for additional details). Importantly, paraphrasing AI output to avoid detection would only evade this detection strategy if participants never paste the AI-generated text into the response box, since we observed minute-by-minute snapshots of responses rather than only the final submission. Using these two methods, we detected potential AI use for 6% of control-group participants based on direct evidence and an additional 7% based on abrupt word-count changes.

Overall, we estimate that 13% of the control-group participants were noncompliant (11% among low-education and 15% among high-education participants), implying a 70 and 71 percentage point difference in AI use between treatment and control in the low- and high-education groups, respectively (Appendix Figure A.5).

These compliance patterns matter for interpretation in two ways. First, our main estimates are intent-to-treat, so any residual control-group use of external AI pushes us toward conservative estimates relative to a counterfactual of zero access. Second, they underscore the distinction between access and adoption: even with embedded access and a guided tutorial, usage is high but not universal. In real workplaces, where access, norms, and incentives differ, adoption may be more unequal across worker types, which can attenuate or reverse the equalizing effects we identify (Imas and Shukla, 2026; Bursztyn et al., 2025; Almog, 2025).

### 3.3 Estimation strategy

The primary goal of this experiment is to evaluate how access to AI-based assistance impacts task performance and whether it reduces performance gaps between participants with high and low educational attainment. Unlike analyses that first estimate average effects and then explore heterogeneity by baseline performance, our design targets the change in the gap between low- and high-education individuals. We will thus compare the intent-to-treat effect of providing the AI tool between both education groups by estimating the following regression:

$$Y_i = \beta_0 + \beta_1 \textit{High education}_i + \beta_2 \textit{Treatment}_i + \beta_3 \textit{High education}_i \times \textit{Treatment}_i + u_i \quad (1)$$

$Y_i$ is a preregistered outcome for participant $i$. $\textit{High education}_i$ is an indicator variable that takes a value of 1 if the participant is in the high-education group (0 if the participant has low education), and $\textit{Treatment}_i$ is an indicator variable equal to 1 if the participant was assigned to the treatment group, and therefore had access to the virtual assistant while performing the task. The treatment effect in the low-education and high-education samples is given by $\beta_2$ and $\beta_2 + \beta_3$, respectively. Our coefficient of interest ($\beta_3$) measures the effect of access to AI on the performance gap between high- and low-education individuals.[15] We use heteroskedasticity-robust standard errors in

[15] An alternative estimand is the effect of AI *use* on the performance gap across education groups, which could be estimated using randomized assignment as an instrument for AI use. Our focus, however, is on the effect of access rather than usage per se. Importantly, the qualitative conclusions

all regressions, since treatment was assigned at the individual level.

# 4 Impact of AI assistance on gaps in task performance

This section presents the results of our experiment for the preregistered outcomes of the experimental task (described in detail in Section 2.3). Table 1 reports the results for the main overall score, and for the other preregistered metrics of performance in the task—namely, the subcomponents of the main performance measure, and task completion time.

## 4.1 Impact on overall task performance

We discuss first the impact of access to the AI assistant on task performance, our main outcome. Figure 1 presents the results on the impact of having access to the AI assistant on this outcome for each education group, as well as the intent-to-treat effect on the difference in performance between the two groups, which maps to the interaction term $\beta_3$ in equation (1). We also express and discuss this change as the share of the baseline control-group gap between high- and low-education participants closed by AI assistance, consistent with our main research question.

The results in Figure 1 reveal a series of important patterns. First, as expected, performing the task without access to the AI assistant is substantially more challenging for low-education participants: their high-education counterparts perform 0.548 standard deviations (SD) higher in the overall score,[16] with the gap reflecting differences in both the content and writing subcomponents (see Section 4.2).

Second, and congruent with the results in previous studies, access to AI assistance substantially increases task performance for participants of both education groups. The overall score increases by 1.242 SD for those in the low-education group, whereas the impact of the assistant for high-education subjects is 0.834 SD (both estimates are statistically significant at the 1% level).

The third pattern, central to our research question, is that the magnitude of these AI-assistant-driven gains differs significantly across education groups. The treatment effect of access to the AI assistant is 0.408 SD higher for the low-education group with respect to that of the high-education group. Crucially, this difference is statistically significant at the 1% level. These findings indicate that AI assistance not only raises overall task performance for all participants, but it does so in an equalizing manner:

are unchanged, as first-stage differences in AI use between treated and control participants are very similar across education groups, as shown in Section 3.2.

[16] As described in Section 2.3, the overall score is standardized with respect to the mean and standard deviation of the low-education participants in the control group, resulting in a null average score for this group.

the treatment closes 75% of the performance gap between low- and high-education participants with no access to the AI assistant. At the same time, the results point to an important qualification: this convergence, while relatively strong, remains incomplete. High-education individuals with access to the AI assistant still outperform their low-education counterparts by 0.139 SD, a large and economically meaningful difference (although marginally insignificant, with p-value=0.110). This remaining gap suggests that while AI can substitute for some inputs that are scarcer in the low-education group, it does not fully eliminate the role of human capital, as participants must still evaluate, select, and integrate the AI-generated content into their answer.[17]

We have so far discussed average effects, but the distribution of the scores by treatment and education group provides insights as to how the AI assistant affected performance in both groups. The different panels in Figure 2 compare the distribution of the overall task score, before standardizing, between treatment and education groups. Several interesting patterns emerge from this additional information beyond the mean. First, as shown in Panel C, the distribution of the performance in the control group for both education groups indicates that the task was well calibrated in terms of difficulty: there is substantial dispersion in scores for those without access to the AI assistant in both education groups. The absence of strong ceiling or floor effects implies that there was substantial variability in the capacity to address the task in both groups. Furthermore, as discussed earlier and as intended by design, the task is more challenging for low-education control participants. Their score distribution is visibly to the left of that of high-education subjects in the control group, and the distribution is skewed to the left—that of the high-education group seems more uniform across the range of scores. Finally, as shown in Panels A-B, the dispersion in scores for treated participants indicates that the larger gains observed among low-education participants are not mechanically driven by a general convergence from a lower baseline toward a perfect score. The AI assistance generates a pronounced rightward shift in the score distribution for both education groups, but only about 13% of treated low-education participants and about 21% of treated high-education subjects end up with a score in the top category (9.75 to 10). If the task were too easy to solve with the AI assistant, we would have expected a much stronger clustering around those levels.

[17] A final comparison helps illustrate the magnitude of the treatment effect for low-education participants. Their average score in the treatment group is more than twice that of high-education participants in the control group (1.242 SD versus 0.548 SD, difference significant at the 1% level). Thus, in our setting, access to AI assistance more than offsets the baseline performance handicap faced by low-education individuals and raises their performance above that of (non-AI-assisted) high-education participants.

## 4.2 Impact on task performance subcomponents

Besides the main overall score, we specified in our preregistration that we would also report the effects on its subcomponents and some additional related outcomes. Although these were preregistered as primary outcomes, they are components of the overall task score or outcomes that are contingent on participants' performance in the task, thereby following the logic of hierarchical outcome families discussed in Calónico and Galiani (2025). We provide a brief summary of the results for the other task performance outcomes here, with a more detailed discussion in Appendix C.1.

The two main components of the overall score are content and writing. As shown in the respective columns in Table 1, high-education participants in the control group perform significantly better than low-education participants in both writing and content. Furthermore, the equalizing effect of AI assistance operates through improvements in both dimensions of performance in our task. Taken together, these additional findings indicate that our main results are not solely driven by writing quality (which is probably the easiest performance boost from the use of the AI assistant) nor by the quality of the responses' content, which reflects more involved engagement with the information provided and the problem at hand. However, convergence remains incomplete among both dimensions. Among treated participants, high-education individuals still outperform low-education individuals by 0.122 SD in writing and 0.136 SD in content, although we are underpowered to detect these differences (p-values are 0.127 and 0.122, respectively).

We find a similar pattern for the preregistered "detected root cause" indicator, which is a component of the content score and is equal to one if the participant correctly identified the root cause of the problem in their assigned task, and zero otherwise. In the control group, high-education participants are also significantly more likely to correctly detect the root cause. Access to the AI assistant substantially raises the probability of correctly identifying the root cause for both education groups, with larger gains for low-education participants, and near-equalization across groups among treated participants.

Finally, we examine the impact of the AI assistant on the time to complete the task. Table 1 shows that access to the AI assistant reduces task completion time in both education groups from baseline levels that are, perhaps surprisingly, very similar for both types of participants in the control group (high-education control participants finish only 0.274 minutes faster than low-education participants, and this difference is not statistically significant). Among low-education subjects, time falls by 0.961 minutes relative to a control-group average of 10.697 minutes (significant at the 10% level). Among high-education participants, the reduction is larger at 1.514 minutes relative to a control-group average of 10.423 minutes (significant at the 1% level). However, the difference in treatment effects across groups is not statistically significant. These patterns imply that the faster completion time observed among treated high-education participants reflect the combination of a slightly faster baseline and a numerically larger, but statistically indistinguishable, treatment effect. The combination of both factors implies that high-education treated participants finish the task 0.827 minutes faster on average than low-education treated participants (marginally insignificant, with p-value=0.113). Overall, task duration is the only outcome for which we do not observe an equalizing pattern, and if anything, the estimates suggest modest divergence rather than convergence.[18]

Overall, these results establish the first main finding: AI substantially but incompletely narrows the education-based productivity gap while assistance is available. The next question is how to interpret these AI-induced gains. They could reflect productive use of the assistant, or mere delegation to AI-generated output with no carry-over. Section 5 examines whether the gains carry over to a non-AI-assisted follow-up module, and whether carry-over depends on participants' engagement with the task. Section 6 then uses the chat logs to explain the distributional pattern: why the gap narrows substantially, and why it does not disappear.

# 5 Performance after AI is removed

## 5.1 Carry-over of gains in the follow-up module

The purpose of this follow-up module is to examine whether the AI-induced gains observed in the main task carry over to a setting without assistance, distinguishing gains that are solely contemporaneous with AI access from gains that partly survive once the assistant is removed. As discussed in Section 2.2, all participants completed a set of follow-up questions about the content of the task and their proposed root cause of the problem after finishing the experimental task, with no access to the AI assistant.[19]

We report the results for the pre-registered overall follow-up score in Table 1. Several noteworthy patterns emerge. First, baseline performance on the follow-up questions in the control group is higher among high-education participants, who score

[18]Consistent with these objective performance gains, perceived task difficulty, a preregistered secondary outcome, also declines with AI access. As seen in Panels A and B of Appendix Figure A.1, treated participants in both education groups are about 7 percentage points more likely to report that the task was "very easy" or "easy" (statistically significant at the 10% level), with no differential effect by education. However, AI access does not eliminate education-based differences in perceived difficulty, which remain visible among treated participants (Panel D of Appendix Figure A.1).

[19]We use the term carry over rather than learning because the design does not require participants to complete a new unassisted task requiring the same underlying skills, and the follow-up is administered immediately after the task. The estimates therefore capture short-run unassisted performance after AI exposure, not durable skill formation.

0.300 SD higher than their low-education counterparts (a difference that is statistically significant at the 1% level), consistent with underlying differences in human capital.

Second, some of the performance gains from having access to the AI assistant carry over to the follow-up exercise, and we find no evidence that prior AI use harms subsequent performance once the assistant is removed for either education group. In particular, access to AI during the task leads to modest improvements in follow-up performance for low-education participants. As shown in Table 1, their follow-up score rises by 0.171 SD (significant at the 5% level). For high-education individuals, the estimated treatment effect is smaller (0.071 SD) and not statistically significant.[20]

Third, despite the positive effect of access to AI on the follow-up scores of low-education participants, they still lag behind their high-education counterparts by 0.200 SD (significant at the 5% level). While the point estimates indicate larger gains for the low-education group, the estimated difference in treatment effects across education groups of −0.100 is not statistically distinguishable from zero (p-value=0.398). In other words, the gains from prior AI use are not large enough to offset the baseline advantage of the high-education group. In fact, high-education participants in the control group score higher than low-education *treated* individuals, although this difference is not statistically significant (p-value=0.131).

Overall, the follow-up results are informative in two complementary ways. The absence of average harm rules out the strongest version of a pure-delegation interpretation in this setting, while the modest but statistically significant carry-over for lower-education participants indicates that the productivity gains are not purely contemporaneous. The persistent education gap in the follow-up nonetheless confirms that underlying human-capital differences remain important once the tool is unavailable. The next subsection sharpens this picture by exploring the patterns associated with more carry-over.

## 5.2 Engagement, delegation, and the carry-over of gains

To further interpret the follow-up results, we analyze whether the gains among treated participants reflect genuine engagement with the task, rather than merely exposure to a better AI-generated answer. This exercise is motivated by recent evidence that AI-related harms to subsequent unassisted performance are concentrated among users who delegate problem-solving rather than engage with the task (Shen and Tamkin, 2026; Liu et al., 2026). For this purpose, we conduct an exploratory analysis on the relationship between follow-up performance and two dimensions of participants' behavior during the task for participants in the treatment group. The first captures the extent to which participants were exposed to AI-generated task solutions. We measure this as the share of task components for which each treated participant requested help from the AI assistant,[21] and classify participants as above or below the treatment-group median. Participants who asked for help on a larger share of task components were more likely to have seen the assistant identify the relevant diagnosis and propose an appropriate solution. The second captures participants' own engagement with the task, which we proxy with whether the participant spent more or less time on the task than the treatment-group median.[22] Combining these two binary measures yields four groups of treated participants: high engagement/high assistance, low engagement/high assistance, high engagement/low assistance, and low engagement/low assistance.[23] This classification allows us to compare participants who were more likely to have been exposed to a correct AI-generated answer but spent relatively little time on the task (consistent with delegation to the assistant without substantial internalization of its response) with those who combined such exposure with more sustained engagement.

[20] These results are driven by the open-ended follow-up question, as there are no effects of prior AI use for either education group in the multiple choice questions. For further details, see Appendix C.2.

Table 2 shows the results of regressions of treated participants' task score and follow-up score against the binary group indicators, with participants with low task engagement and low AI assistance as the omitted category. We conduct these estimations pooling low- and high-education participants, as well as separately for each education group. As expected, participants who received high levels of AI solution assistance obtain substantially higher task scores than those who received less assistance, regardless of whether they had high or low levels of engagement with the task, with similar results for both education groups. In the pooled specification, the task scores of the high-engagement/high-assistance and low-engagement/high-assistance groups are statistically indistinguishable (1.878 and 1.972 SD, respectively; p-value for equality = 0.206). Intensive AI assistance is therefore sufficient to generate strong submitted answers, even when participants spend relatively little time engaging with the task.

The pattern is different in the follow-up module, where participants no longer had access to the assistant. Despite achieving similarly high task scores, participants with low engagement and high assistance perform substantially worse than those who combined high engagement with high assistance. In the pooled specification, the follow-up score for the high-engagement/high-assistance group is 0.732 SD, compared with 0.065 SD for the low-engagement/high-assistance group (p-value of the

[21] For details on the construction of this variable, see Section E. Treated participants who did not use the assistant are assigned a value of zero on this measure.

[22] We also tested alternative thresholds (33rd and 66th percentiles), rather than splitting at the treatment-group medians, and obtained similar results.

[23] The distribution of participants across these four groups is similar across education groups. Among low-education treated participants, 27% are in the high engagement/high assistance group, 21% in the low engagement/high assistance group, 26% in the high engagement/low assistance group, and 26% in the low engagement/low assistance group. Among high-education treated participants, the corresponding shares are 28%, 24%, 20%, and 28%, respectively.

difference<0.01). Follow-up performance is also high among participants with high engagement but low AI assistance.[24] These results are consistent with the interpretation that exposure to a correct AI-generated answer is not by itself sufficient for follow-up performance: carry-over is concentrated among participants who combined intensive AI assistance with sustained task engagement.

# 6 Patterns of AI use

As discussed in Section 4.1, AI assistance raises overall task performance for all participants, but more so for low-education individuals. Among treated participants, however, high-education individuals continue to outperform their low-education counterparts. Having examined in Section 5 whether the AI-induced gains carry over once the tool is removed, we now turn to the chat logs to explain the distributional pattern behind the main result: in particular, whether differences in how low- and high-education participants use the AI assistant help account for why the gap narrows substantially but does not disappear.[25]

We analyze the recorded conversations with the AI assistant for the 471 treated participants who used the assistant (out of a total of 562 in the combined treatment group) and classify patterns of AI use along three dimensions. First, we capture how participants use the assistant to work through the task, including both the extent to which they request assistance across task components, as well as the quality of their prompts aimed at guiding the assistant's reasoning. Second, we characterize how participants rely on the assistant in generating the final output, including the type of drafting assistance requested, the specificity of instructions provided to guide the output, and the extent to which they incorporated the AI-generated content into their task response. Third, we measure how participants structure and manage their interaction with the assistant over the course of the conversation, capturing differences in initial specificity, iterativity, and conversational structuring. We designed and implemented an LLM-based procedure to tag each conversation along these dimensions. Appendix E provides further details on these measures, along with the prompt used for this analysis.

[24] Reassuringly, the differences across these categories of participants do not seem to merely reflect selection, as the patterns presented in Table 2 are robust to controlling for participants' education group, age, gender, employment status, and work experience. In particular, the differences across the four engagement/assistance groups in both task and follow-up scores remain quantitatively similar and maintain their statistical significance.

[25] We use these measures to characterize how the tool was used and to contextualize the residual performance gap among treated participants, focusing on magnitudes and qualitative patterns rather than formal hypothesis tests. Relatedly, the correlations between AI use patterns and task scores reported in Appendix Table A.6 should be interpreted as associations, not as causal effects of particular interaction strategies on performance.

Overall, our measures of AI use correlate positively with overall task scores, and together explain almost 49% of the variation in overall scores within the group of treated participants who used the AI assistant to complete the task (see Appendix Table A.6 for details).[26]

Table 3 summarizes the average patterns of AI use, distinguishing between the low- and high-education participants who used the assistant. Overall, there are no meaningful differences between the two groups in the intensity of their interaction with the assistant. Participants in both groups send a similar number of messages (almost 3 on average) and a similar amount of text (measured by number of characters), and they request AI help in working through the task for a comparable share of task components (around 61% in both groups). The majority of participants explicitly request the assistant to generate the output: around 67% directly ask the AI to draft the response, and only a minor share ask the assistant to extend or edit a user-written draft.

Differences instead emerge along some qualitative dimensions of use. High-education participants provide more detailed instructions aimed at guiding how the assistant thinks through the response, scoring 0.174 SD higher on the index capturing this. High-education participants are slightly more likely to provide content-specific instructions for the AI assistant to generate the output, scoring 0.066 SD higher on an index measuring this. Furthermore, high-education participants better organize their interaction with the assistant, with a higher likelihood of having an initial prompt that is highly specific (6.7 percentage points), and a higher likelihood of having a structured workflow (7.5 percentage points), consistent with a more systematic orchestration of the conversation. Besides underlying differences in skills and human capital, this can also be attributed to experience: high-education participants are somewhat more likely to report that they had previously used large language models, as discussed in Section 3.1. However, notably, perceptions of the usefulness of the assistant and comfort with its use are very similar across education groups (Appendix Figure A.9).

Regarding patterns of reliance on AI-generated content in the final output, there are similarities and differences between the two groups. Most participants rely on AI-generated content in their final output: 64% of final submissions include a full copy-paste of AI-generated text, and 34% partially rely on AI-generated text; only a small minority rely exclusively on self-generated content. In this margin of partial reliance on AI-generated text, individual skills likely play a role in shaping answer quality, potentially explaining part of the residual performance gap observed between low- and high-education participants. Low-education participants are 10 percentage points more likely to fully copy and paste AI-generated text into their final answer,

[26]Specifically, in a regression of task scores on the full set of AI-use measures, the $R^2$ is approximately 0.49.

whereas high-education participants are 9.3 percentage points more likely to only partially incorporate AI-generated content, either by combining it with their own writing (6.5 percentage points more likely), or by paraphrasing AI-generated content (2.8 percentage points more likely). This pattern is consistent with higher-education participants using the assistant as an input into their own production process rather than as a substitute for it.

The overall picture is one in which differences in outcomes among treated participants are not driven primarily by the intensity of assistant use, but by how the assistant is employed and how its outputs are integrated into participants' own work. Lower-education participants, despite less formal training and somewhat less prior AI experience, obtain substantial assistance from the tool. This helps explain why AI closes so much of the baseline education gap. Higher-education participants nonetheless interact with the assistant in a more structured way, provide more specific instructions to guide its reasoning, and are more likely to use AI-generated content as an input into their own response rather than fully copying it. These patterns help explain why the education gap narrows substantially but does not disappear: effective use of AI remains partly shaped by underlying human capital.

# 7 Robustness checks

Our main results presented above are robust to a series of robustness and specification checks. The main checks refer to robustness of our findings with respect to alternative approaches to measure the overall task score, our main outcome of interest. As discussed in Section 2.3, the dependent variable was obtained by running the subjects' responses through OpenAI's GPT-5-mini model with our preregistered criteria and more detailed task-specific rubrics as inputs, and iterating 10 times to average out the inherent randomness of AI-generated scores across runs introduced by the stochastic nature of these LLMs. Appendix Figure A.6 reports the treatment effects for low- and high-education participants, as well as the difference between these effects, each with 95% confidence intervals, using every individual iteration as the final overall score. For comparison, the figure also plots the estimates based on the averaged score (our main outcome discussed so far). The results show that the estimated treatment effects and their differences are virtually identical across iterations, indicating that our conclusions are not driven by random variation in the scoring process.[27]

Secondly, and as specified in our preregistration, we requested two Economics students (one third year undergraduate, one Masters student) from the Universidad de San Andrés to independently grade a 10% random sample of our subjects' responses, representative by treatment and education group. The students were requested to manually grade the responses according to our preregistered criteria and the detailed task-specific rubric. They were blind to participants' treatment status and education group, and were not informed about the broader project design (most importantly, that there were two education groups, and that some participants had access to an AI assistant), although they were told that there could be ample variation in response quality and that they should not become suspicious nor reduce the grades of responses that were too good—as if generated by AI. Reassuringly, the correlations between the students' manual grading and our own were all above 0.9, as discussed in Appendix D.3.

[27] For further details about the consistency of grades across iterations, see Appendix D.2.

Third, we examine the robustness of our results to variations in our grading procedure, and report the results in Appendix Figure A.7. In particular, we generated alternative overall task performance gradings based on single runs of alternative state-of-the-art LLMs at the time of writing (Anthropic's Claude Haiku 4.5 and Google's Gemini Flash 3.0). While the estimated treatment effects for each education group are somewhat larger under Haiku 4.5 and somewhat smaller under Gemini Flash 3.0, the difference in treatment effects across education groups is highly stable and remains very close to our main estimate.

Fourth, we commissioned an independent fellow researcher not linked to this project and with experience in AI-based evaluations of written material to develop and execute three alternative AI-assisted grading strategies, using only the preregistered scoring structure and the task materials, but with no access to our task-specific detailed grading rubric. Importantly, the researcher had participants' answers, but was blind to their treatment status and education group. The three grading strategies are: (i) a grading approach based on an AI-generated rubric; (ii) the same rubric augmented with task-specific fact sheets to reduce ambiguity in scoring; and (iii) a relative grading approach based on pairwise comparisons aggregated using an Elo rating system.[28] See Appendix D.4 for further details of these procedures and for pointers to all the prompts used to obtain these gradings. Reassuringly, our main results are consistent throughout these alternative grading procedures, as shown in Appendix Figure A.7.[29] Specifically, using all three grading strategies, access to the AI assistant increases performance for our subjects, and it significantly narrows the performance gap between the two groups. The treatment effects appear larger when using more detailed grading criteria (i.e., the independent researcher's second more involved measure or our own, based on detailed task-specific rubrics). Perhaps most notably, we find qualitatively and quantitatively similar results with the Elo based grading, which is substantially different to the other grading strategies. While the correlation of our score to the scores from the first two alternative gradings is high (above 0.9), this might be expected given the common nature of all these measures. The correlation between our main score and the Elo-based score is still very high at 0.78, which gives us confidence that we are truly capturing some underlying latent performance.

[28]We thank Santiago Afonso, the independent researcher, for suggesting and implementing this third alternative criterion.

[29]The scores for 29 observations from the independent researcher grading are missing because these were inadvertently not sent for external grading. All of our results are robust to imputing these missing observations with the minimum, maximum, mean, or median score of the corresponding education/treatment status group.

Finally, we conduct a series of additional specification and sample robustness checks which we report in Appendix Figure A.8. First, we show the results from our main regression controlling for task fixed effects, age, gender, employment status, educational attainment, and work experience. The figure also presents the estimated effects after removing from the sample 25 responses classified as very low quality (about 2% of the sample). Finally, we present results that illustrate the robustness of our findings to alternative education-group definitions—there is substantial variation in incomplete postsecondary education in Argentina, and we establish that our results do not depend on our preferred classification. We first re-estimate the effects after excluding participants with incomplete postsecondary education. We also present results based on two alternative definitions of the high-education group: having more than a complete high school education, and having a complete postsecondary education. Across all checks, the estimated treatment effects and their difference across education groups remain stable and consistent with the main results.

# 8 Conclusion

This paper studies whether generative artificial intelligence reinforces or reduces productivity differences across individuals with substantially different levels of formal education. Using a randomized online experiment conducted outside firms, we obtain four main findings. First, on whether AI narrows the education-based performance gap, we find that AI closes about three-quarters of the baseline gap, with larger gains for lower-education participants. Second, on the nature of these gains, the follow-up results do not support a pure temporary-delegation interpretation: treated participants do not perform worse than controls once AI is removed, and lower-education participants retain part of their gain, although a sizable education gap remains. Third, on when gains carry over, intensive AI assistance predicts strong main-task performance, but follow-up performance is substantially higher when assistance is combined with sustained task engagement. Fourth, on why the gap narrows but does not disappear, lower-education participants obtain substantial assistance from AI, while higher-education participants use the tool somewhat more effectively across several margins.

Our interpretation emphasizes a distinction between underlying human capital and effective productivity in task execution. Generative AI does not eliminate the role of human capital, nor does it equalize fundamental abilities. Instead, it relaxes constraints related to problem structuring, organization, and written communication that are more binding for some individuals than for others. In this sense, AI operates as a task-level skill-leveling technology for complex but general activities. Viewed through a task-based lens, this is closely related to recent frameworks in which technology changes the expertise barriers embedded in tasks and can reduce the expertise required to complete work even when tasks remain human-performed (Autor and Thompson, 2025; Althoff and Reichardt, 2026; Hosseini and Lichtinger, 2026; Acemoglu et al., 2026). Our findings are therefore consistent with the skill-democratizing potential emphasized by Autor (2024) and Acemoglu et al. (2026).

Our experimental design deliberately abstracts from firms, wages, and organizational task allocation. This is best viewed as a trade-off rather than a limitation. By conducting the experiment outside firms, we recover the between-education-group variation that organizational data cannot cleanly identify, and isolate a task-level capability effect under equal access and basic onboarding. Whether such capability gains translate into narrower inequality in practice depends not only on what workers can do with AI, but also on who adopts it and under what conditions. These considerations connect to broader questions about whether experimental findings scale to policy-relevant settings (Al-Ubaydli et al., 2020). Recent evidence suggests that adoption responds to social interactions and perceived peer use (Bursztyn et al., 2025; Almog, 2025), and that AI usage is already more common among more educated workers and among managers and professional workers (Imas and Shukla, 2026). Our own results add a further dimension: even under equal access, the quality of AI use varies with education. Accordingly, we do not interpret our estimates as direct predictions about wage effects or aggregate equilibrium inequality. Our results are specific to a self-contained workplace-style task that combines reading comprehension, problem diagnosis, and written communication. Tasks of this kind fall within the part of the jagged frontier where current AI is capable of substituting for execution-oriented inputs (Dell'Acqua et al., 2026). The equalizing pattern documented here may attenuate or reverse on tasks that rely more heavily on interpersonal coordination or other skills that current AI cannot easily replicate.

These results suggest a broader view of AI's impact on inequality. AI may be strongly equalizing at the level of assisted task execution under equal access, while leaving substantial room for inequality to persist or even widen through differences in adoption, engagement, and effective use. Our findings are specific to a particular generation of AI capabilities. The equalizing pattern reflects a model that is highly capable at the inputs most binding for lower-education users (structuring information, organizing arguments, producing fluent prose). As capabilities expand, the relative gains for the two groups could shift, potentially attenuating or reversing the pattern. Tracking how distributional effects evolve with AI capabilities is an important direction for future work, and one that the literature on AI inequality has only begun to address (Benzell and Myers, 2026).

In that sense, the central question is not only whether AI can substitute for scarce inputs on a fixed task, but also whether institutions, organizations, and policy environments will make that capability broadly accessible and productively usable. Understanding how the task-level effects documented here interact with adoption frictions, organizational design, policy initiatives and equilibrium adjustments remains an important direction for future research.

# References

Acemoglu, Daron (2024) "The Simple Macroeconomics of AI," *Economic Policy*, 40 (121), 13–58.

Acemoglu, Daron and David Autor (2011) "Skills, Tasks and Technologies: Implications for Employment and Earnings," in Ashenfelter, Orley and David Card eds. *Handbook of Labor Economics*, 4, 1043–1171: Elsevier.

Acemoglu, Daron, David Autor, and Simon Johnson (2026) "Building Pro-Worker Artificial Intelligence," NBER Working Paper No. 34854.

Agrawal, Ajay, Joshua Gans, and Avi Goldfarb (2024) "The Turing Transformation: Artificial Intelligence, Intelligence Augmentation, and Skill Premiums," *Harvard Data Science Review* (Special Issue 5).

Al-Ubaydli, Omar, John A. List, and Dana Suskind (2020) "2017 Klein Lecture: The Science Of Using Science: Toward An Understanding Of The Threats To Scalability," *International Economic Review*, 61 (4), 1387–1409.

Almog, David (2025) "Barriers to AI Adoption: Image Concerns at Work," *arXiv preprint arXiv:2511.18582*.

Althoff, Lukas and Hugo Reichardt (2026) "Task-Specific Technical Change and Comparative Advantage," Working Paper.

Anderson, Michael L. (2008) "Multiple inference and gender differences in the effects of early intervention: A reevaluation of the Abecedarian, Perry Preschool, and Early Training Projects," *Journal of the American Statistical Association*, 103 (484), 1481–1495.

Autor, David H. (2024) "Applying AI to Rebuild Middle Class Jobs," NBER Working Paper No. 32140.

Autor, David and Neil Thompson (2025) "Expertise," *Journal of the European Economic Association*, 23 (4), 1203–1271.

Bastani, Hamsa, Osbert Bastani, Alp Sungu, Haosen Ge, Özge Kabakcı, and Rei Mariman (2025) "Generative AI Without Guardrails Can Harm Learning: Evidence from High School Mathematics," *Proceedings of the National Academy of Sciences*, 122 (26), e2422633122.

Benzell, Seth Gordon and Kyle R Myers (2026) "Automation Experiments and Inequality," NBER Working Paper No. 34668.

Bermejo, Vicente J., Andrés Gago, Ramiro H. Gálvez, and Nicolás Harari (2025) "LLMs Outperform Outsourced Human Coders on Complex Textual Analysis," *Scientific Reports*, 15 (1), 40122.

Brynjolfsson, Erik, Danielle Li, and Lindsey Raymond (2025) "Generative AI at Work," *Quarterly Journal of Economics*, 140 (2), 889–942.

Bursztyn, Leonardo, Alex Imas, Rafael Jiménez-Durán, Aaron Leonard, and Christopher Roth (2025) "Social Dynamics of AI Adoption," NBER Working Paper No. 34488.

Calónico, Sebastian and Sebastian Galiani (2025) "Beyond Bonferroni: Hierarchical Multiple Testing in Empirical Research," NBER Working Paper No. 34050.

Chen, Zenan and Jason Chan (2024) "Large Language Model in Creative Work: The Role of Collaboration Modality and User Expertise," *Management Science*, 70 (12), 9101–9117.

Choi, Jonathan H., Amy B. Monahan, and Daniel Schwarcz (2024) "Lawyering in the Age of Artificial Intelligence," *Minnesota Law Review*, 109, 147.

Cui, Zheyuan, Mert Demirer, Sonia Jaffe, Leon Musolff, Sida Peng, and Tobias Salz (Forthcoming) "The Effects of Generative AI on High-Skilled Work: Evidence from Three Field Experiments with Software Developers," *Management Science*.

Daniotti, Simone, Johannes Wachs, Xiangnan Feng, and Frank Neffke (2026) "Who is Using AI to Code? Global Diffusion and Impact of Generative AI," *Science*, 0 (0), eadz9311.

Dell'Acqua, Fabrizio, Edward McFowland, Ethan Mollick et al. (2026) "Navigating the Jagged Technological Frontier: Field Experimental Evidence of the Effects of Artificial Intelligence on Knowledge Worker Productivity and Quality," *Organization Science*, 37 (2), 403–423.

Freund, Lukas and Lukas Mann (2025) "Job Transformation, Specialization, and the Labor Market Effects of AI," RFBerlin Discussion Paper Series 25117, ROCKWOOL Foundation Berlin (RFBerlin).

Gilardi, Fabrizio, Meysam Alizadeh, and Maël Kubli (2023) "ChatGPT Outperforms Crowd Workers for Text-Annotation Tasks," *Proceedings of the National Academy of Sciences*, 120 (30), e2305016120.

Goldin, Claudia and Lawrence F. Katz (2008) *The Race Between Education and Technology*, Cambridge, MA: Harvard University Press.

Harrison, Glenn W. and John A. List (2004) "Field Experiments," *Journal of Economic Literature*, 42 (4), 1009–1055.

Haslberger, Matthias, Jane Gingrich, and Jasmine Bhatia (2025) "No Great Equalizer: Experimental Evidence on AI in the UK Labor Market," Working Paper.

Hosseini, Seyed M. and Guy Lichtinger (2026) "Generative AI, Expertise, and Effective Labor Supply," Working Paper.

Imas, Alex (2026) "What Is the Impact of AI on Productivity?," Substack post, available at https://aleximas.substack.com/p/what-is-the-impact-of-ai-on-productivity.

Imas, Alex and Soumitra Shukla (2026) "Who Uses AI (and How)?," Substack post, available at https://aleximas.substack.com/p/who-uses-ai-and-how.

Kanazawa, Kyogo, Daiji Kawaguchi, Hitoshi Shigeoka, and Yasutora Watanabe (2025) "AI, Skill, and Productivity: The Case of Taxi Drivers," *Management Science*.

Kreitmeir, David and Paul Raschky (2024) "The Heterogeneous Productivity Effects of Generative AI: Evidence from Italy's ChatGPT Ban," Working Paper.

Liu, Grace, Brian Christian, Tsvetomira Dumbalska, Michiel A. Bakker, and Rachit Dubey (2026) "AI Assistance Reduces Persistence and Hurts Independent Performance," *arXiv preprint arXiv:2604.04721*.

Marsal, Ales and Patryk Perkowski (2025) "Generative AI as Routine-Biased Technical Change? Evidence from a Field Experiment in Central Banking," Working Paper.

Noy, Shakked and Whitney Zhang (2023) "Experimental Evidence on the Productivity Effects of Generative Artificial Intelligence," *Science*, 381 (6654), 187–192.

Otis, Nicholas, Rowan Clarke, Solène Delecourt, David Holtz, and Rembrand Koning (Forthcoming) "The Uneven Impact of Generative AI on Entrepreneurial Performance," *Management Science*.

Pack, Austin, Alex Barrett, and Juan Escalante (2024) "Large Language Models and Automated Essay Scoring of English Language Learner Writing: Insights into Validity and Reliability," *Computers and Education: Artificial Intelligence*, 6, 100234.

Roldan, Antonio (2024) "When GenAI Increases Inequality: Evidence from a University Debating Competition," POID Working Paper 096, Centre for Economic Performance, London School of Economics and Political Science.

Shen, Judy Hanwen and Alex Tamkin (2026) "How AI Impacts Skill Formation," *arXiv preprint arXiv:2601.20245*.

Stantcheva, Stefanie (2023) "How to Run Surveys: A Guide to Creating Your Own Identifying Variation and Revealing the Invisible," *Annual Review of Economics*, 15 (1), 205–234.

# Tables and Figures

Table 1: Main results

| | Task | | | | | |
|---|---|---|---|---|---|---|
| | Overall (z-score) | Writing (z-score) | Content (z-score) | Detected root cause (0/1) | Duration (minutes) | Follow-up (z-score) |
| **Low education** | | | | | | |
| $\overline{Y}_L^C$ | 0.000 | 0.000 | 0.000 | 0.495*** | 10.697*** | 0.000 |
| | (0.060) | (0.060) | (0.060) | (0.029) | (0.367) | (0.060) |
| $\overline{Y}_L^T$ | 1.242*** | 1.201*** | 1.166*** | 0.841*** | 9.736*** | 0.171*** |
| | (0.065) | (0.059) | (0.066) | (0.022) | (0.397) | (0.063) |
| $\overline{Y}_L^T - \overline{Y}_L^C$ | 1.242*** | 1.201*** | 1.166*** | 0.346*** | -0.961* | 0.171** |
| | (0.088) | (0.084) | (0.089) | (0.036) | (0.541) | (0.087) |
| **High education** | | | | | | |
| $\overline{Y}_H^C$ | 0.548*** | 0.503*** | 0.525*** | 0.647*** | 10.423*** | 0.300*** |
| | (0.061) | (0.056) | (0.062) | (0.025) | (0.319) | (0.058) |
| $\overline{Y}_H^T$ | 1.382*** | 1.323*** | 1.302*** | 0.881*** | 8.909*** | 0.371*** |
| | (0.059) | (0.054) | (0.059) | (0.018) | (0.339) | (0.056) |
| $\overline{Y}_H^T - \overline{Y}_H^C$ | 0.834*** | 0.820*** | 0.777*** | 0.234*** | -1.514*** | 0.071 |
| | (0.084) | (0.078) | (0.085) | (0.031) | (0.465) | (0.080) |
| **High vs. Low education** | | | | | | |
| $\overline{Y}_H^C - \overline{Y}_L^C$ | 0.548*** | 0.503*** | 0.525*** | 0.152*** | -0.274 | 0.300*** |
| | (0.086) | (0.082) | (0.086) | (0.038) | (0.486) | (0.083) |
| $\overline{Y}_H^T - \overline{Y}_L^T$ | 0.139 | 0.122 | 0.136 | 0.040 | -0.827 | 0.200** |
| | (0.087) | (0.080) | (0.088) | (0.028) | (0.522) | (0.084) |
| $(\overline{Y}_H^T - \overline{Y}_H^C) - (\overline{Y}_L^T - \overline{Y}_L^C)$ | -0.408*** | -0.382*** | -0.389*** | -0.112** | -0.553 | -0.100 |
| | (0.122) | (0.114) | (0.123) | (0.048) | (0.713) | (0.118) |
| Observations | 1174 | 1174 | 1174 | 1174 | 1174 | 1174 |

Note: The first three columns report the results of estimations where the dependent variable is the task score (average across 10 grading iterations), standardized relative to the low-education control group: overall score (column 1), writing score (column 2), and content score (column 3). In column 4, the dependent variable is the average across ten grading iterations of a dummy for whether the respondent detected the root cause of the problem. In column 5, the dependent variable is time taken to complete the task (in minutes). In the last column, the dependent variable is the overall score in the follow-up questions, standardized relative to the low-education control group. The top and central panels show the mean of the control and treatment group and the difference between the treatment and control group for the low- and high-education groups, respectively. The bottom panel shows the difference between control and treatment means for the high- and low-education groups, as well as the difference between the treatment effects of both education groups. * significant at 10%; ** significant at 5%; *** significant at 1%.

Table 2: Relationship between scores and AI assistance and task engagement

| | Task (z-score) | | | Follow-up (z-score) | | |
|---|---|---|---|---|---|---|
| | Pooled | Low education | High education | Pooled | Low education | High education |
| High engagement + High assistance | 1.324$^{***}$ | 1.286$^{***}$ | 1.351$^{***}$ | 0.886$^{***}$ | 0.743$^{***}$ | 0.992$^{***}$ |
| | (0.098) | (0.145) | (0.132) | (0.106) | (0.168) | (0.134) |
| Low engagement + High assistance | 1.418$^{***}$ | 1.342$^{***}$ | 1.469$^{***}$ | 0.219 | 0.254 | 0.193 |
| | (0.097) | (0.148) | (0.129) | (0.112) | (0.176) | (0.146) |
| High engagement + Low assistance | 0.372$^{**}$ | 0.244 | 0.502$^{**}$ | 0.644$^{***}$ | 0.591$^{***}$ | 0.724$^{***}$ |
| | (0.118) | (0.169) | (0.165) | (0.114) | (0.170) | (0.154) |
| Constant | 0.554$^{***}$ | 0.555$^{***}$ | 0.553$^{***}$ | -0.154$^{*}$ | -0.235 | -0.096 |
| | (0.082) | (0.122) | (0.110) | (0.077) | (0.122) | (0.100) |
| Observations | 562 | 244 | 318 | 562 | 244 | 318 |
| $R^2$ | 0.356 | 0.356 | 0.359 | 0.130 | 0.091 | 0.172 |
| *P-values* | | | | | | |
| High/High=Low/High | 0.206 | 0.629 | 0.229 | 0.000 | 0.005 | 0.000 |
| High/High=High/Low | 0.000 | 0.000 | 0.000 | 0.029 | 0.357 | 0.069 |
| Low/High=High/Low | 0.000 | 0.000 | 0.000 | 0.000 | 0.055 | 0.001 |

Note: This table reports the results of estimations where the dependent variable is the overall score in the task (columns 1-3) and follow-up questions (columns 4-6), standardized relative to the low-education control group. We conduct these regressions pooling low- and high-education participants, as well as separately for each group, as specified in the column headers. The regressors are indicator variables for four mutually exclusive groups defined by whether participants spent above- or below-median time on the task (engagement) and whether they required above- or below-median AI solution assistance. Task time is measured relative to the treatment-group median. AI solution assistance is measured as the share of task components for which the participant requested help from the AI assistant. The omitted category is participants with below-median time on the task and below-median AI solution assistance. * significant at 10%; ** significant at 5%; *** significant at 1%.

Table 3: AI use by education level

| | Low education | High education | Difference |
|---|---|---|---|
| **Number and length of messages sent** | | | |
| Number of messages sent to AI assistant | 2.813 | 2.971 | 0.158 |
| Total characters sent to AI assistant | 794.172 | 847.396 | 53.224 |
| **Assistance with working through the task** | | | |
| Share of task components with any AI assistance | 0.608 | 0.615 | 0.007 |
| Detailed instructions for thinking through task (std. index) | -0.000 | 0.174 | 0.174* |
| **Assistance with generating the output** | | | |
| Explicitly requested AI to draft answer | 0.702 | 0.652 | -0.050 |
| Requested AI to extend or complete user output | 0.020 | 0.040 | 0.020 |
| Requested AI to edit user output | 0.035 | 0.022 | -0.013 |
| No explicit request for AI assistance for output generation | 0.242 | 0.286 | 0.043 |
| Detailed instructions for drafting response (std. index) | -0.000 | 0.066 | 0.066 |
| **Use of AI output** | | | |
| Final output full copy/paste of AI content | 0.702 | 0.601 | -0.101** |
| Final output partial copy/paste of AI content | 0.242 | 0.308 | 0.065 |
| Final output paraphrases AI content | 0.045 | 0.073 | 0.028 |
| Final output does not rely on AI content | 0.010 | 0.018 | 0.008 |
| **Workflow/orchestration of the conversation** | | | |
| High initial specificity | 0.288 | 0.355 | 0.067 |
| Iterative engagement | 0.465 | 0.473 | 0.008 |
| Structured workflow | 0.141 | 0.216 | 0.075** |
| Observations | | | |

Notes: This table reports mean values of different measures of AI use for participants with low (column 1) and high education (column 2), all of whom used the AI assistant to complete the task. Column 3 reports the difference between both groups. All variables are constructed from detailed logs of interactions with the AI assistant during the task. Measures of assistance with thinking through the response capture the share of task components (diagnostic questions and solution) for which participants requested AI help. The index for detailed instructions for thinking through the task captures the degree to which participants requested disciplined, evidence-based reasoning from the assistant, and provided relevant contextual information. Measures of assistance with generating the output classify the type of drafting assistance requested (generate, extend, or edit); each category is coded as a binary indicator. The index on detailed instructions for drafting the response captures the extent to which participants guided the writing process through stylistic and content-specific constraints. Use of AI-generated output classifies final submissions according to whether they incorporate AI-generated text through full copy-paste, partial copy-paste, paraphrasing, or no direct use; these categories are also coded as indicator variables. Workflow measures capture three dimensions of conversation orchestration using indicator variables that measure whether the first message provides sufficient context and direction (initial specificity), whether participants actively refine the interaction beyond a single exchange (iterative engagement), and whether they deliberately distribute their request across multiple substantive turns (structured workflow). All indices are standardized relative to the low-education group. See Appendix E for further details on the classification methodology. * significant at 10%; ** significant at 5%; *** significant at 1%.

Figure 1: Effects of AI assistance on overall task score by education level

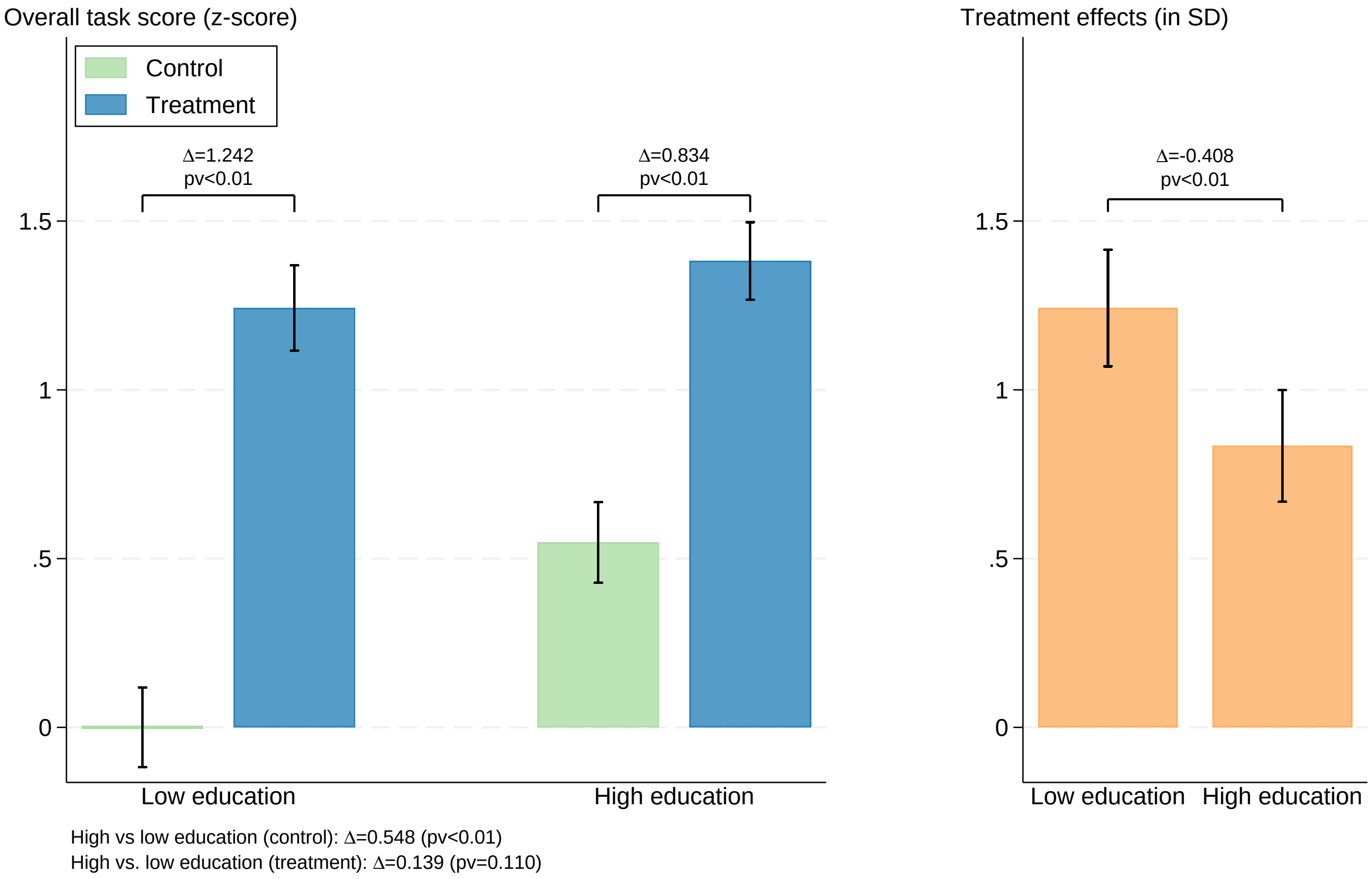

Notes: The left panel shows the average score on the task (average across 10 grading iterations), standardized relative to the low-education control group, along with 95% confidence intervals, separately for high- and low-education participants in the treatment and control groups. Above each education group, we report the difference between the treatment and control groups within that education category. The right panel presents the estimated treatment effects for high- and low-education participants. Above, we report the difference between the treatment effects of high- and low-education participants.

Figure 2: Distribution of overall task score by treatment and education group

(a) Low-education sample: treatment vs. control

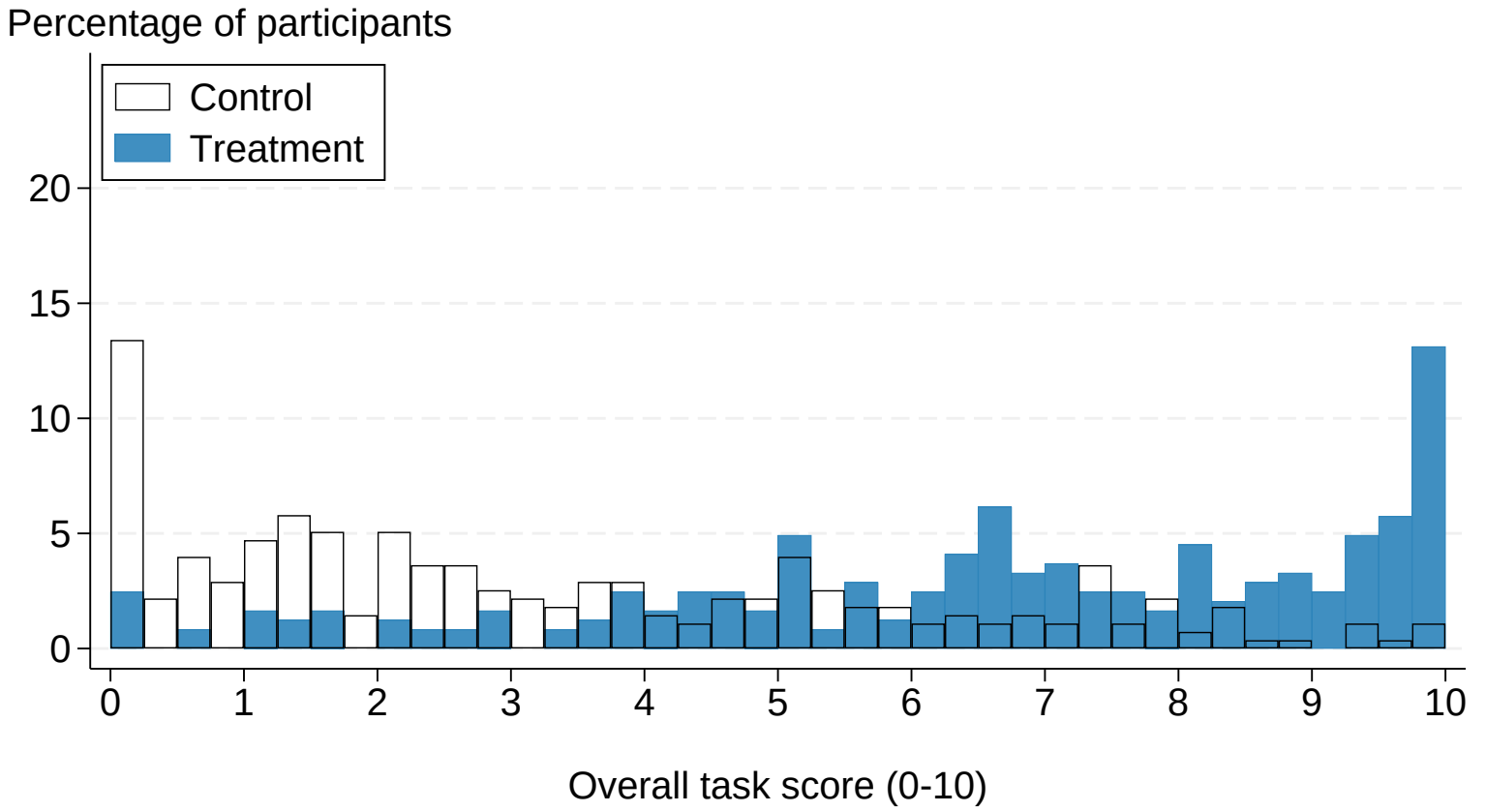

(b) High-education sample: treatment vs. control

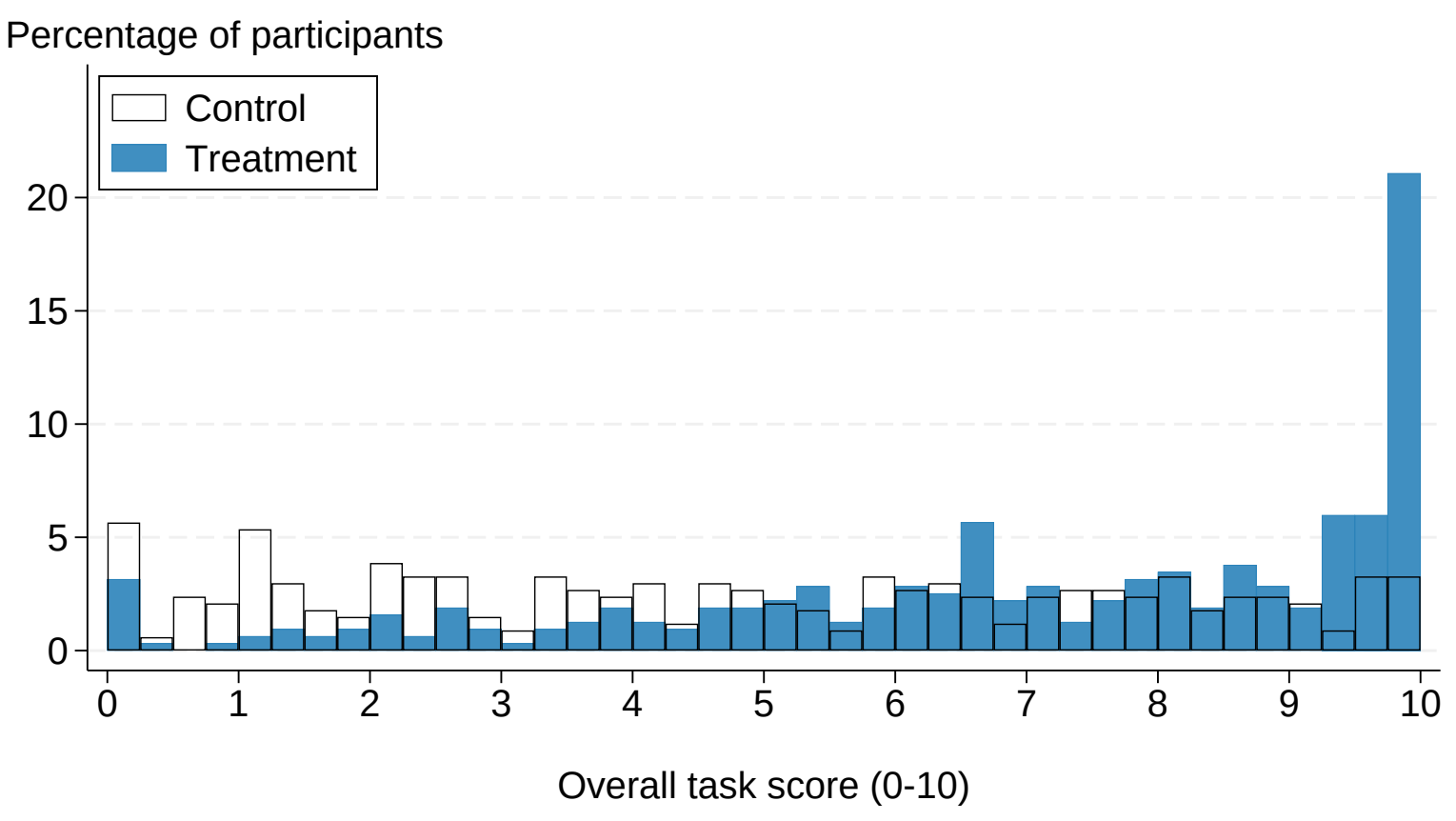

(c) Control group: high- vs. low-education

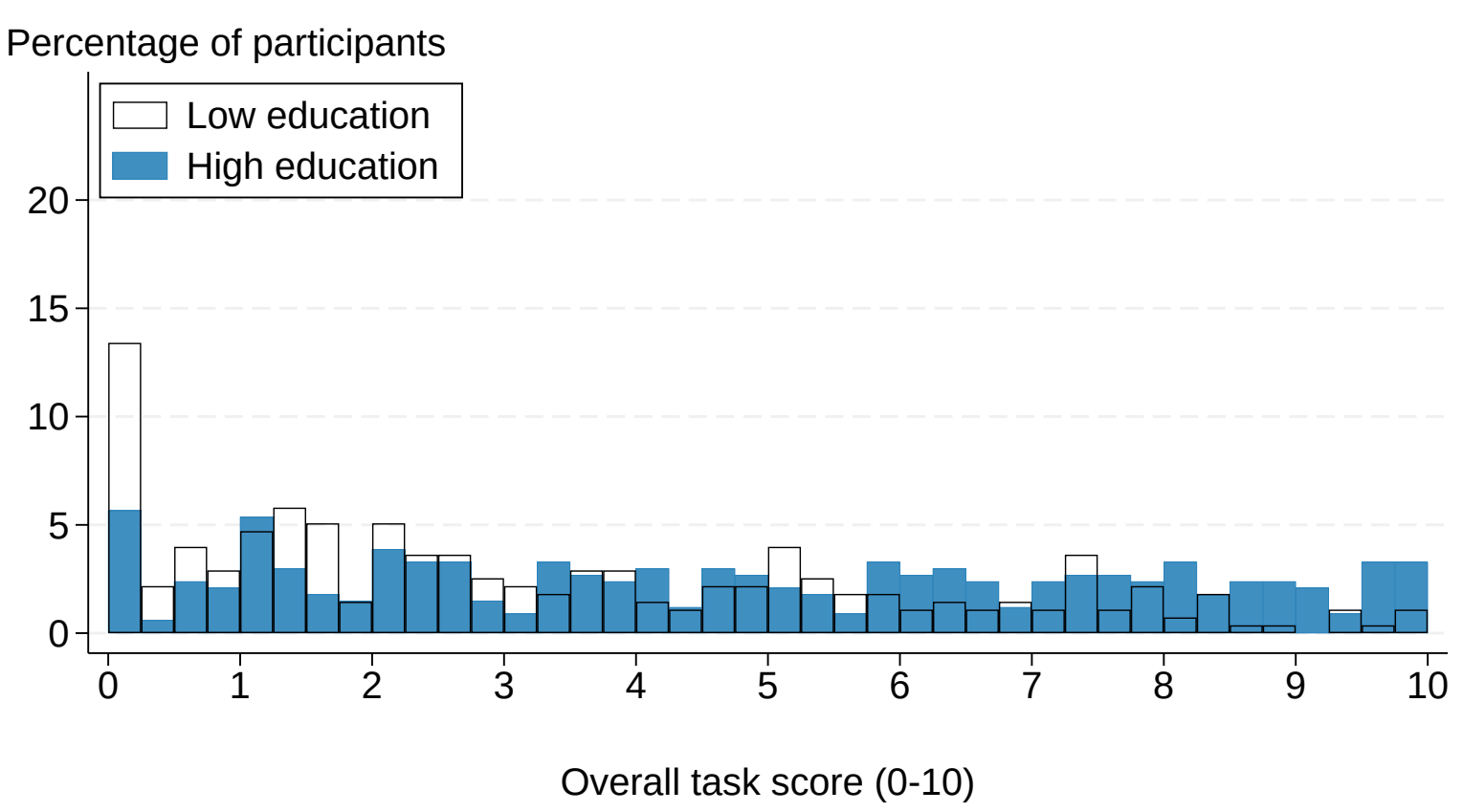

(d) Treatment group: high- vs. low-education

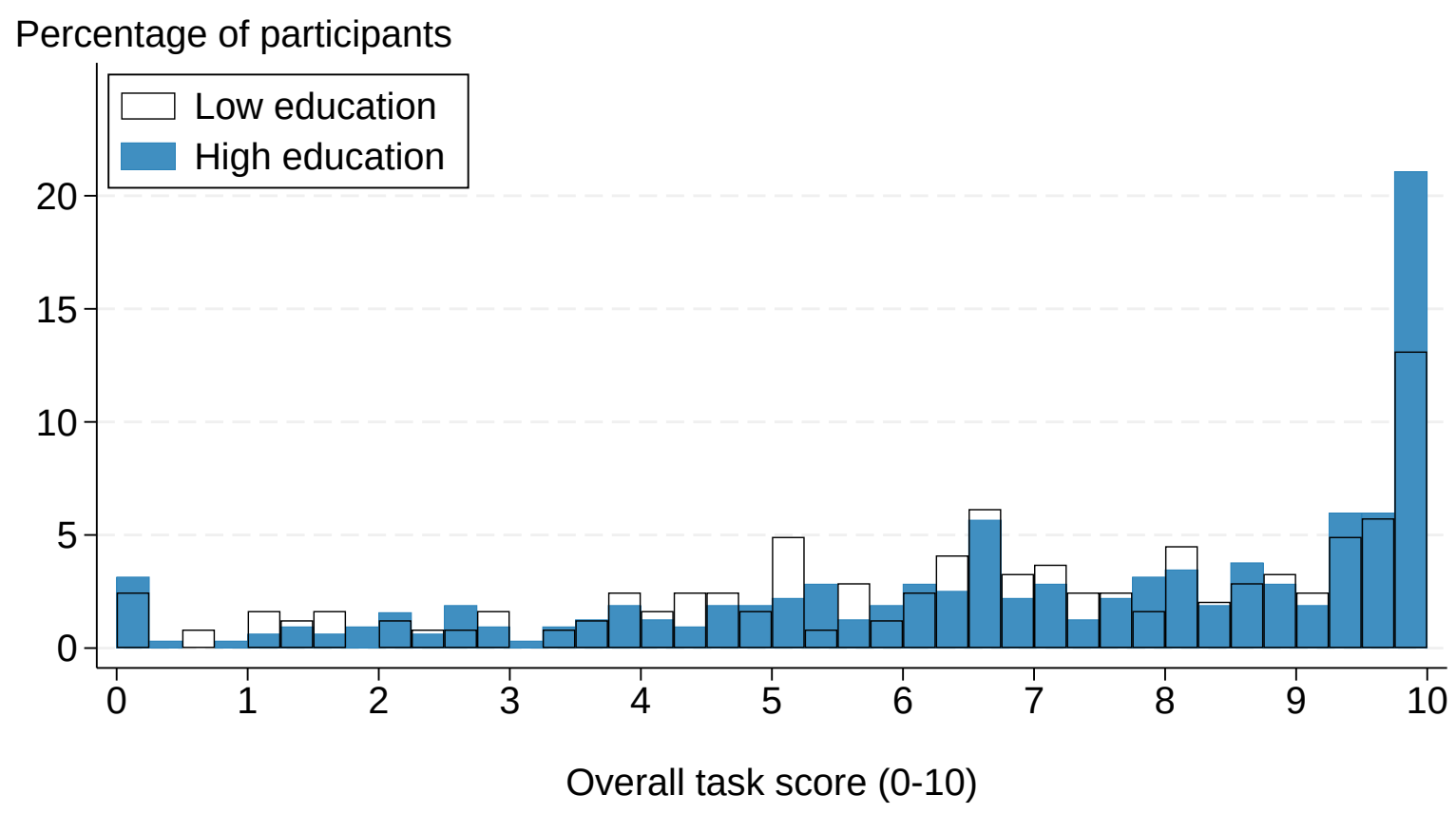

Notes: The top figures show the distribution of overall task score (on a 0 to 10 scale, prior to standardizing) by treatment group for the sample of low-education (left) and high-education individuals (right). The bottom figures show the analogous distribution by education group for the sample of control group (left) and treatment group individuals (right).

# Online Appendix for: "Does generative AI narrow education-based productivity gaps? Evidence from a randomized experiment"

# A Appendix Tables and Figures

Table A.1: Balance between treatment and control

| Variable | Control | Treatment | Difference | P-value (diff.) |
|---|---|---|---|---|
| *Panel A: Initial sample* | | | | |
| Age | 35.234 | 35.239 | 0.005 | 0.985 |
| Female | 0.511 | 0.501 | -0.010 | 0.664 |
| Male | 0.485 | 0.493 | 0.008 | 0.743 |
| Another gender | 0.004 | 0.007 | 0.003 | 0.474 |
| High school | 0.311 | 0.249 | -0.063*** | 0.003 |
| Less than half postsecondary | 0.140 | 0.171 | 0.031* | 0.072 |
| More than half postsecondary | 0.151 | 0.170 | 0.019 | 0.279 |
| Complete postsecondary | 0.349 | 0.367 | 0.017 | 0.445 |
| Graduate degree | 0.049 | 0.045 | -0.004 | 0.678 |
| Self employed or employer | 0.155 | 0.136 | -0.019 | 0.259 |
| Employee | 0.692 | 0.702 | 0.010 | 0.638 |
| Not employed | 0.153 | 0.162 | 0.009 | 0.618 |
| Years of work experience | 10.980 | 11.257 | 0.276 | 0.376 |
| Observations | 922 | 873 | | |
| P-value (joint significance) | | | | 0.172 |
| *Panel B: Sample that finished experiment* | | | | |
| Age | 35.124 | 34.945 | -0.179 | 0.601 |
| Female | 0.490 | 0.504 | 0.013 | 0.648 |
| Male | 0.503 | 0.491 | -0.012 | 0.677 |
| Another gender | 0.007 | 0.005 | -0.001 | 0.789 |
| High school | 0.306 | 0.260 | -0.046* | 0.082 |
| Less than half postsecondary | 0.145 | 0.174 | 0.029 | 0.177 |
| More than half postsecondary | 0.154 | 0.167 | 0.014 | 0.525 |
| Complete postsecondary | 0.346 | 0.359 | 0.013 | 0.641 |
| Graduate degree | 0.049 | 0.039 | -0.010 | 0.410 |
| Self employed or employer | 0.144 | 0.141 | -0.003 | 0.875 |
| Employee | 0.696 | 0.705 | 0.009 | 0.750 |
| Not employed | 0.160 | 0.155 | -0.005 | 0.803 |
| Years of work experience | 10.853 | 11.215 | 0.362 | 0.341 |
| Observations | 612 | 562 | | |
| P-value (joint significance) | | | | 0.616 |

Notes: This table shows the average characteristics at baseline of the control (column 1) and treatment (column 2) groups. Column 3 reports the difference between both groups, and column 4 reports the p-value of the difference in means test. In Panel A, the analysis is conducted for the entire sample, and in Panel B, for the sample that finished the experiment.

Table A.2: Comparison of participants that finished and did not finish the experiment

| | Finished experiment | | | |
|---|---|---|---|---|
| | No | Yes | Difference | P-value (diff.) |
| *Panel A: Low education* | | | | |
| Age | 35.767 | 34.852 | -0.915** | 0.039 |
| Female | 0.557 | 0.494 | -0.063* | 0.096 |
| Male | 0.439 | 0.502 | 0.063* | 0.096 |
| Another gender | 0.004 | 0.004 | 0.000 | 0.995 |
| High school | 0.653 | 0.640 | -0.012 | 0.734 |
| Less than half postsecondary | 0.347 | 0.360 | 0.012 | 0.734 |
| Self employed or employer | 0.137 | 0.150 | 0.013 | 0.634 |
| Employee | 0.611 | 0.633 | 0.022 | 0.551 |
| Not employed | 0.252 | 0.217 | -0.035 | 0.286 |
| Years of work experience | 10.626 | 10.310 | -0.316 | 0.548 |
| Observations | 262 | 520 | | |
| P-value (joint significance) | | | | 0.260 |
| *Panel B: High education* | | | | |
| Age | 35.499 | 35.187 | -0.312 | 0.418 |
| Female | 0.499 | 0.498 | -0.000 | 0.997 |
| Male | 0.496 | 0.494 | -0.002 | 0.953 |
| Another gender | 0.006 | 0.008 | 0.002 | 0.690 |
| More than half postsecondary | 0.276 | 0.287 | 0.012 | 0.692 |
| Complete postsecondary | 0.635 | 0.633 | -0.002 | 0.948 |
| Graduate degree | 0.089 | 0.080 | -0.010 | 0.601 |
| Self employed or employer | 0.164 | 0.136 | -0.028 | 0.234 |
| Employee | 0.749 | 0.754 | 0.005 | 0.874 |
| Not employed | 0.086 | 0.110 | 0.024 | 0.218 |
| Years of work experience | 11.760 | 11.596 | -0.164 | 0.697 |
| Observations | 359 | 654 | | |
| P-value (joint significance) | | | | 0.913 |

Notes: This table compares the average characteristics at baseline of participants who did not complete and who completed the entire experiment. Column 3 reports the difference between both groups, and column 4 reports the p-value of the difference in means test. Panel A presents statistics for the low-education group, and Panel B for the high-education group.

Table A.3: Balance between treatment and control by education level – sample that finished experiment

| | Control | Treatment | Difference | P-value (diff.) |
|---|---|---|---|---|
| *Panel A: Low education* | | | | |
| Age | 34.797 | 34.914 | 0.117 | 0.821 |
| Female | 0.478 | 0.512 | 0.034 | 0.440 |
| Male | 0.518 | 0.484 | -0.035 | 0.433 |
| Another gender | 0.004 | 0.004 | 0.000 | 0.931 |
| High school | 0.678 | 0.598 | -0.079* | 0.061 |
| Less than half postsecondary | 0.322 | 0.402 | 0.079* | 0.061 |
| Self employed or employer | 0.149 | 0.152 | 0.003 | 0.922 |
| Employee | 0.627 | 0.639 | 0.013 | 0.768 |
| Not employed | 0.225 | 0.209 | -0.016 | 0.667 |
| Years of work experience | 9.931 | 10.738 | 0.807 | 0.176 |
| Observations | 276 | 244 | | |
| P-value (joint significance) | | | | 0.436 |
| *Panel B: High education* | | | | |
| Age | 35.393 | 34.969 | -0.424 | 0.355 |
| Female | 0.500 | 0.497 | -0.003 | 0.936 |
| Male | 0.491 | 0.497 | 0.006 | 0.883 |
| Another gender | 0.009 | 0.006 | -0.003 | 0.698 |
| More than half postsecondary | 0.280 | 0.296 | 0.016 | 0.655 |
| Complete postsecondary | 0.631 | 0.635 | 0.004 | 0.910 |
| Graduate degree | 0.089 | 0.069 | -0.020 | 0.341 |
| Self employed or employer | 0.140 | 0.132 | -0.008 | 0.771 |
| Employee | 0.753 | 0.755 | 0.002 | 0.959 |
| Not employed | 0.107 | 0.113 | 0.006 | 0.805 |
| Years of work experience | 11.610 | 11.582 | -0.028 | 0.954 |
| Observations | 336 | 318 | | |
| P-value (joint significance) | | | | 0.959 |

Notes: This table shows the average characteristics at baseline of the control (column 1) and treatment (column 2) groups, for the sample that finished the experiment. Column 3 reports the difference between both groups, and column 4 reports the p-value of the difference in means test. Panel A presents statistics for the low-education group, and Panel B for the high-education group.

Table A.4: Descriptive statistics by education group

| | Low education | High education |
|---|---|---|
| *Demographics* | | |
| Age | 34.852 | 35.187 |
| Female | 0.494 | 0.498 |
| Male | 0.502 | 0.494 |
| Another gender | 0.004 | 0.008 |
| *Educational attainment* | | |
| High school | 0.640 | 0.000 |
| Less than half postsecondary | 0.360 | 0.000 |
| More than half postsecondary | 0.000 | 0.287 |
| Complete postsecondary | 0.000 | 0.633 |
| Graduate degree | 0.000 | 0.080 |
| *Labor market characteristics* | | |
| Self employed or employer | 0.150 | 0.136 |
| Employee | 0.633 | 0.754 |
| Not employed | 0.217 | 0.110 |
| Years of work experience | 10.310 | 11.596 |
| Observations | 520 | 654 |

Notes: This table shows the average characteristics of the low- and high-education participants.

Table A.5: Effect on follow-up scores

| | Overall (z-score) | Open ended (z-score) | Multiple choice (z-score) |
|---|---|---|---|
| **Low education** | | | |
| $\overline{Y}_L^C$ | 0.000 | 0.000 | 0.000 |
| | (0.060) | (0.060) | (0.060) |
| $\overline{Y}_L^T$ | 0.171*** | 0.237*** | -0.078 |
| | (0.063) | (0.061) | (0.069) |
| $\overline{Y}_L^T - \overline{Y}_L^C$ | 0.171** | 0.237*** | -0.078 |
| | (0.087) | (0.086) | (0.091) |
| **High education** | | | |
| $\overline{Y}_H^C$ | 0.300*** | 0.260*** | 0.208*** |
| | (0.058) | (0.057) | (0.057) |
| $\overline{Y}_H^T$ | 0.371*** | 0.378*** | 0.133** |
| | (0.056) | (0.055) | (0.060) |
| $\overline{Y}_H^T - \overline{Y}_H^C$ | 0.071 | 0.118 | -0.075 |
| | (0.080) | (0.079) | (0.083) |
| **High vs. Low education** | | | |
| $\overline{Y}_H^C - \overline{Y}_L^C$ | 0.300*** | 0.260*** | 0.208** |
| | (0.083) | (0.083) | (0.083) |
| $\overline{Y}_H^T - \overline{Y}_L^T$ | 0.200** | 0.141* | 0.211** |
| | (0.084) | (0.082) | (0.091) |
| $(\overline{Y}_H^T - \overline{Y}_H^C) - (\overline{Y}_L^T - \overline{Y}_L^C)$ | -0.100 | -0.119 | 0.003 |
| | (0.118) | (0.116) | (0.124) |
| Observations | 1174 | 1174 | 1174 |

Note: This table reports the results of estimations where the dependent variable is the score in the follow-up questions, standardized relative to the low-education control group: overall follow-up score (column 1), average across ten grading iterations of the open-ended question score (column 2), and multiple-choice score (column 3). The top and central panels show the mean of the control group and the difference between the treatment and control group for the low- and high-education groups, respectively. The bottom panel shows the difference between control and treatment means for the high- and low-education groups, as well as the difference between the treatment effects of both education groups. * significant at 10%; ** significant at 5%; *** significant at 1%.

## Table A.6: Overall task score and AI use

| | (1) | (2) | (3) | (4) | (5) | (6) |
|---|---|---|---|---|---|---|
| **Thinking through response** | | | | | | |
| Share of task components with any AI assistance | 1.495*** | | | | | |
| | (0.089) | | | | | |
| Detailed instructions for thinking through task (std. index) | | 0.250*** | | | | |
| | | (0.039) | | | | |
| **Generating the output** | | | | | | |
| Explicitly requested AI to draft answer | | | 0.569*** | | | |
| | | | (0.095) | | | |
| Requested AI to extend or complete user output | | | 0.328 | | | |
| | | | (0.242) | | | |
| Requested AI to edit user output | | | 0.586** | | | |
| | | | (0.246) | | | |
| Detailed instructions for drafting response (std. index) | | | | 0.173*** | | |
| | | | | (0.029) | | |
| **Use of AI output** | | | | | | |
| Final output full copy/paste of AI content | | | | | 1.672*** | |
| | | | | | (0.577) | |
| Final output partial copy/paste of AI content | | | | | 1.560*** | |
| | | | | | (0.579) | |
| Final output paraphrases AI content | | | | | 0.923 | |
| | | | | | (0.591) | |
| **Workflow and orchestration of conversation** | | | | | | |
| High initial specificity | | | | | | 0.577*** |
| | | | | | | (0.078) |
| Iterative engagement | | | | | | 0.289*** |
| | | | | | | (0.080) |
| Structured workflow | | | | | | 0.311*** |
| | | | | | | (0.095) |
| Observations | 471 | 471 | 471 | 471 | 471 | 471 |
| $R^2$ | 0.422 | 0.105 | 0.082 | 0.043 | 0.090 | 0.151 |

Notes: This table reports the result of several regressions where the dependent variable is the average score on the task (average across 10 grading iterations), standardized relative to the low-education control group. The sample is composed of treatment group participants who interacted with the AI assistant. All regressors are constructed from detailed logs of interactions with the AI assistant during the task. Measures of assistance with thinking through the response capture the share of task components (diagnostic questions and solution) for which participants requested AI help, as well as an index capture the degree to which participants requested disciplined, evidence-based reasoning from the assistant, and provided relevant contextual information. Measures of assistance with generating the output classify the way in which the participant requested type of drafting assistance requested (generate, extend, or edit), with no explicit assistance for drafting as the omitted category. The index measuring detailed instructions for drafting response captures the extent to which participants guided the writing process through stylistic and content-specific constraints. Use of AI-generated output are binary indicators of whether the final response incorporates AI-generated text through full copy-paste, partial copy-paste, or paraphrasing (with no direct use as the omitted category). The workflow measures are indicator variables that capture three dimensions of conversation orchestration: whether the first message provides sufficient context and direction (initial specificity), whether participants actively refine the interaction beyond a single exchange (iterative engagement), and whether they deliberately distribute their request across multiple substantive turns (structured workflow). All indices are standardized relative to the low-education group. See Appendix E for further details on the classification methodology. significant at 10%; ** significant at 5%; *** significant at 1%.

Figure A.1: Perceived task difficulty by treatment and education group

(a) Low-education sample: treatment vs. control

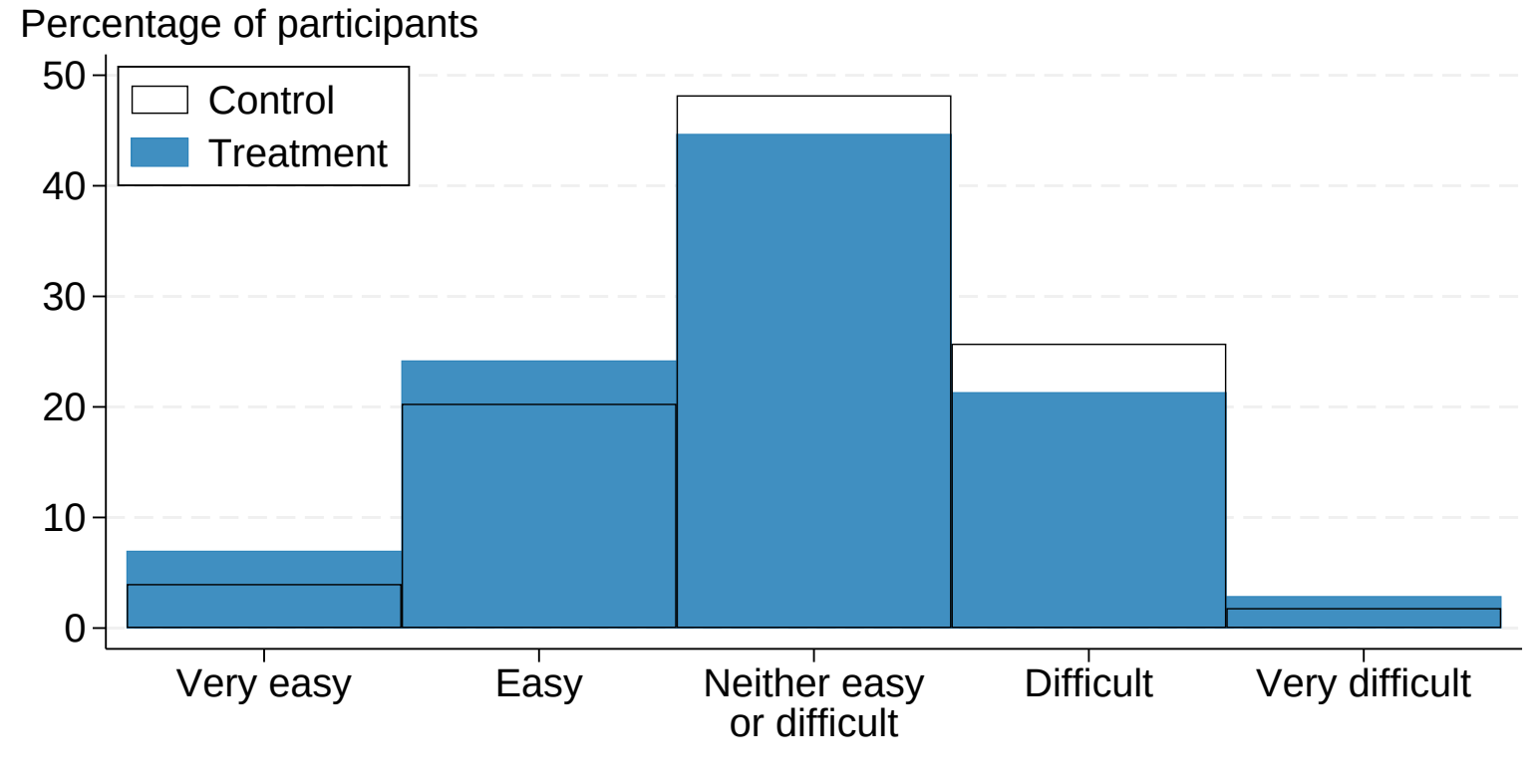

(b) High-education sample: treatment vs. control

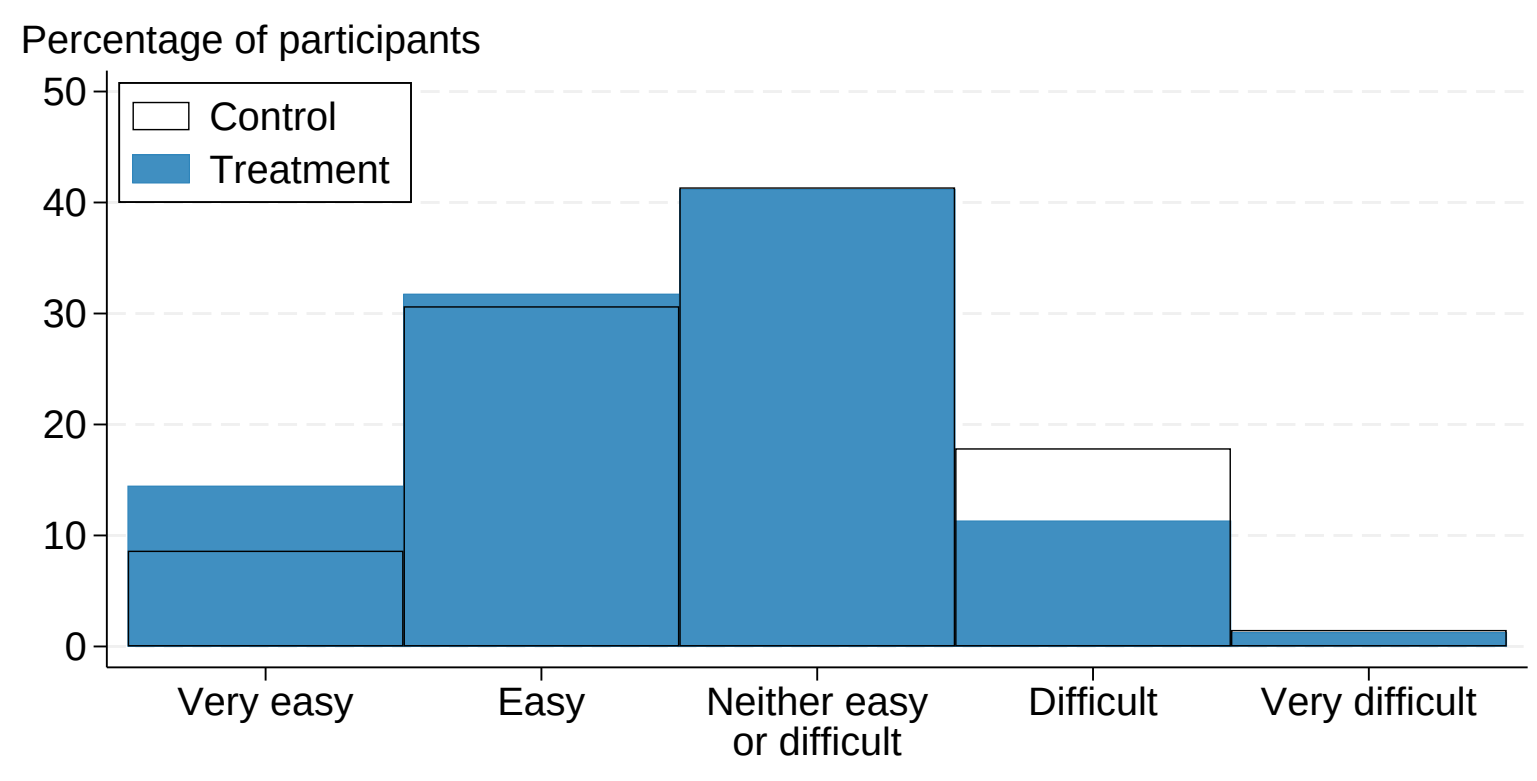

(c) Control group: high- vs. low-education

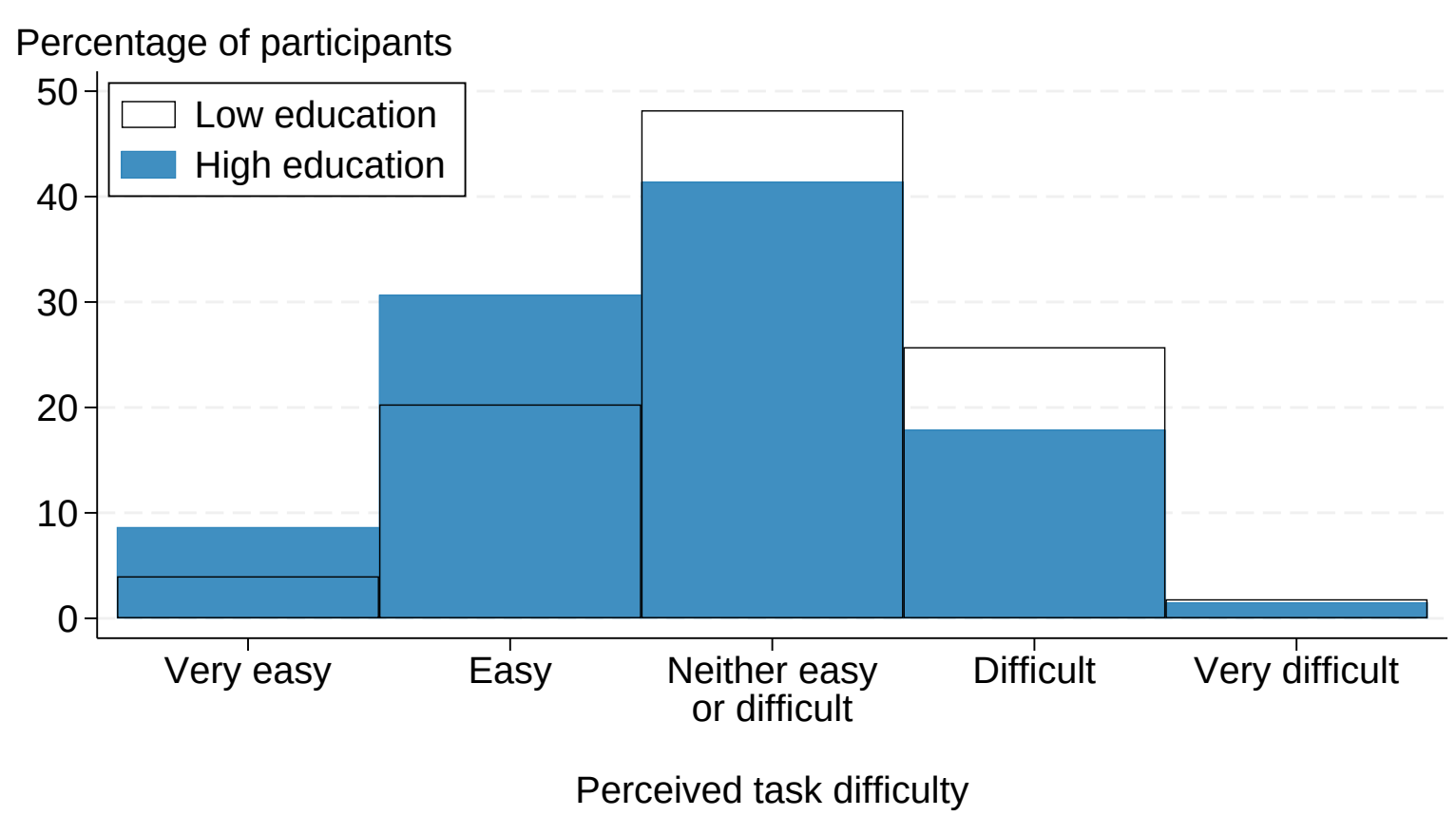

(d) Treatment group: high- vs. low-education

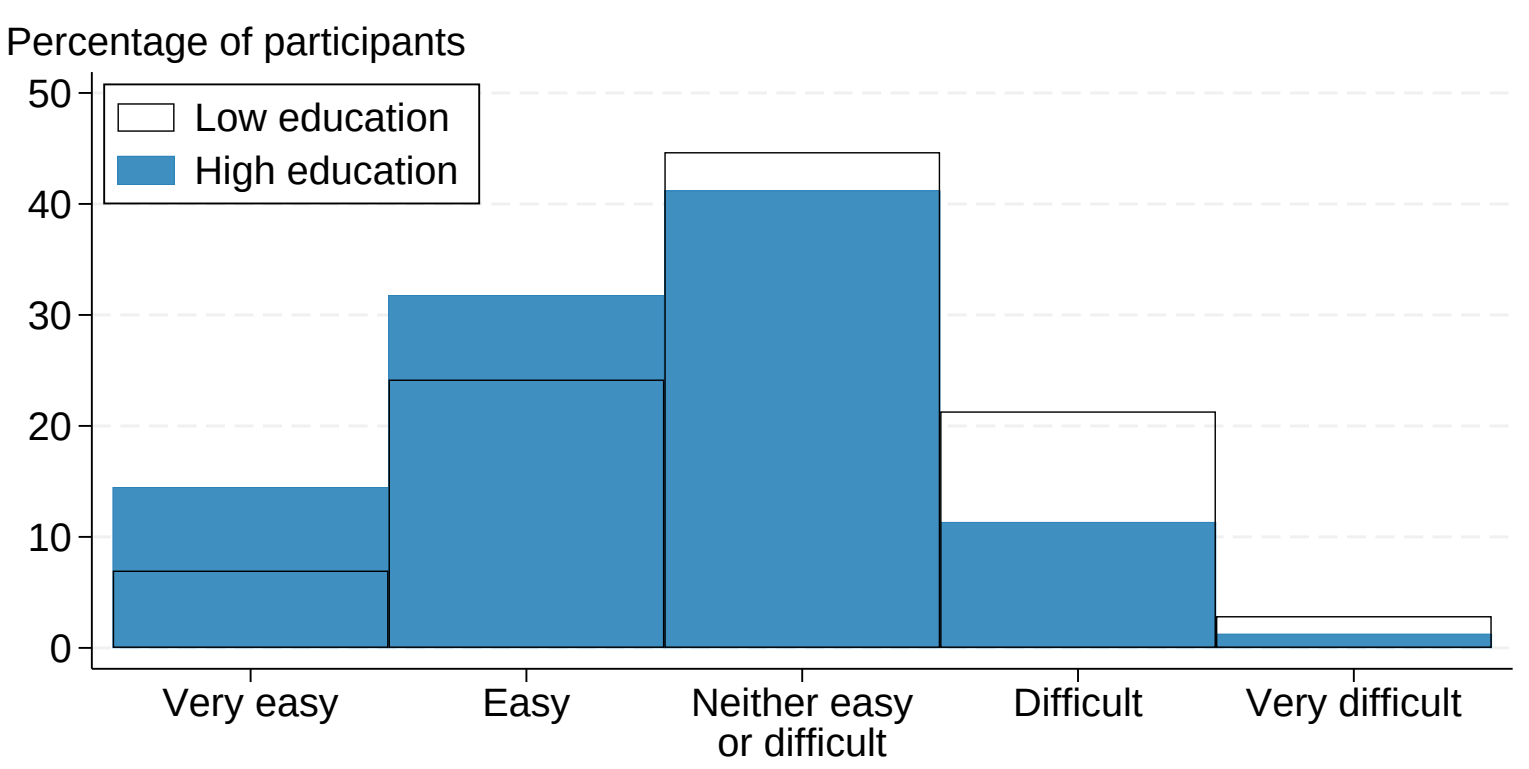

Notes: The top figures show the distribution of the perceived difficulty of the task by treatment group for the sample of low-education (left) and high-education individuals (right). The bottom figures show the analogous distribution by education group for the sample of control group (left) and treatment group individuals (right).

Figure A.2: Familiarity with the task in the control group by education level

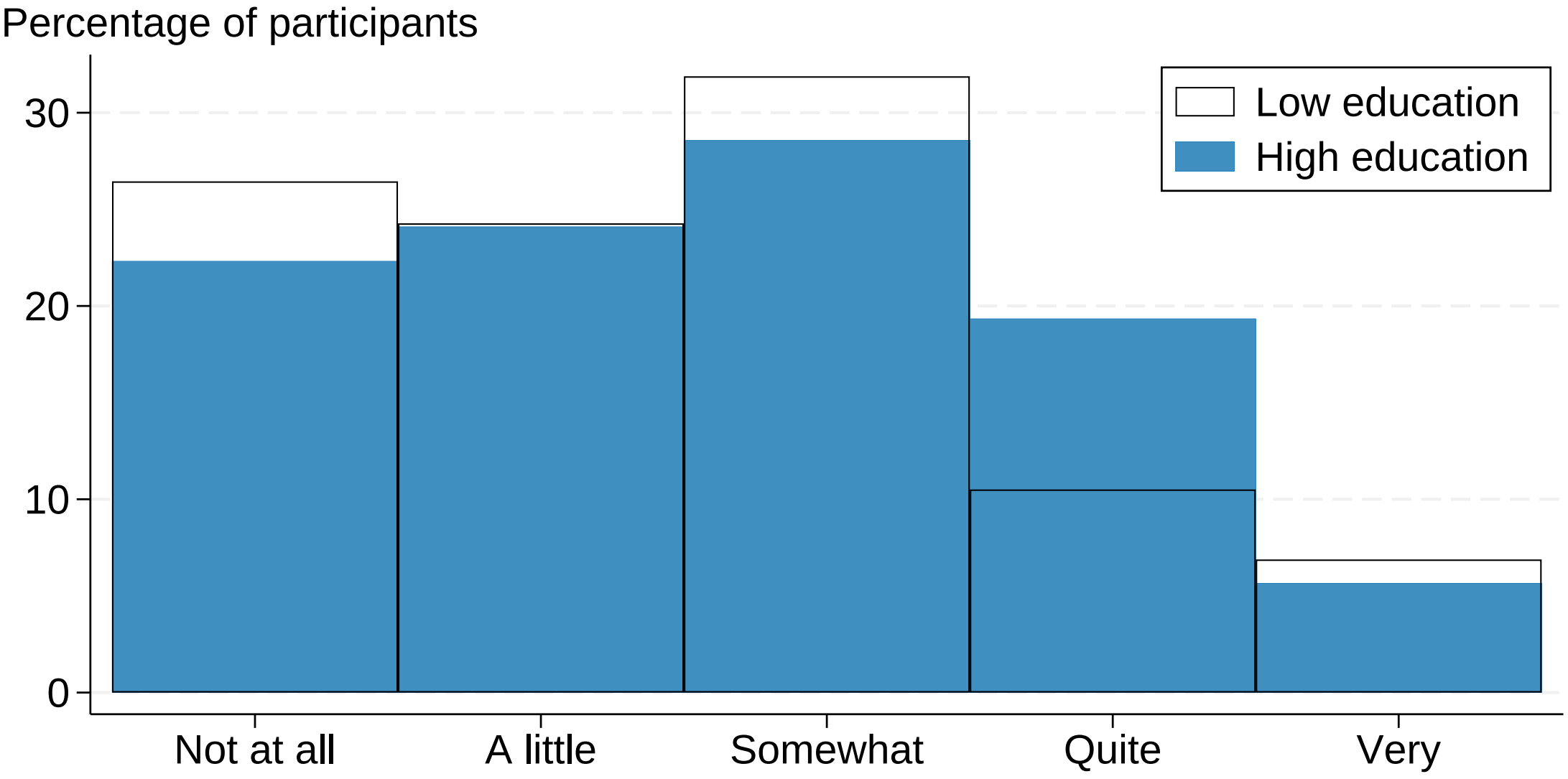

Notes: This figure shows the distribution of participants' familiarity with the task by education level for the sample of individuals in the control group.

Figure A.3: Experiment completion rate by education and treatment group

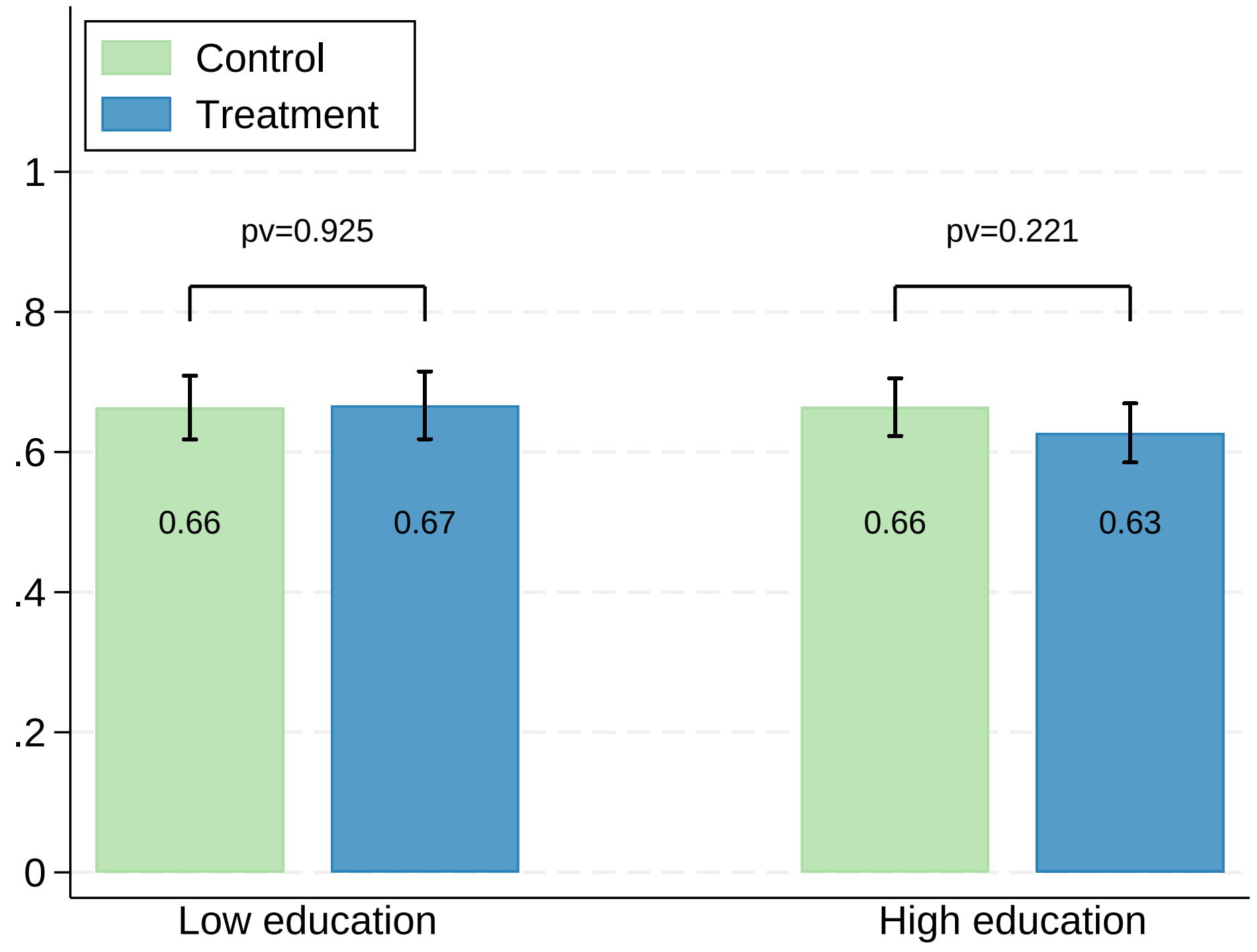

Notes: This figure shows the share of participants that finished the experiment, for each education and treatment group, along with the 95% confidence intervals. Above each education group, we report the p-value for the difference between the treatment and control groups within that education category.

Figure A.4: Frequency of previous AI use in the control group by education level

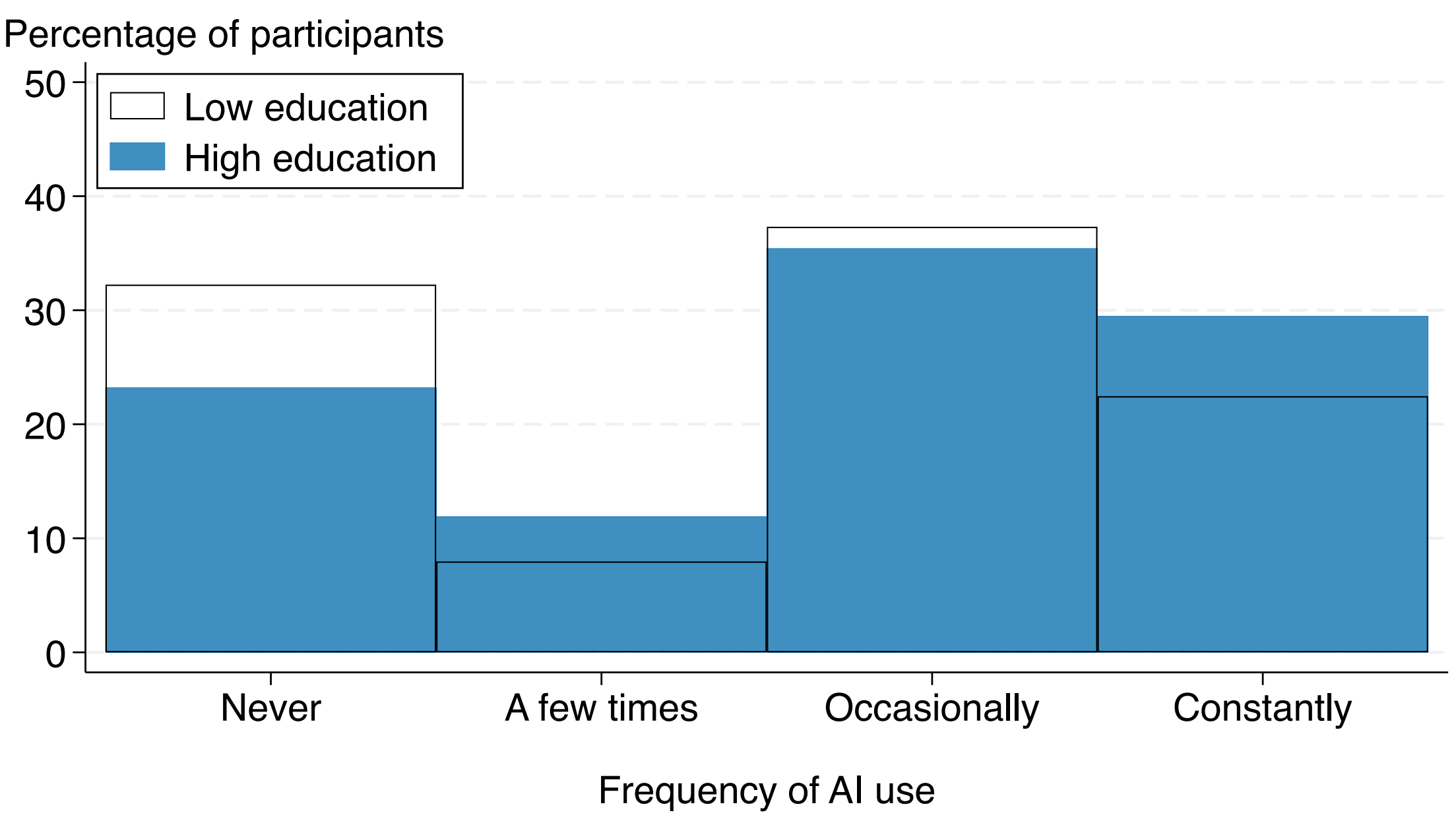

Notes: This figure plots the frequency of prior AI use for low- and high-education participants in the control group.

Figure A.5: AI use by education and treatment group

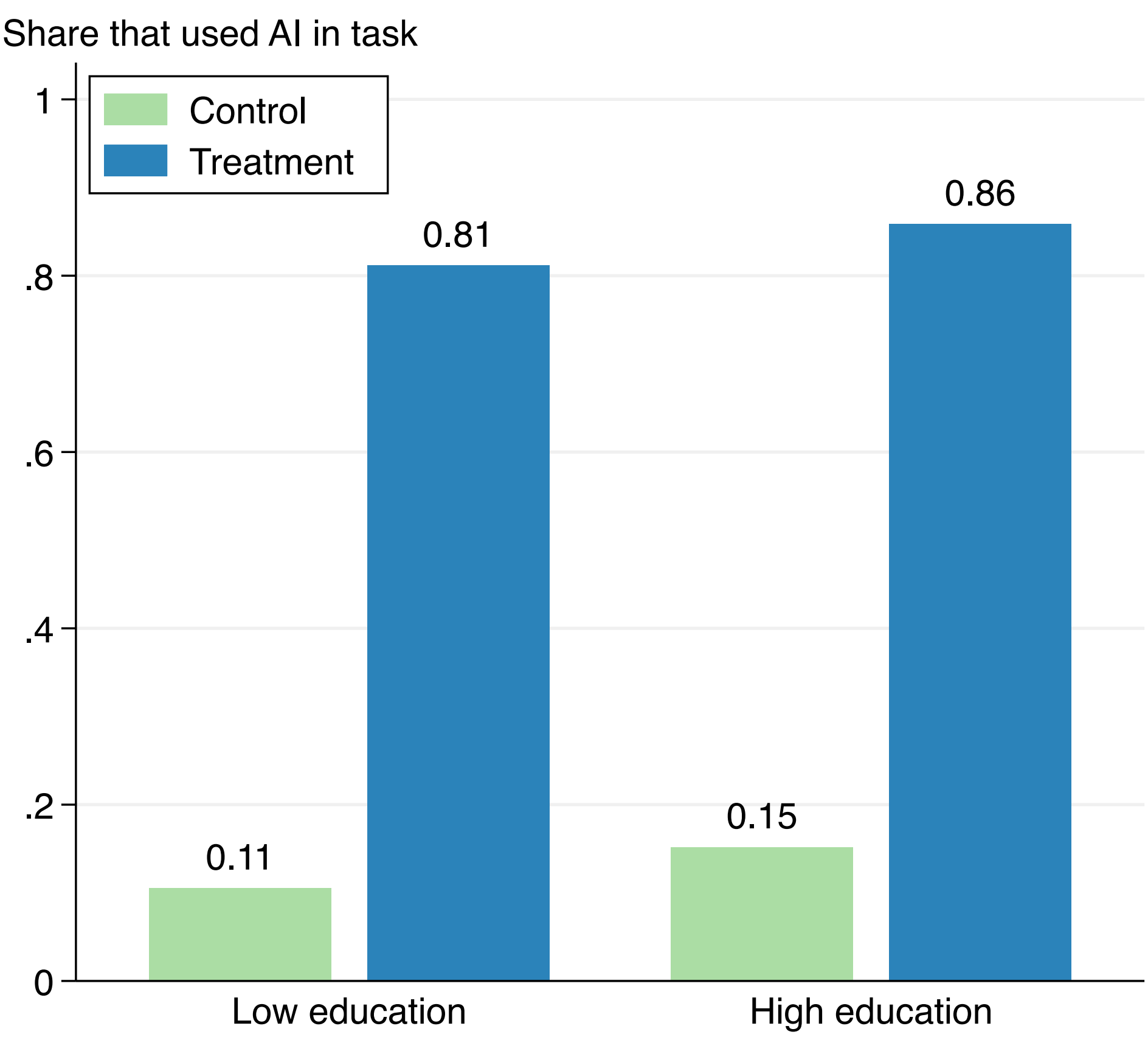

Notes: This figure shows the share of participants that used AI to complete the experimental task, for each education and treatment group. In the treatment group, a respondent is considered to have used AI if they had at least one interaction with the virtual assistant. In the control group, we assume a respondent used AI if they exited the page during the task and there is direct evidence of AI use in their final responses or snapshots of their response, or if they exhibited a maximum word count between consecutive snapshots of their task response larger than 75 words (the 25th percentile among treated AI users).

Figure A.6: Treatment effects by education group – Sensitivity of results to grading iterations

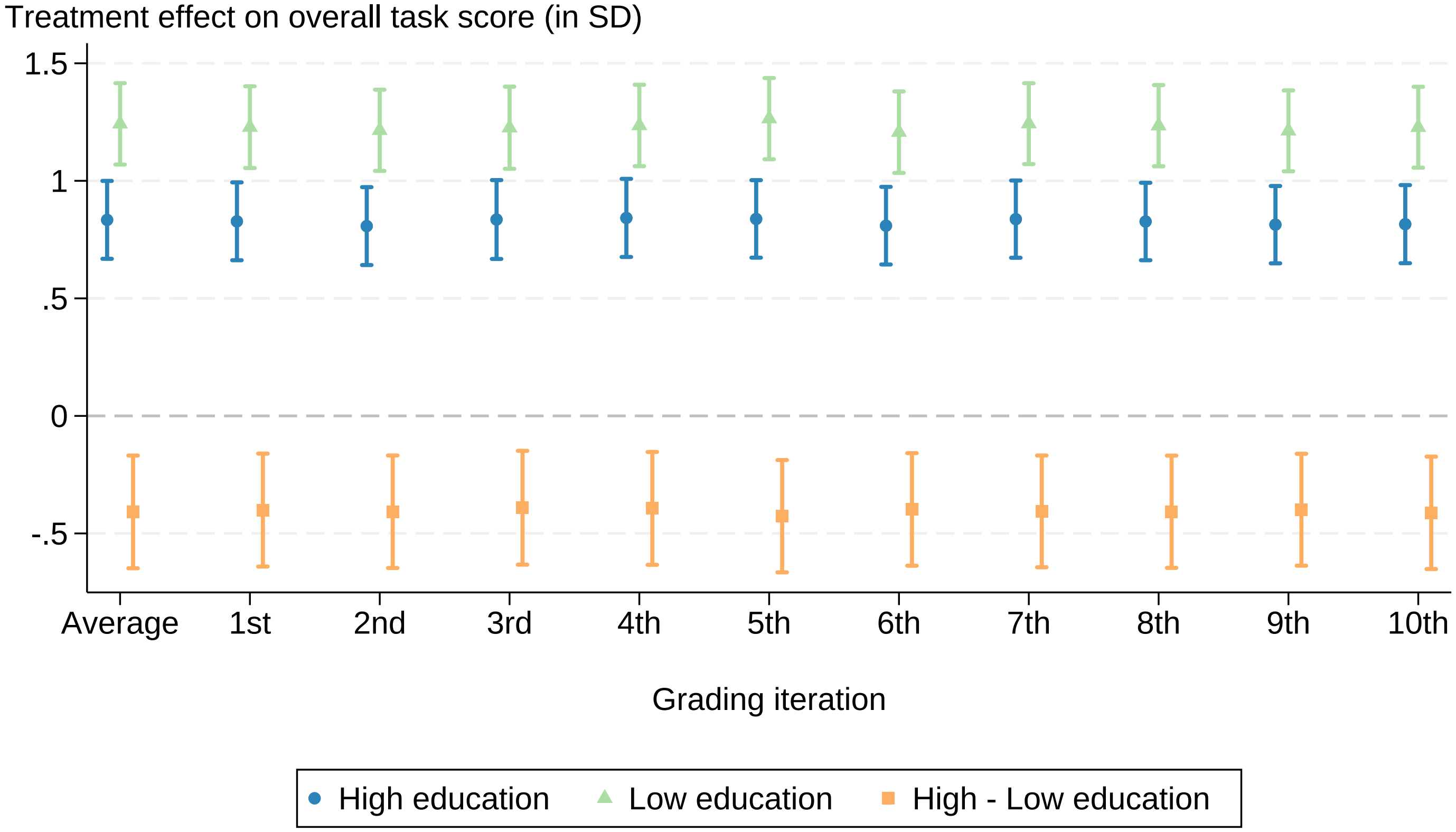

Notes: This figure plots treatment effects on the overall task score (standardized relative to the low-education control group) for low-education participants, high-education participants, and the difference between these effects, estimated separately using each individual grading iteration as the final score. Point estimates are shown with 95% confidence intervals. The first set of estimates corresponds to the average of ten independent grading iterations used in the main analysis. The remaining sets report results for each individual iteration.

Figure A.7: Treatment effects by education group – Robustness to alternative grading procedures

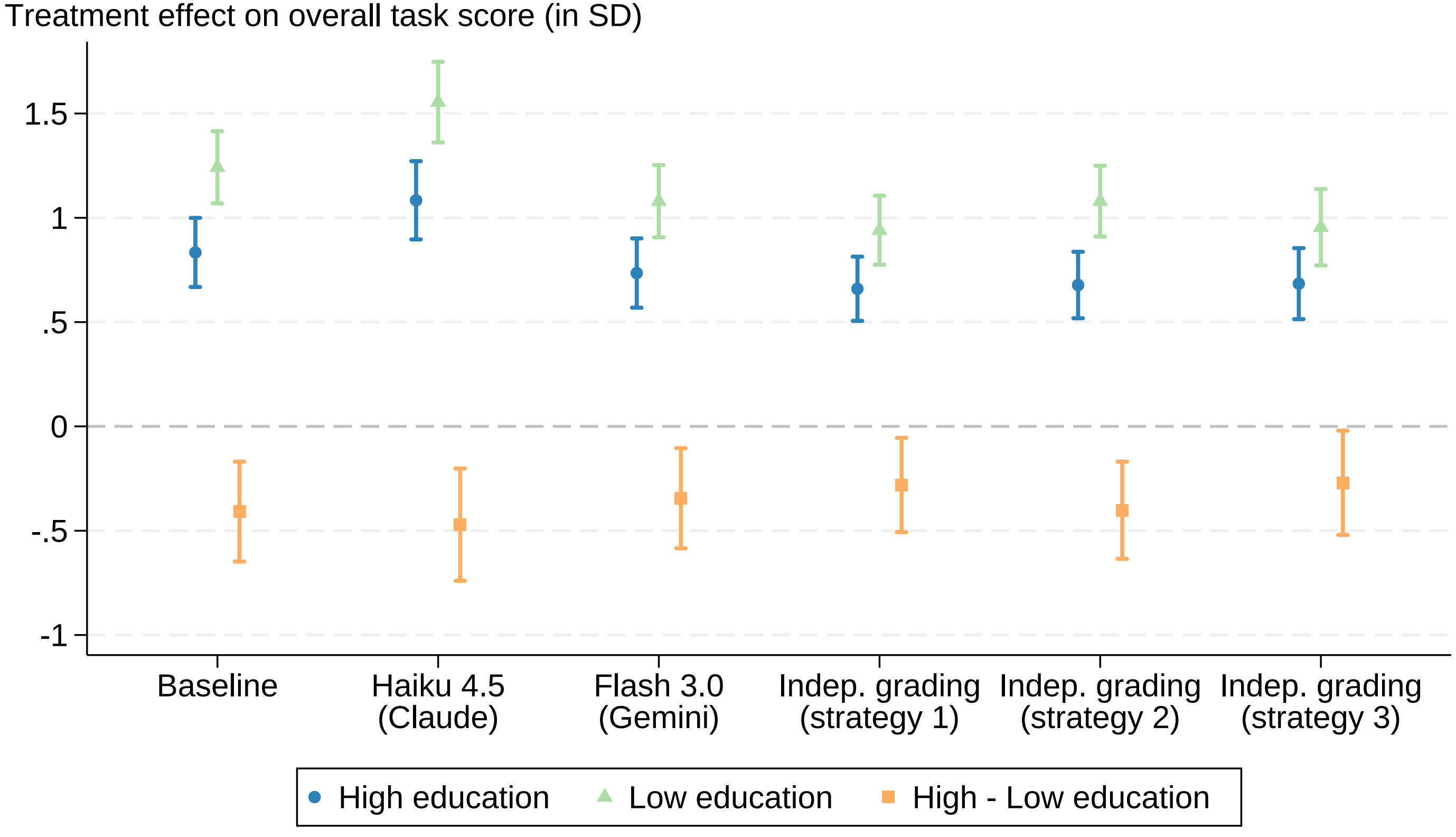

Notes: This figure summarizes a series of robustness checks for the estimated treatment effects on the overall task score (standardized relative to the low-education control group). The figure reports the estimated treatment effects for low-education participants, high-education participants, and the difference between these effects, together with 95% confidence intervals. "Baseline" corresponds to the estimates reported in the main analysis. "Haiku 4.5" and "Flash 3.0" report estimates obtained by grading all responses in a single run using Anthropic's Haiku 4.5 and Google's Gemini-Flash 3.0 models, respectively. "Independent grading" reports the estimates using the three AI-assisted grading strategies developed by an independent researcher not involved in the project.

Figure A.8: Treatment effects by education group – Additional robustness checks

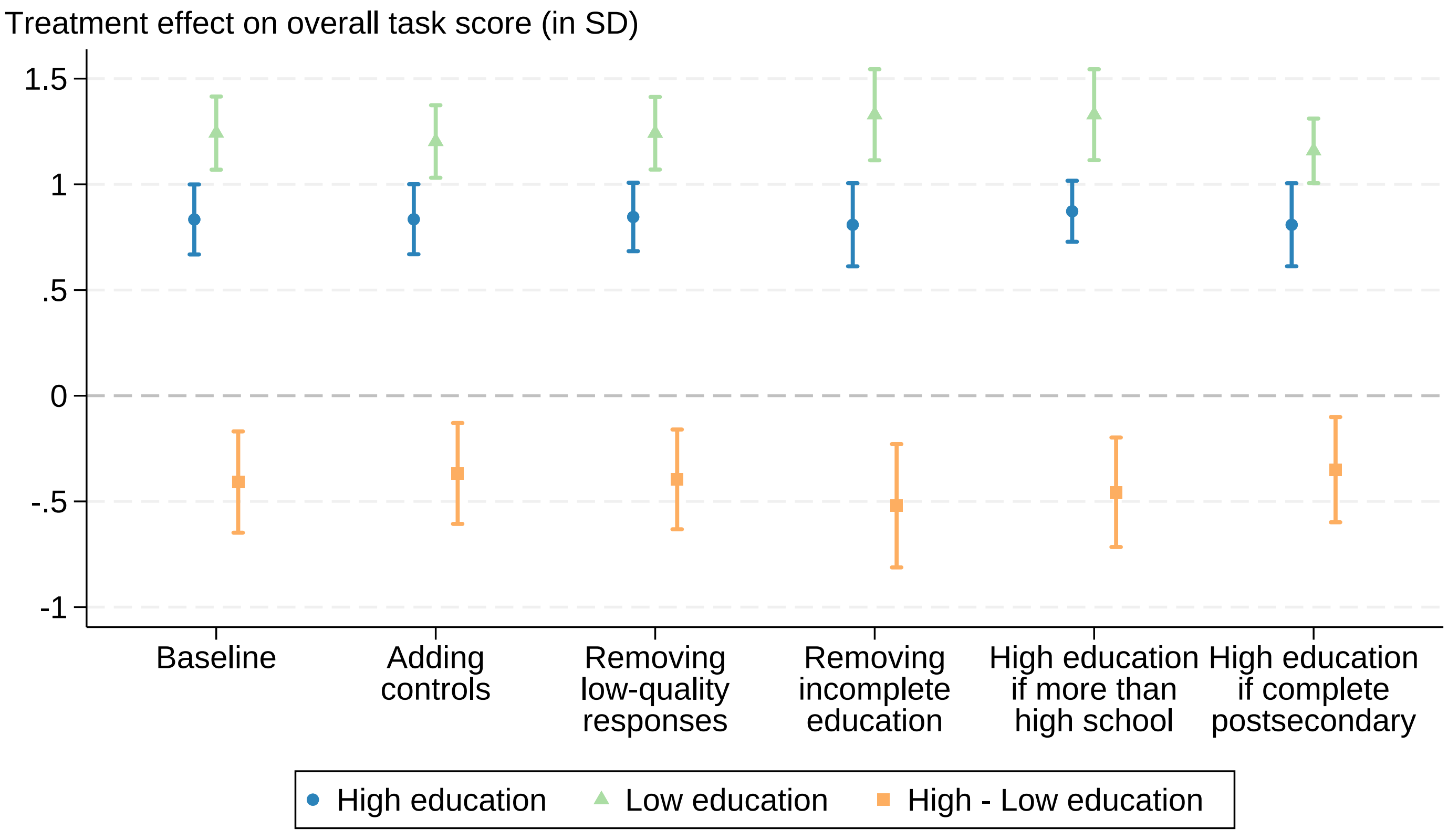

Notes: This figure summarizes a series of robustness checks for the estimated treatment effects on the overall task score (standardized relative to the low-education control group). The figure reports the estimated treatment effects for low-education participants, high-education participants, and the difference between these effects, together with 95% confidence intervals. "Baseline" corresponds to the estimates reported in the main analysis. "Adding controls" reports estimates from regressions that include task fixed effects, age, gender, employment status, educational attainment, and work experience. "Removing low-quality responses" excludes from the estimation sample the 25 individuals whose responses are identified as low quality. "Removing incomplete education" excludes participants with incomplete postsecondary education. "High education if more than high school" and "High education if complete postsecondary" present results using alternative definitions of the high-education group.

Figure A.9: Self-reported usefulness and comfort of using the AI assistant in the treatment group by education group

(a) Usefulness of AI assistant

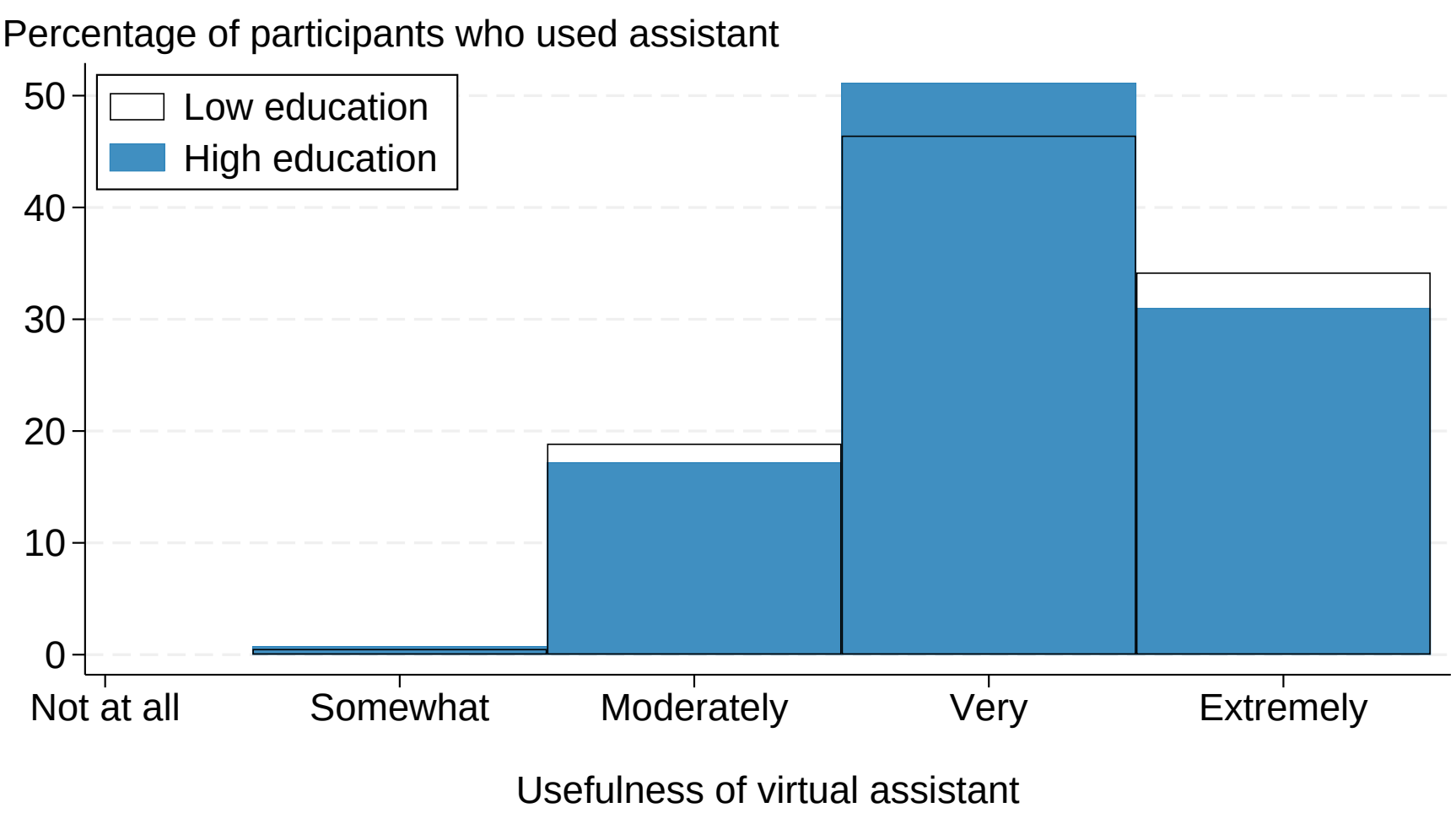

(b) Comfort while using AI assistant

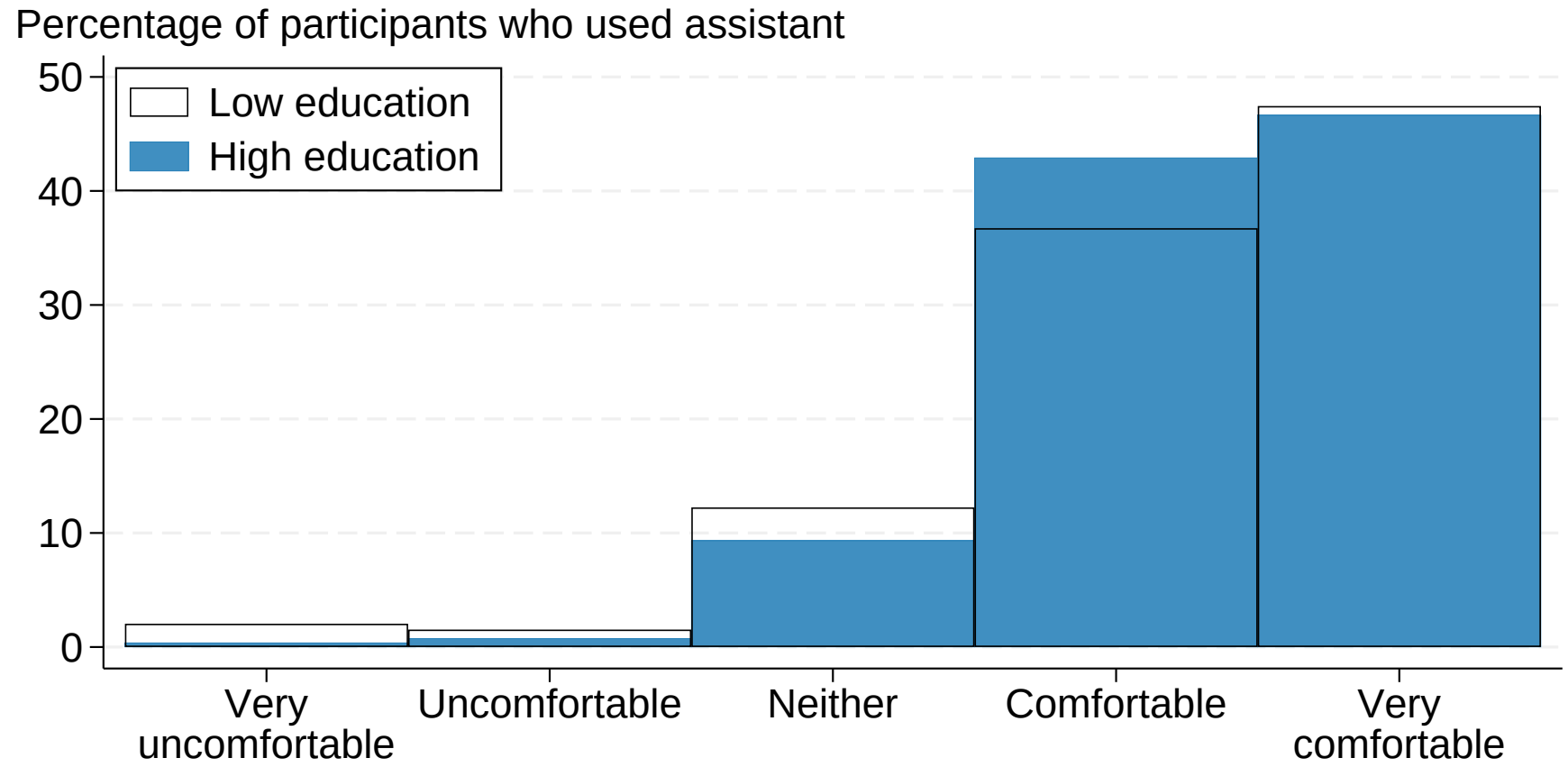

Notes: The top figure shows the distribution of perceived usefulness of the AI assistant for participants in the treatment group that used it during the task, for low- and high-education individuals. The bottom figure shows the distribution of self-reported comfort in using the AI assistant, for the same sample.

# B Further details of the experiment

## B.1 Overview of the experiment

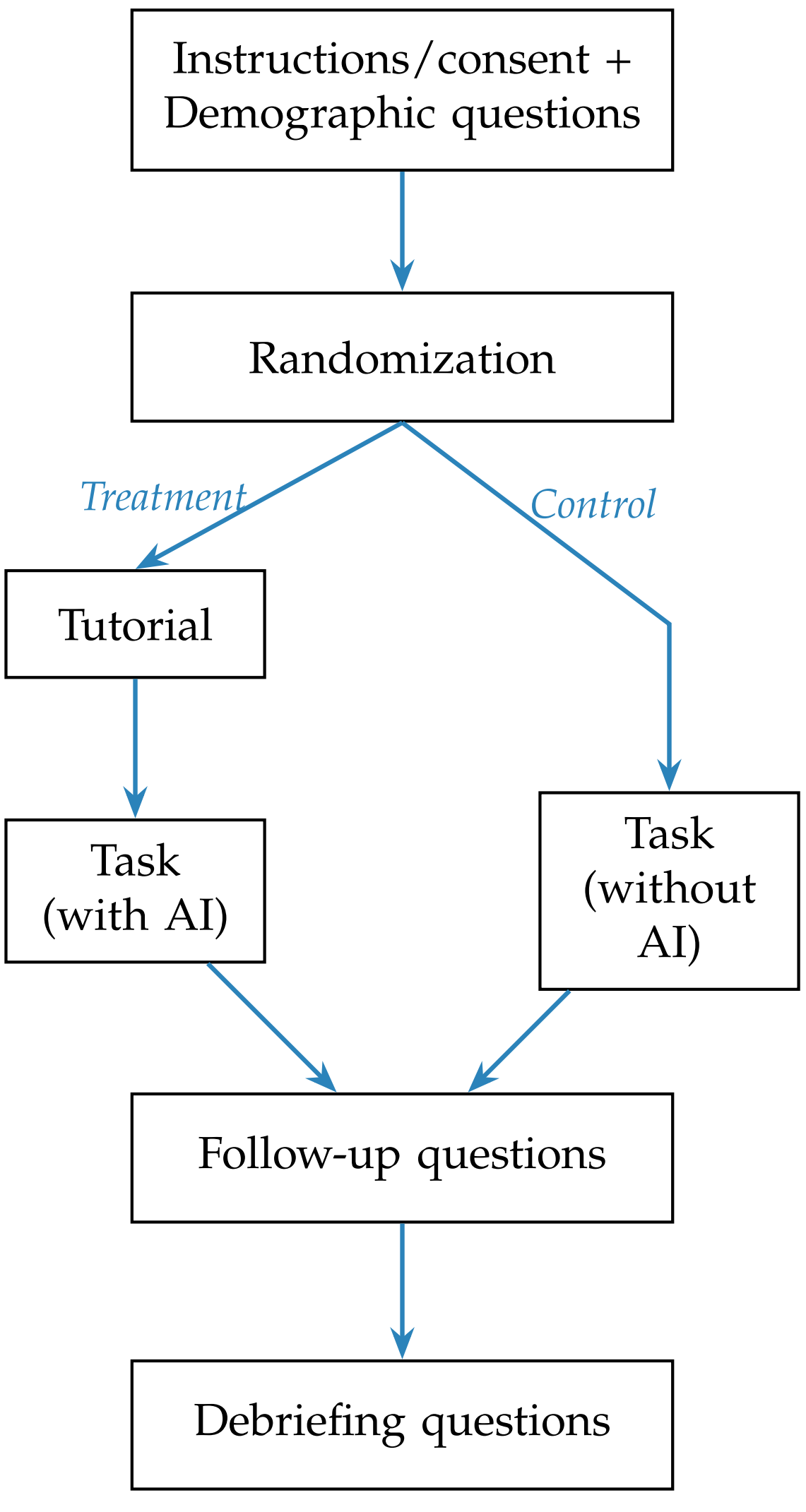

## B.2 Recruitment and sample restrictions

Table B.1: Sample Sizes

| Stage | | N |
|---|---|---|
| Recruitment and advancement throughout the experiment | Individuals invited to participate | 74,449 |
| | Entered the experimental platform | 3,002 |
| | Entered using a computer | 2,445 |
| | Consented to participate | 1,858 |
| | Completed demographic questions | 1,802 |
| | Finished the experiment | 1,188 |
| Exclusions | Excluding 1 copied response | 1,187 |
| | Excluding 3 submissions when the AI assistant was not functional | 1,184 |
| | Excluding 10 unintelligible submissions | 1,174 |
| Final sample | | 1,174 |

Table B.1 summarizes the flow of participants from recruitment to the final sample. In total, 74,449 individuals were invited to participate in the experiment (pooling recruitment across the three panel survey companies). Invitations relied on prescreening based on age and educational attainment. Specifically, the panel companies were instructed to recruit individuals aged 25–45 with at least a completed high-school degree.

There were 3,662 entries to the experimental platform, of which 3,002 correspond to unique IP addresses. To avoid potential contamination across observations from the same IP address, we adopt a uniform rule and keep only the first observation for each IP address. In addition, when the same individual appears multiple times in the recruitment records of a survey company (i.e., the same panelist ID) but with different IP addresses, we keep only the first observation. Of these 3,002 unique IP observations, 2,445 entered the experimental platform using a personal computer. Among those, 1,858 individuals consented to participate, and 1,802 completed the demographic questions. A total of 1,188 participants completed the experiment.

We then apply additional data-quality exclusions to arrive at our final experimental sample. We exclude 1 copied response, identified as a submission from an IP address associated with other submissions containing identical answers, and 3 submissions recorded during a short period when the AI assistant was temporarily unavailable.[1] Finally, we classify 10 submissions with unintelligible responses (e.g., "wfwrfrffewqwew" or "yteg3wv") as non-completions. After these restrictions, we arrive at our final sample of 1,174 participants.

Our preregistration specified a conservative target sample size of 600 (300 low-and 300 high-education). However, our main objective of estimating differential effects between the high- and low-education groups is demanding in terms of statistical power. Accordingly, we noted in the preregistration that recruitment might exceed the conservative target if fieldwork proceeded smoothly despite the logistical constraints of requiring participants to perform the task using a personal computer—a high fraction of most panel surveys' subjects respond on mobile devices, which were not suitable for the nature of our tasks. In practice, data collection progressed better than anticipated, resulting in the optimistic scenario of larger samples for both education groups.

[1]The AI assistant was hosted on Render. Due to a hosting-related change affecting DNS-based database connections, the assistant could not connect to its database and was unavailable from October 24, 2025 at 12:32:52 to October 27, 2025 at 12:23:26 (Buenos Aires time).

### B.3 Detecting AI use in the control group using changes in word count

To detect AI use that is less apparent, we computed the largest word-count increase between consecutive response snapshots taken every minute for each participant.[2] Figure B.1 below shows the distribution of these maximum word-count changes. As expected, treated participants who used the AI assistant exhibit much larger word-count jumps than those who did not (median 198 vs. 28), as seen in the top panel. The bottom panel compares control participants who exited the page during the task (31%) with those who did not (69%). Most control group participants who exited the page show minor word count changes, similar to those who did not exit the page. However, a small subset of those who exited display sharp word-count increases. We assume that control group participants sought external AI assistance if they exited the task page and have direct evidence of AI use or a maximum word-count change above 75 words (the 25th percentile among treated AI users).

Concretely, we classify control-group participants as noncompliant if they exited the task page and either (i) display direct evidence of AI use in their submission (e.g., emojis, bullet-point formatting, or excerpts of an AI conversation), or (ii) exhibit a maximum word-count increase above 75 words (the 25th percentile of the distribution among treated AI users). Using only direct textual evidence, we identify 6% of control participants as noncompliant. Using only the snapshot-based proxy, this share rises to 12%. Combining both sources of evidence yields an overall noncompliance rate of 13%.

[2] If no snapshot was available (because the participant submitted the task before the first snapshot was taken), we used the word count of the submitted task as the maximum change.

Figure B.1: Comparison of maximum change in word count between consecutive snapshots during task

(a) Used vs. did not use the AI assistant (treatment group)

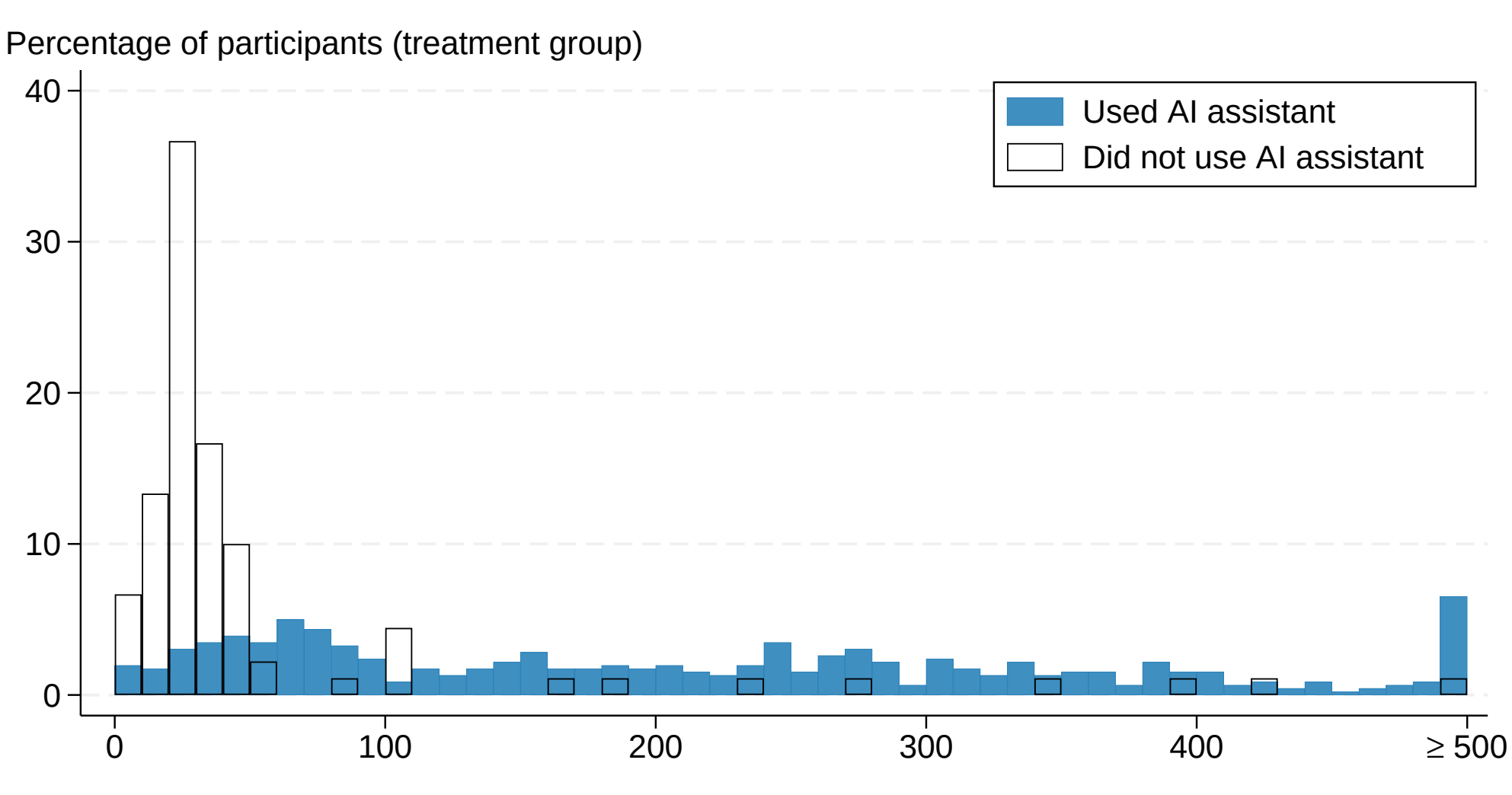

(b) Exited vs. did not exit the page during the task (control group)

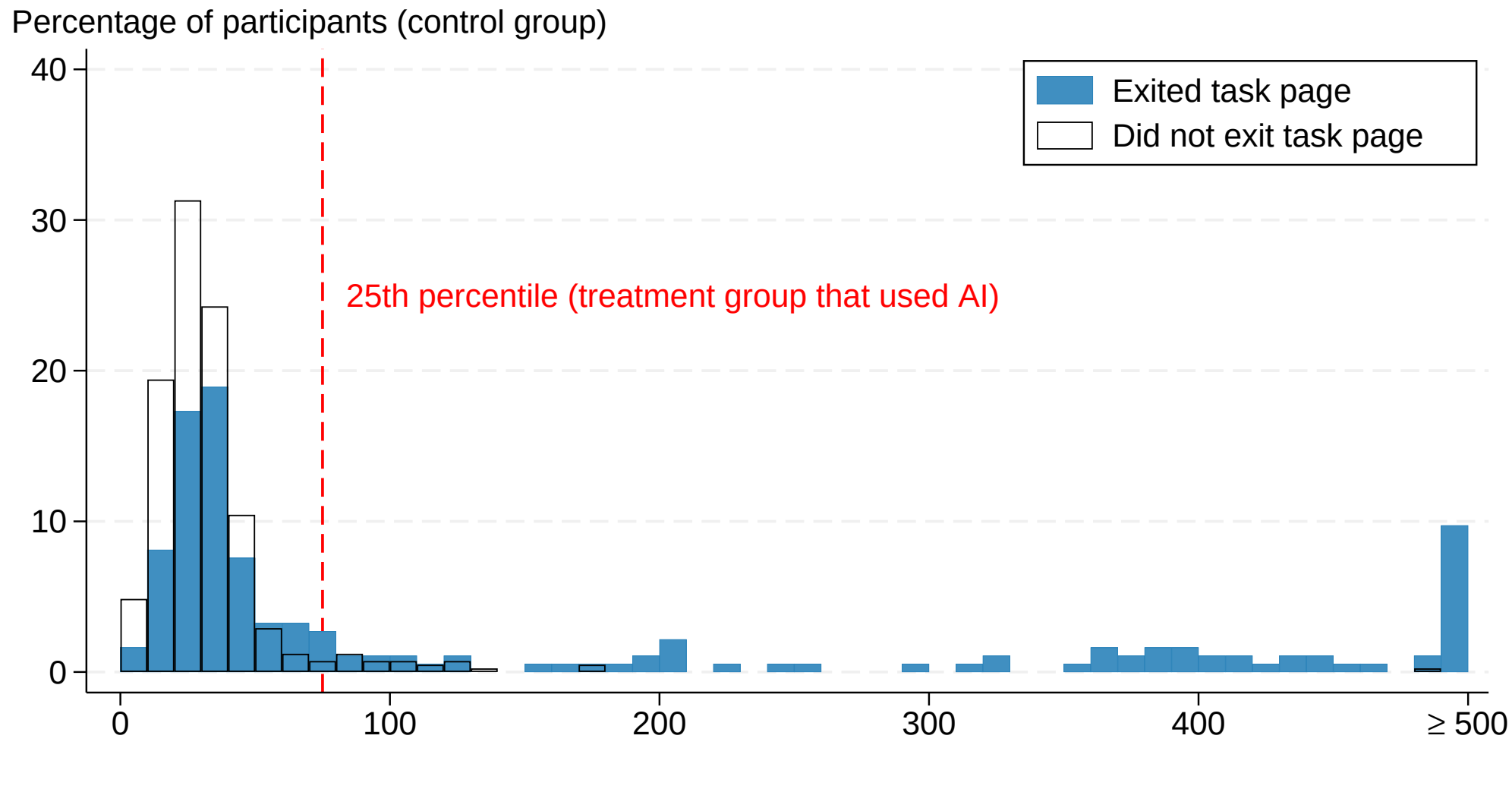

Notes: These figures show the distribution in the maximum word count change between consecutive snapshots of participants' task responses. For each participant, we computed the difference in word count between each consecutive snapshot and recorded the largest change observed. The top figure shows the distribution of the maximum word count change for individuals in the treatment group that used and did not use the AI assistant. The bottom figure shows the same for the individuals in the control group that exited the page during the task and those that did not exit.

# C Further analysis

## C.1 Effects of AI assistance on task performance subcomponents

In this section, we discuss in more detail the results for the two main components of the overall score (content and writing), as well as an additional outcome measuring whether participants correctly identified the root cause of the problem.

The two main components of the overall score are content and writing. As shown in the respective columns in Table 1, we find similar effects across both dimensions. Consistent with the intended difficulty gradient of the task, baseline differences are also present in both components. In the control group, high-education participants score 0.503 SD higher in writing and 0.525 SD higher in content (both gaps statistically significant at the 1% level). Regarding AI assistant-based performance convergence between the groups, the treatment effect for low-education participants exceeds that for high-education participants by 0.382 SD in writing and 0.389 SD in content, and both differences are statistically significant at the 1% level. Despite these larger gains for low-education participants, convergence remains incomplete. Among treated participants, high-education individuals still outperform low-education individuals by 0.122 SD in writing and 0.136 SD in content, although we are underpowered to detect these differences (p-values are 0.127 and 0.122, respectively). Taken together, these additional findings indicate that our main results are not solely driven by writing quality (which is probably the easiest performance boost from the use of the AI assistant) nor by the quality of the responses' content, which reflects more involved engagement with the information provided and the problem at hand. The equalizing effect of AI assistance operates through improvements in both dimensions of performance in our task.

This is further reflected in the additional results for the preregistered "detected root cause" indicator, which is a component of the content score and is equal to one if the subject correctly identified the root cause of the problem in their assigned task, and zero otherwise. The pattern for this coarser and more limited outcome is similar to that for the overall score and its subcomponents: for those without access to the AI assistant, 49.5% of low-education participants and 64.7% of high-education participants correctly detected the root cause of the problem in the task, whereas this proportion increases in treated subjects to 84.1% and 88.1% respectively. Again, low education individuals seem to gain more in their performance from access to the AI assistance: the difference in treatment effects across education groups is -11.2 percentage points, statistically significant at the 5% level. For this specific sub-outcome, performance seems to be equalized by the access to the AI assistant: the proportion of individuals correctly identifying the root cause of the problem is only four percentage points higher for the high-education group, and this difference is not statistically significant.

## C.2 Impact on (non-AI-assisted) follow-up performance subcomponents

As discussed in Section 2.2, all participants completed a set of follow-up questions about the content of the task and their proposed solutions after finishing the experimental task and with no access to the AI assistant. We can delve deeper into these differences in follow-up performance by analyzing the treatment effects on its subcomponents, the open-ended question and the multiple-choice questions. We present the results in Appendix Table A.5.

For the open-ended question, we only find a positive and statistically significant treatment effect among low-education individuals, with a point estimate of 0.237 SD (significant at the 1% level). For high-education individuals, the estimated effect is smaller (0.118 SD) and not statistically different from zero. Although the point estimates suggest larger gains for the low-education group, the estimated difference in treatment effects across education groups of −0.119 is again not statistically significant. These results closely mirror those for the aggregate follow-up score. Furthermore, in this case we also reject the null of equality in the performance in the open-ended question between treated high- and low-education participants, as the estimated gap of 0.141 SD is statistically significant at the 10% level.

Finally, for the multiple-choice questions, we do not find evidence of treatment effects in either education group. The estimated effects, equal to −0.078 SD for low-education participants and −0.075 SD for high-education participants, are relatively smaller, statistically insignificant, and nearly identical across groups. As a result, the baseline performance advantage of high-education individuals of 0.208 SD, which is statistically significant at the 5% level, remains largely unchanged after treatment. Among treated participants, the performance gap is 0.211 SD, and it is statistically significant at the 5% level.

Overall, the follow-up results paint a coherent picture. AI assistance generates modest but statistically meaningful improvements in post-task performance only among low-education individuals, with effects concentrated in the open-ended question. These gains, however, are not large enough to offset the substantial baseline advantage of high-education individuals.

# D AI-assisted grading

In this appendix, we describe the AI-assisted grading procedure used to construct the task and follow-up outcomes analyzed in the paper (Appendix D.1), and analyze the stability of the AI-assisted scores across grading iterations (Appendix D.2). Furthermore, we compare the scores obtained using our AI-assisted procedure with scores obtained by manually applying our grading rubrics to a random sample of responses (Appendix D.3). Finally, we describe an alternative AI-assisted grading procedure implemented by an independent researcher, and compare the scores obtained in that procedure with those obtained using our own AI-assisted grading (Appendix D.4).

## D.1 Implementation of AI-assisted grading

This appendix describes the AI-assisted grading procedure used to construct the task and follow-up outcomes analyzed in the paper.

### D.1.1 Task

We graded participants' responses using an AI-based scoring pipeline powered by OpenAI's GPT-5-mini model. The goal of this system is to implement our preregistered grading scheme in a transparent and reproducible manner, while ensuring that scoring relies exclusively on (i) the task materials and (ii) the participant's response.

During the piloting phase, and strictly following the preregistered scoring structure, we developed task-specific grading rubrics, one for each of the three task versions. Each rubric is stored as a structured JSON object that includes (a) the full task context (email, attached report components, and instructions) and (b) detailed scoring criteria for both the content and writing dimensions. In addition to these criteria, the JSON rubrics include short, illustrative examples for each score level within each sub-item. These examples are designed to leverage the in-context, few-shot learning capabilities of large language models by providing concrete anchors that map rubric language to typical response patterns.[3]

For content, the rubric specifies three diagnostic sub-questions (A1–A3), each scored on a 0–2 scale (maximum 6 points), and a solution component (maximum 4 points). The three diagnostic items correspond to the three diagnosis questions posed in the task email. A1 captures whether the participant directly addresses the first hypothesized mechanism mentioned in the email (the mechanism differs by task version). A2 captures whether the participant addresses the second hypothesized mechanism mentioned in the email (again task-specific). A3 is a general diagnostic question that asks whether the participant identifies other relevant evidence from the attached report beyond the first two hypotheses, for example additional patterns by branch, time, or product segment that help explain what is happening.

The solution rubric explicitly allows participants to propose one or more concrete actions. Each action is evaluated separately under three criteria: whether it addresses the root cause (B1), how specific it is (B2), and whether it is realistic (B3).

[3] "In-context learning" refers to a model's ability to adapt its behavior to a task using instructions and examples provided directly in the prompt at inference time, without updating model parameters. "Few-shot learning" is the special case in which the prompt includes a small number of labeled examples that illustrate the desired input–output mapping.

For writing, the rubric evaluates four components that sum to 10 points: orthography/grammar (0–2), clarity/legibility (0–3), organization (0–4), and register/tone (0–1).

The model is invoked with a fixed system prompt (in Spanish, matching the language of the tasks and answers) that instructs it to behave as an extremely rigorous grader, to avoid inferring missing information, and to return only a valid JSON object in a fixed schema. The system prompt also enforces an "anti-hallucination" rule: if a content element is not present in the participant's response, it must be marked as absent and assigned 0 points. A key feature of the prompt is that, if a response contains multiple proposed actions, the model must (i) detect all the proposed actions and (ii) output B1/B2/B3 scores for each action separately.

In each call, the model receives two inputs: (1) the participant's response and (2) the relevant task-specific rubric, which includes the full task context and the scoring guidelines (including the illustrative examples). The model returns a single JSON object containing (i) diagnostic scores for A1–A3, (ii) an action-level list with one entry per proposed solution with its corresponding B1/B2/B3 scores, and (iii) writing subscore

Because participants may propose more than one action, we separate two steps. First, the model detects and scores each action. Second, we use the model's action-level outputs to mechanically construct the single solution score used in the analysis. The model is explicitly instructed not to apply cutoff rules and not to select a "best" action. All aggregation is performed by deterministic code outside the model, following the preregistered logic.

Let actions be indexed by $j \in \{1, \ldots, J_i\}$ for participant $i$. For each action, the model outputs $B1_{ij} \in \{0, 1\}$, $B2_{ij} \in \{0, 1, 2\}$, and $B3_{ij} \in \{0, 1\}$. We compute the participant's solution score $\text{Sol}_i \in \{0, 1, 2, 3, 4\}$ as:

$$\text{Sol}_i = \begin{cases} 0, & \text{if } \max_j B1_{ij} = 0, \\ \max_{j \in \mathcal{J}_i} \left(1 + B2_{ij} + B3_{ij}\right), & \text{otherwise}, \end{cases} \tag{A1}$$

where $\mathcal{J}_i = \{j : B1_{ij} = 1\}$ is the set of actions that address the root cause. This rule implements the preregistered requirement that the solution component receives 0 points if no proposed action addresses the root cause, while allowing participants to receive credit for their best root-cause-addressing action when multiple actions are proposed.

We then compute the content total mechanically as $A1_i + A2_i + A3_i + \text{Sol}_i$ and the writing total mechanically as the sum of the four writing sub-scores returned by the model. Finally, to prevent very short, incorrect responses from receiving high writing scores, we apply the preregistered deterministic rule that responses with a content score of 0 and length below 200 characters are assigned a writing score of 0. This adjustment is implemented by code outside the model using the model's content and writing outputs.

For transparency and replicability, we provide the full system prompt and the complete task-specific JSON rubrics used for AI-assisted grading in our public GitHub repository at https://github.com/ramirogalvez/ai-skill-gap-experiment-data.

### D.1.2 Follow-up question

This subsection describes grading of the open-ended follow-up question. The two multiple-choice follow-up items are graded deterministically using the answer key. We graded the open-ended follow-up question using the same general approach as for the task, but with a simpler rubric and scoring scale. During the piloting phase, we developed one follow-up grading rubric for each task version. Each rubric is stored as a structured JSON file that contains (i) the follow-up question participants answered, (ii) a benchmark correct answer, and (iii) a three-level rubric defining what constitutes a fully correct answer, a partially correct answer, and an incorrect or irrelevant answer. The rubric also includes short examples for each score level to illustrate how the criteria map into actual responses.

We then asked GPT-5-mini to apply this rubric. We used a fixed system prompt in Spanish that instructs the model to behave as an extremely rigorous grader, to rely only on the rubric and the participant's response, and to avoid inferring information that is not explicitly stated in the response. In each call, the model receives two inputs: (1) the participant's follow-up response and (2) the relevant task-specific follow-up rubric. The model returns a single JSON object containing an integer score in $\{0, 1, 2\}$ and a brief explanation.

In our analysis, we use the model-returned integer score directly as the follow-up open-ended score. Any subsequent transformations, such as weighting, aggregation with the multiple-choice items, or standardization, are computed mechanically in code and are not performed by the model.

For transparency and replicability, we provide the follow-up system prompt and the task-specific follow-up JSON rubrics in our public GitHub repository at: https://github.com/ramirogalvez/ai-skill-gap-experiment-data

## D.2 Variation of scores across grading iterations

As noted in Section 2.3, because AI-generated scores exhibit some variation across runs, we grade each response ten times and use the average score across the ten iterations as our final measure. This appendix examines the degree of variation in scores across these iterations.

Figure D.1 below shows the distribution of the within-participant difference between the minimum and maximum score obtained across the ten grading iterations. The variation is modest: 83% of responses have less than a 1.5-point-difference (out of a maximum of 10 points) between their minimum and maximum scores across iterations. This indicates that while individual grading runs contain some inherent randomness, the scores are generally stable across iterations for most participants.

Figure D.1: Distribution of the variation in scores across AI-grading iterations

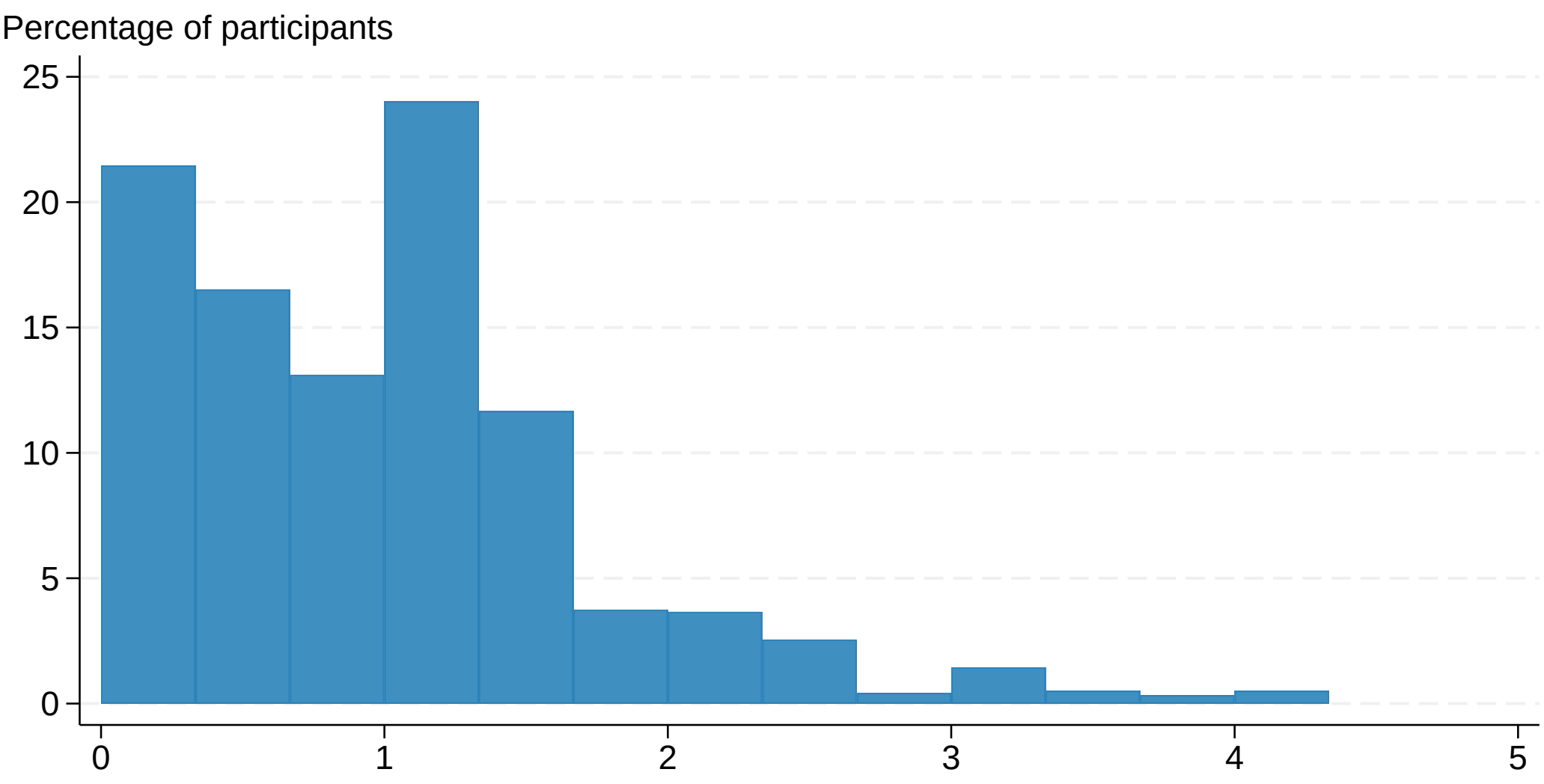

Notes: This figure plots the distribution of the difference between the minimum and maximum score for each response across ten AI-grading iterations.

Table D.1 below reports the pairwise correlations between scores across the ten iterations. The correlations are consistently high, ranging from 0.981 to 0.986, indicating strong consistency across different grading runs. Overall, these patterns confirm that averaging across iterations produces a reliable measure of task performance.

Table D.1: Correlation of scores between AI-grading iterations

| | Iteration 1 | Iteration 2 | Iteration 3 | Iteration 4 | Iteration 5 | Iteration 6 | Iteration 7 | Iteration 8 | Iteration 9 | Iteration 10 |
|---|---|---|---|---|---|---|---|---|---|---|
| Iteration 1 | — | 0.984 | 0.984 | 0.981 | 0.984 | 0.983 | 0.983 | 0.983 | 0.984 | 0.981 |
| Iteration 2 | 0.984 | — | 0.984 | 0.984 | 0.981 | 0.985 | 0.982 | 0.984 | 0.984 | 0.983 |
| Iteration 3 | 0.984 | 0.984 | — | 0.982 | 0.985 | 0.985 | 0.984 | 0.985 | 0.985 | 0.985 |
| Iteration 4 | 0.981 | 0.984 | 0.982 | — | 0.983 | 0.982 | 0.985 | 0.984 | 0.983 | 0.982 |
| Iteration 5 | 0.984 | 0.981 | 0.985 | 0.983 | — | 0.983 | 0.984 | 0.984 | 0.983 | 0.982 |
| Iteration 6 | 0.983 | 0.985 | 0.985 | 0.982 | 0.983 | — | 0.986 | 0.984 | 0.985 | 0.985 |
| Iteration 7 | 0.983 | 0.982 | 0.984 | 0.985 | 0.984 | 0.986 | — | 0.986 | 0.984 | 0.985 |
| Iteration 8 | 0.983 | 0.984 | 0.985 | 0.984 | 0.984 | 0.984 | 0.986 | — | 0.984 | 0.983 |
| Iteration 9 | 0.984 | 0.984 | 0.985 | 0.983 | 0.983 | 0.985 | 0.984 | 0.984 | — | 0.984 |
| Iteration 10 | 0.981 | 0.983 | 0.985 | 0.982 | 0.982 | 0.985 | 0.985 | 0.983 | 0.984 | — |

Note: This table reports the correlation across iterations of the AI-generated task scores.

## D.3 Comparison of AI-assisted grading and manual grading

To assess the reliability of our AI grading approach, we compare the scores obtained in our AI procedure with human-codified scores for a sample of 117 responses (approximately 10% of our sample). We randomly selected these responses, stratifying by task, education group, treatment group, and whether the grade obtained using AI-grading was above or below the median. The manual grading was independently done by two Economics university (one third year undergraduate, one Masters student) from the Universidad de San Andrés, who used the same task-specific rubrics given to the AI-grader. Importantly, students were only told that participants had to complete a task as part of an experiment, but had no information about the hypothesis of the experiment or the characteristics of the sample.

Figure D.2 below plots the scores given by the human graders (average across both graders) average against the scores obtained with our AI-grading procedure (average of 10 iterations). We find a strong alignment between AI-generated and human-coded scores, with most observations clustering close to the the 45-degree line.

Figure D.2: Relationship between human-codified scores and AI-obtained scores

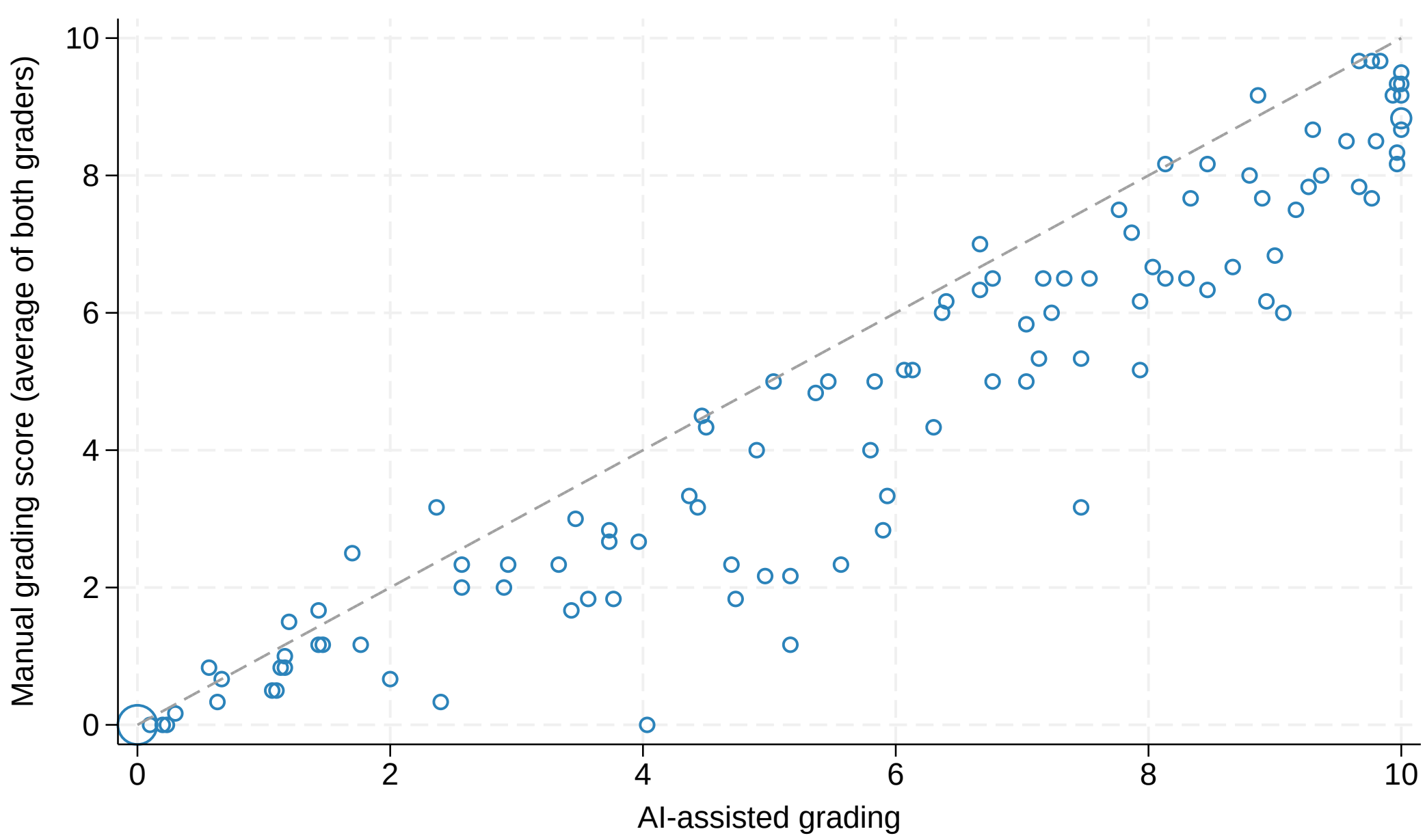

Notes: This figure plots the human-coded scores (average across both graders) against the AI-generated task scores for a random sample of 117 responses, along with a 45 degree line. The size of each marker is proportional to the number of observations at that coordinate.

Finally, we calculated the Pearson correlation and rank correlation (Spearman's $\rho$) between the AI-obtained scores and the scores given by the human graders. As shown in Table D.2 below, we find a Pearson correlation of 0.951 and a rank-correlation of 0.957 between the AI-generated grades and the average of the manually generated grades, with little variation across graders. Overall, our AI-based grading procedure is reliable and well aligned with human judgment: in the 10% manually graded sample, correlations between our AI-based scores and the two human graders are above 0.9. Section 7 provides the full validation results.

Table D.2: Correlation between human-codified scores and AI-obtained scores

| | Pearson | Spearman |
|---|---|---|
| Average across graders | 0.951 | 0.957 |
| Grader 1 | 0.918 | 0.939 |
| Grader 2 | 0.950 | 0.951 |

Note: This table reports the Pearson and Spearman rank correlations between AI-generated task scores and human-coded scores for a random sample of 117 responses. "Average across graders" refers to the average of the two human graders' scores.

## D.4 Alternative procedures for AI-assisted grading

To assess whether our findings depend on the specific implementation choices of our AI-assisted grading pipeline, we hired an independent researcher (not involved in this project) to design and run alternative LLM-based grading procedures. The researcher worked from the experiment preregistration (which defines the content and writing components, their weights, and the diagnosis + solution logic) and from the task materials, but did not have access to our detailed correction guidelines. Although we provided the researcher with participants' answers, he was blind to participants' treatment status and education group. These alternative grading procedures were therefore developed independently from ours and provide a robustness check based on different prompts, schemas, and grading workflows.

The independent researcher implemented three strategies, which we describe below.[4] We then compare the scores obtained in our AI-grading procedure to those obtained by the independent researcher.

### D.4.1 Strategy 1: LLM-generated rubric and single-step grading

In the first strategy, the researcher used an LLM (GPT-5.1-Pro) to translate the preregistered scoring design and the task instructions into a detailed, LLM-ready grading rubric with explicit anchors and a structured output schema. This translation preserves the preregistered structure but makes it directly usable for large-scale, consistent grading by requiring precise score definitions and standardized outputs. The rubric decomposes performance into content and writing scores. For content, it first scores the three diagnosis questions on a 0/1/2 scale, with criteria specified for each score. Second, it scores the solution component by separately assessing whether the proposed action(s) address the root cause, their specificity, and their realism, and it implements a strict "pass/fail" rule that assigns zero points to the entire solution component if the response does not address the root cause. For writing quality, scores are granted among the four preregistered subdimensions: orthography/grammar, clarity, organization, and tone. The rubric provides specific criteria for each score category, and requires brief justifications for each subscore.

[4]We provide the independent researcher's prompts, schemas, and task-specific materials used to implement these strategies in our public GitHub repository at https://github.com/ramirogalvez/ai-skill-gap-experiment-data.

Given this rubric, strategy 1 grades each participant response in a single model call in which the evaluator model must both interpret the task materials (including tables and figures) and apply the rubric to the participant response. The researcher implemented the evaluator using GPT-5.1-thinking-high (verbosity set to high) via the API chat completions interface.

The key strength of this strategy is that it provides a transparent, end-to-end implementation of the preregistered scoring logic with minimal manual steering. The corresponding tradeoff is that it places a relatively high cognitive burden on the evaluator model because it must re-derive the key facts from the task materials for every graded response.

#### D.4.2 Strategy 2: Same rubric as Strategy 1, plus task-specific "fact sheets" to reduce ambiguity

The second strategy keeps the rubric, schema, score ranges, and aggregation logic exactly the same as in Strategy 1, but changes the information provided to the evaluator model in order to reduce ambiguity and improve stability. In addition to the participant response and the general rubric, the researcher supplied a task-specific fact sheet derived from each task's text, figure, and table (generated with GPT-5.1-Pro). These task-specific materials explicitly list the key facts embedded in the report, including the root cause and other facts relevant for the diagnosis questions, and they highlight common reading errors.

The purpose of this strategy is to avoid requiring the evaluator model to repeatedly rediscover what the chart implies or which numbers matter in the table for each response. Instead, the model can focus on applying the same rubric more consistently, particularly for diagnosis and root-cause identification, where occasional misreadings of the task materials could otherwise add noise. The researcher implemented Strategy 2 using the same evaluator model as Strategy 1 (GPT-5.1-thinking-high, verbosity high). Relative to Strategy 1, the practical advantage of Strategy 2 is improved scoring stability while preserving the same conceptual mapping from responses to points.

#### D.4.3 Strategy 3: "Grading on a curve" via pairwise comparisons aggregated with Elo (and within-task $z$-scores)

The third strategy departs from absolute point assignment and instead elicits comparative judgments. An evaluator model is asked to rank groups of responses from the same task version in a strict order (no ties), using the preregistered notion of quality as guidance (content weighted more than writing, diagnosis + solution logic, and the root-cause requirement). The implementation forms groups of 10 responses at a time. From each strict ranking, it constructs a conservative set of pairwise comparisons using only adjacent pairs in the ordering (so a group of size 10 yields 9 binary comparisons). Groups are constructed to balance coverage so that each response participates in approximately 20 pairwise comparisons within its task.

These win/loss outcomes are then aggregated into Elo ratings within each task version. Elo is a standard rating system originally developed for chess that estimates a latent "skill" (here, response quality) from a sequence of head-to-head outcomes. Intuitively, responses are treated as "players": when response $i$ is ranked above response $j$, $i$ is recorded as winning a pairwise comparison against $j$. Elo then updates ratings sequentially after each comparison based on (i) the observed outcome (win/loss) and (ii) the expected probability of winning implied by the current rating difference. Ratings therefore drift upward for responses that consistently win and drift downward for responses that consistently lose. The size of each update is governed by a tuning parameter $K$, which controls how quickly ratings react to new comparisons.

In the implementation used here, all responses begin with the same initial rating (1000). Expected win probabilities are computed using the standard Elo logistic mapping, and ratings are updated sequentially with a fixed step size (set to 32 by the researcher). Because Elo levels are not directly comparable across separate "tournaments" (here, task versions), the outcome variable will have to be standardized with respect to the low-education control group separately for each task.

The key advantage of Strategy 3 is that it relies on ordinal, comparative evaluation rather than on the absolute calibration of a numeric rubric scale. This can sharpen discrimination among responses that would otherwise bunch together under point-based grading, and it provides a conceptually distinct robustness check in which the primary measurement object is relative quality (and relative distance) rather than exact point attribution by subcomponent.

#### D.4.4 Comparison of our AI-assisted grading to the independently developed AI-assisted grading

In this section, we compare the scores obtained in our AI grading procedure with the scores obtained in the three strategies developed by the independent researcher. For strategies 1 and 2, we compare the overall score on a 0–10 scale with the average overall score across 10 iterations of our own AI-assisted grading, also on a 0–10 scale. As shown in Figures D.3 and D.4 below, we find a strong alignment between our own AI-generated scores and those obtained by the independent researcher using strategies 1 and 2, with most observations clustering close to the the 45-degree line.

Figure D.3: Own and independently developed AI-assisted scores – Strategy 1

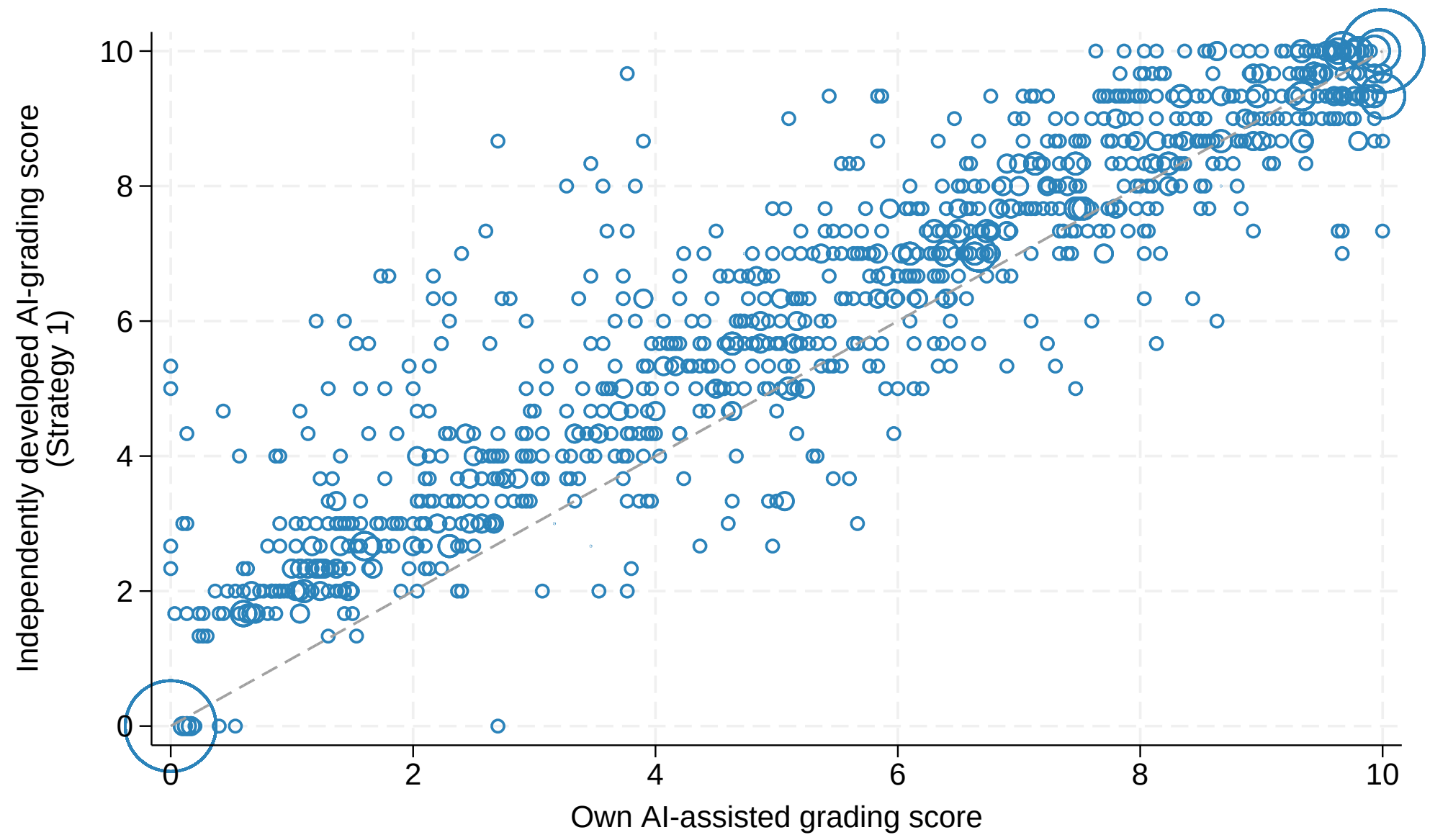

Notes: This figure plots the scores obtained by the independent researcher using Strategy 1 against our own AI-generated task scores, all on a 0–10 scale, along with a 45 degree line. The size of each marker is proportional to the number of observations at that coordinate.

Figure D.4: Own and independently developed AI-assisted scores – Strategy 2

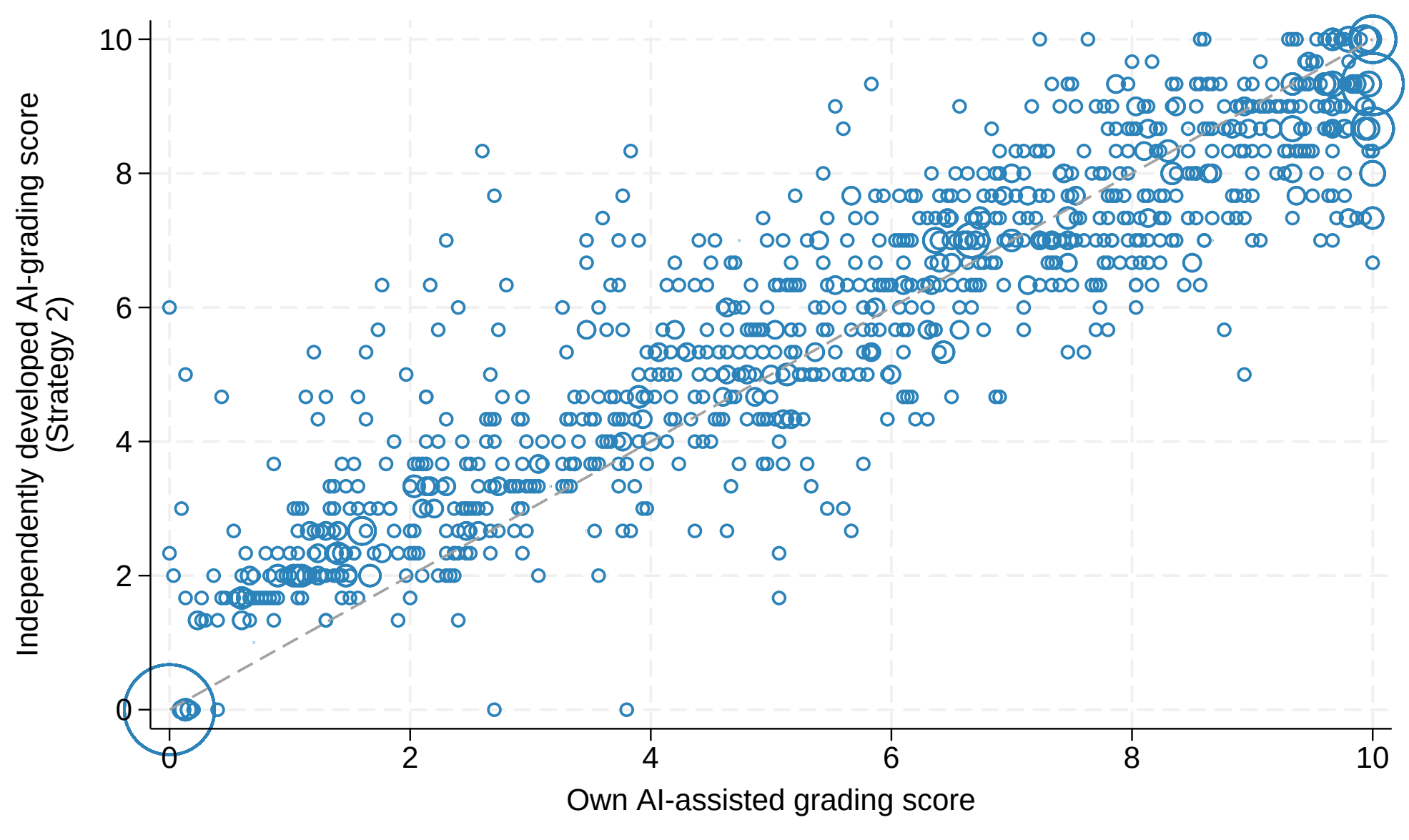

Notes: This figure plots the scores obtained by the independent researcher using Strategy 2 against our own AI-generated task scores, all on a 0–10 scale, along with a 45 degree line. The size of each marker is proportional to the number of observations at that coordinate.

Note that in the case of strategy 3, the final output of the grading procedure is an Elo rating, presented on an arbitrary scale, and not comparable across tasks (as the ranking procedure is conducted within each task). Therefore, to compare the scores under strategy 3 with our own AI-generated score, we first standardized the Elo rating by the mean and standard deviation of the low-education control group (separately within each task), and plot this against our main outcome variable, the standardized overall task score. As shown in Figure D.5 below, the two measures track each other closely, indicating a high degree of agreement across grading approaches. The correspondence is particularly tight over the central part of the distribution. Differences become more visible in the tails, where the Elo-based measure exhibits greater dispersion. This reflects the fact that the Elo rating is not constrained to the 0–10 scale used in our main grading procedure, allowing for more extreme relative distinctions among very low- and very high-quality responses.

Figure D.5: Own and independently developed AI-assisted scores – Strategy 3

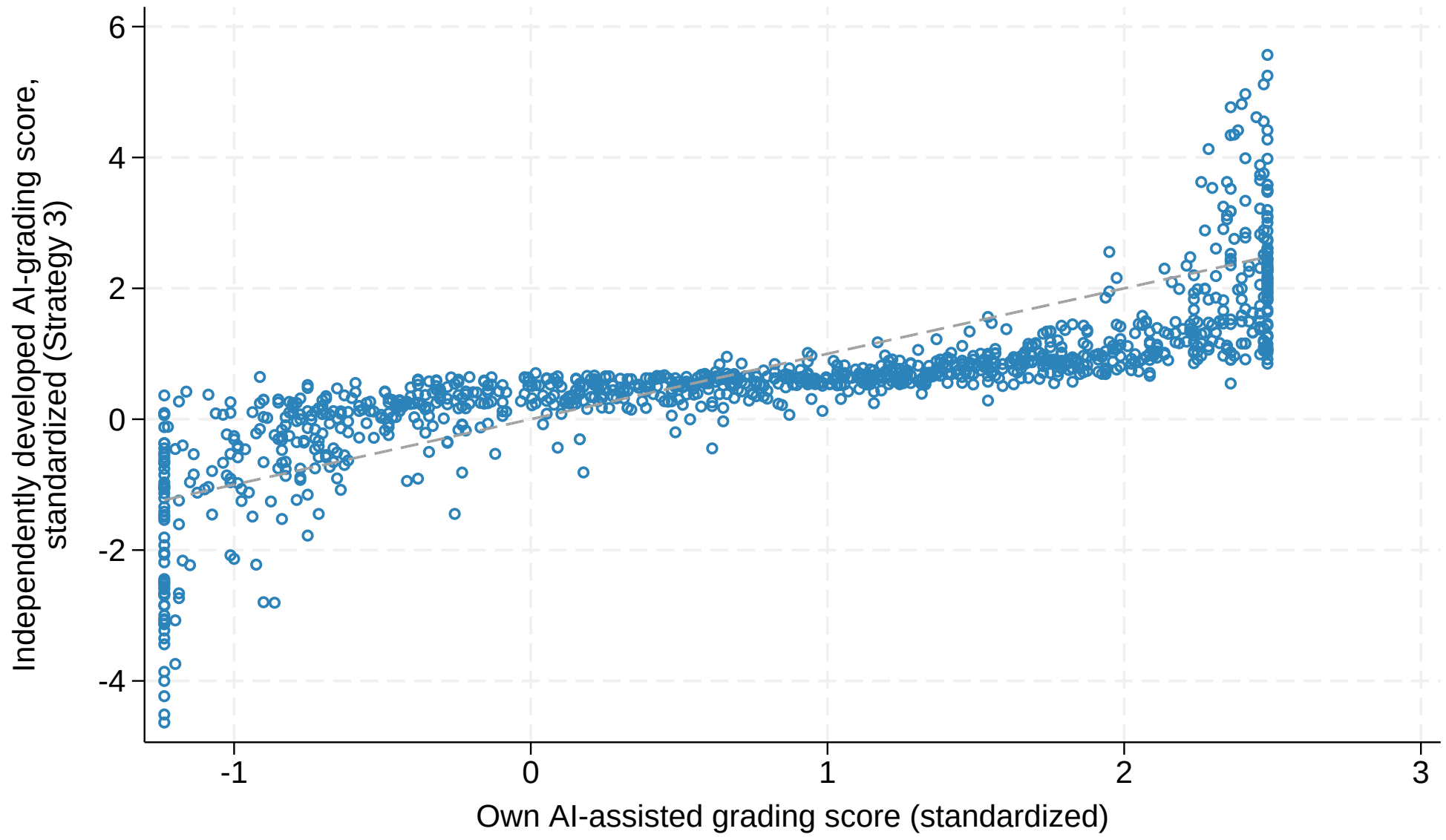

Notes: This figure plots the scores obtained by the independent researcher using Strategy 3 (standardized within task by the mean and standard deviation of the low-education control group) against our own AI-generated task scores (standardized by the mean and standard deviation of the low-education control group), along with a 45 degree line. The size of each marker is proportional to the number of observations at that coordinate.

Table D.3 summarizes these comparisons by reporting pairwise correlations across the different grading measures. Consistent with the close visual correspondence shown in Figures D.3 and D.4, the correlations between our AI-generated score and the scores obtained under strategies 1 and 2 are very high, on the order of 0.9 or above, indicating that the alternative rubric-based implementations generate scores that are very similar to those produced by our own AI-assisted grading procedure. For strategy 3, the linear (Pearson) correlation with our own score is lower but still high, at around 0.784, reflecting the unconstrained and relative nature of the Elo measure. At the same time, the rank correlation is very high (0.934), indicating strong agreement in the ordering of responses across the two approaches. Overall, these results indicate that our findings are not sensitive to the specific AI grading implementation used.

Table D.3: Correlation between independently developed and own AI-assisted scores

| | Pearson | Spearman |
|---|---|---|
| Strategy 1 | 0.936 | 0.939 |
| Strategy 2 | 0.928 | 0.926 |
| Strategy 3 | 0.784 | 0.934 |

Note: This table reports the Pearson and Spearman rank correlations between our own AI-generated task scores and the scores obtained by the independent researcher under all three strategies. For strategy 1 and 2, we compare the overall score for each strategy on a 0–10 scale with the average overall score across 10 iterations of our own AI-assisted grading, also on a 0–10 scale. For strategy 3, we compare the Elo rating obtained in strategy 3 (standardized within task by the mean and standard deviation of the low-education control group) against our own AI-generated task scores (standardized by the mean and standard deviation of the low-education control group).

# E Classification of patterns of AI use

This appendix describes in detail how we classified participants' patterns of interaction with the AI assistant among the 471 treated participants who used the tool during the task. For transparency and replicability, we provide the full system prompt used to classify participants' conversations with the AI assistant in our public GitHub repository at https://github.com/ramirogalvez/ai-skill-gap-experiment-data.

We focus on three dimensions of AI use: how participants relied on the assistant to work through the task, how they used the assistant in generating the final output, and how they orchestrated the interaction with the assistant over the course of the conversation.

**Working through the task.** We capture how participants use the AI assistant to reason about the task along two main dimensions. First, we measure the extent of assistance requested, defined as the share of task components (diagnostic questions and solution) for which participants asked the assistant for help. We compute this by first coding, for each task component, whether the participant requested AI assistance, and then constructing a weighted average across components, where weights correspond to the share of total points assigned to each component in the scoring rubric.

Second, we measure the quality of prompting in this dimension. We construct a standardized index using inverse covariance weighting following Anderson (2008). This index captures the extent to which participants provide the assistant with detailed instructions that support reasoning through the task. It combines two dimensions of prompting: requests for disciplined reasoning rather than surface-level answers, and the extent to which participants provide contextual information that is not available to the AI assistant but is required to reason through the task. For the first dimension, we take into account whether the participant asks for data-based justifications, requests cross-checking across sources, ensures that all task components are addressed, or prompts the assistant to evaluate multiple solutions, their trade-offs, and their viability in context. For the second dimension, we consider whether the participant provides the assistant with the full email from the boss, the opening paragraph describing the general problem, or the initial description of the assignment. The index is standardized using the mean and standard deviation of the low-education group.

**Generating the output.** We characterize the role of the AI assistant in the production of the final output along three dimensions. First, we code the type of drafting assistance requested by participants, distinguishing between explicit requests for the assistant to generate the output, requests to extend or complete a user-generated draft, and requests to edit a draft written by the participant. This classification excludes cases in which participants relied on the assistant to think through the response, perhaps using AI generated text in their answer, but did not explicitly request the assistant to draft any part of their final answer.

Second, we measure the quality of drafting instructions provided to the assistant using an index that captures how tightly participants guide the writing process. This index measures instructions related to output style, including whether participants specify tone, register, length, structure, formatting, or audience, as well as requests that the output sound natural or context-appropriate, as well as constraints imposed on output content, such as requests to preserve key points or evidence, correct mistakes, prioritize information, or adhere to substantive requirements of the task.

Third, we classify the extent to which AI-generated content is incorporated into the final submission. Final responses are coded according to whether they rely on AI-generated text through full copy-paste—defined as final answers composed entirely of AI-generated text, with no original content added by the participant (from a single or multiple messages)—partial copy-paste, paraphrasing of AI-generated content, or no direct use of AI-generated text.

**Workflow orchestration.** Beyond what participants ask for and how they use the assistant's output, we characterize how they orchestrate the conversation itself. We develop a classification system based on three dimensions that capture different aspects of conversation management. The first dimension, high initial specificity, is a binary variable that identifies whether the participant's first message provides sufficient context and direction to the assistant. Participants are identified as having high initial specificity if their first message is substantive in length (more than one sentence) and either establishes an explicit frame of reference to the data or presents relevant task context. The second dimension, iterativity, captures whether participants actively build or refine the interaction beyond a single exchange. Participants are identified as having high iterativity if they send more than two messages and, in messages after the first, request that the assistant add elements or refine previous outputs. The third dimension, structuring, measures whether participants deliberately distribute and develop their request across multiple conversational turns. Participants have a highly structured workflow if the conversation includes at least two user messages that substantively develop or elaborate the task—by providing context, posing analytical questions, requesting new outputs, or adding meaningful requirements—rather than merely refining format or confirming earlier responses.

# F Supplementary materials

In this section, we present additional details of the experiment. Section F.1 presents screenshots of the recruitment emails sent to participants, and Section F.2 has the translation from Spanish of the instructions given to participants. Section F.3 presents the translation of the tutorial on how to use the AI assistant done by treated participants, and Sections F.4 to F.6 contain the exact wording of the task and follow-up questions, also translated to English. Finally, Section F.7 presents the system prompt for the AI assistant.

## F.1 Recruitment materials

We used three reputable survey companies to recruit participants—Wonder, Opinaia, and Offerwise—each of which manages its own survey panel. Each survey company contacted their panel participants inviting them to participate in the experiment. The email mentioned the standard remuneration offered by the panel, that participants had to complete the experiment using a computer, and that the experiment would take approximately 20 minutes.

**Screenshot of the email sent by Wonder:**

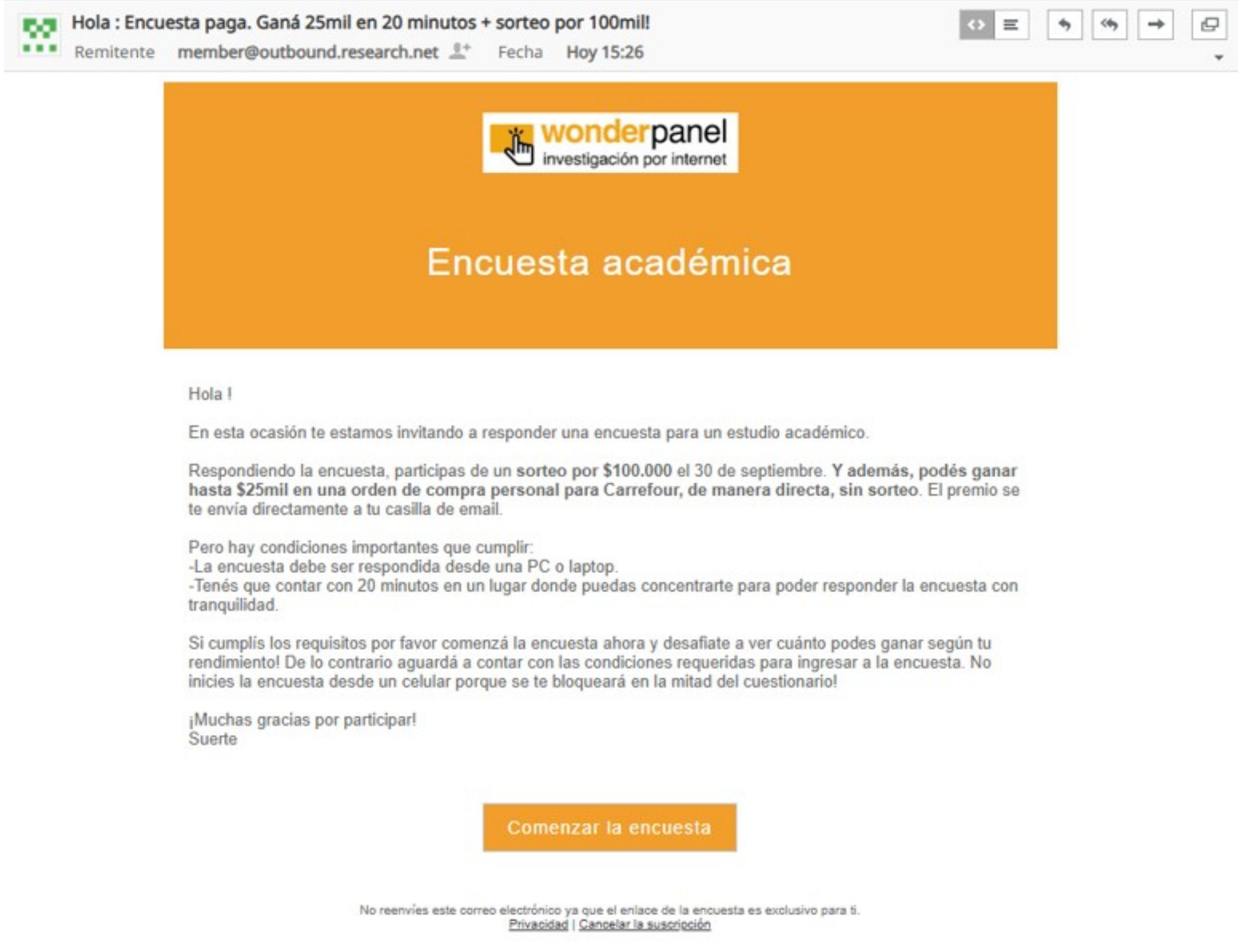

Hola : Encuesta paga. Ganá 25mil en 20 minutos + sorteo por 100mil!
Remitente member@outbound.research.net Fecha Hoy 15:26

wonderpanel
investigación por internet

Encuesta académica

Hola !

En esta ocasión te estamos invitando a responder una encuesta para un estudio académico.

Respondiendo la encuesta, participas de un **sorteo por $100.000** el 30 de septiembre. **Y además, podés ganar hasta $25mil en una orden de compra personal para Carrefour, de manera directa, sin sorteo**. El premio se te envía directamente a tu casilla de email.

Pero hay condiciones importantes que cumplir:
-La encuesta debe ser respondida desde una PC o laptop.
-Tenés que contar con 20 minutos en un lugar donde puedas concentrarte para poder responder la encuesta con tranquilidad.

Si cumplís los requisitos por favor comenzá la encuesta ahora y desafiate a ver cuánto podes ganar según tu rendimiento! De lo contrario aguardá a contar con las condiciones requeridas para ingresar a la encuesta. No inicies la encuesta desde un celular porque se te bloqueará en la mitad del cuestionario!

¡Muchas gracias por participar!
Suerte

Comenzar la encuesta

No reenvíes este correo electrónico ya que el enlace de la encuesta es exclusivo para ti.
Privacidad | Cancelar la suscripción

**Screenshot of the email sent by Opinaia:**

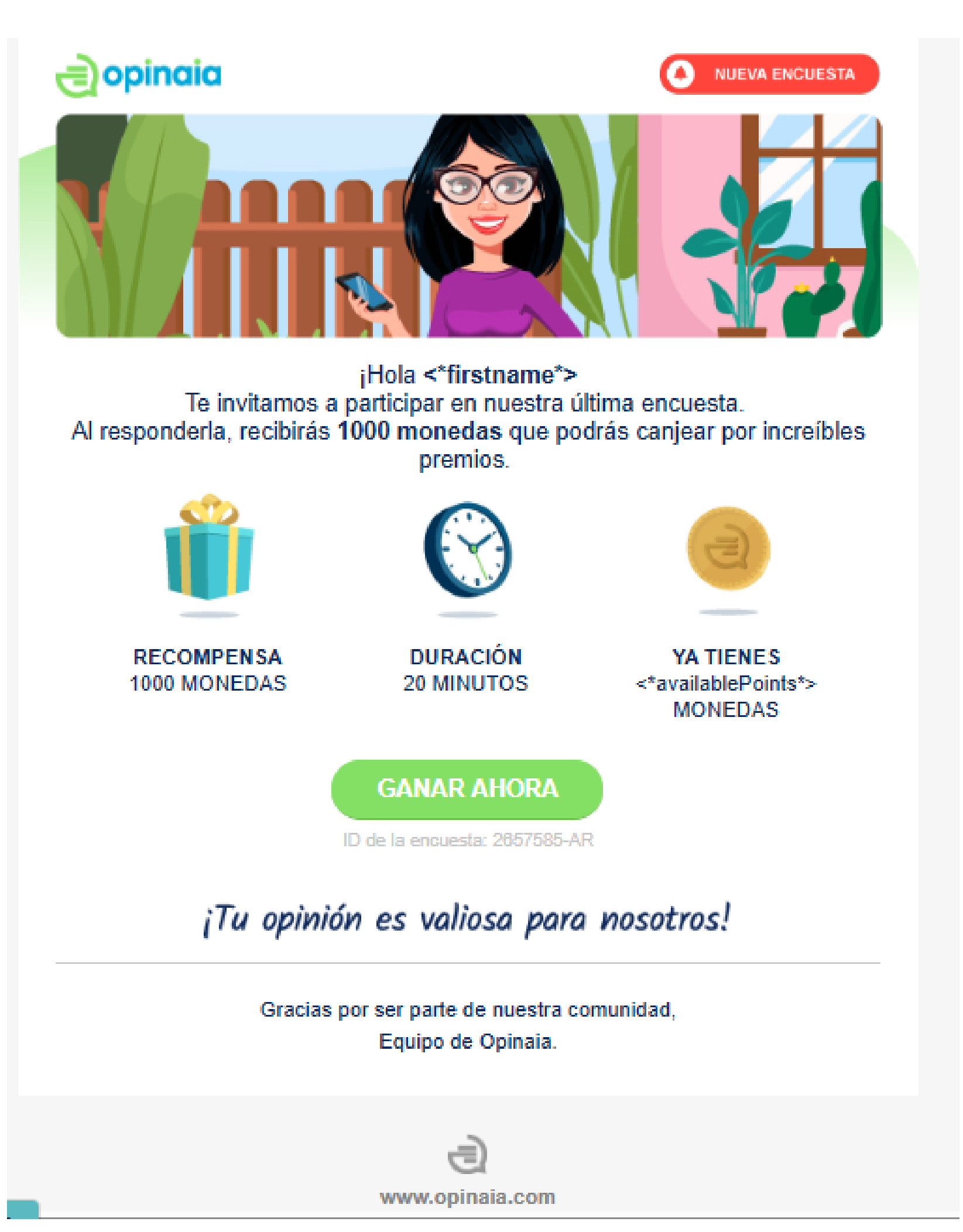

**Screenshot of the email sent by Offerwise:**

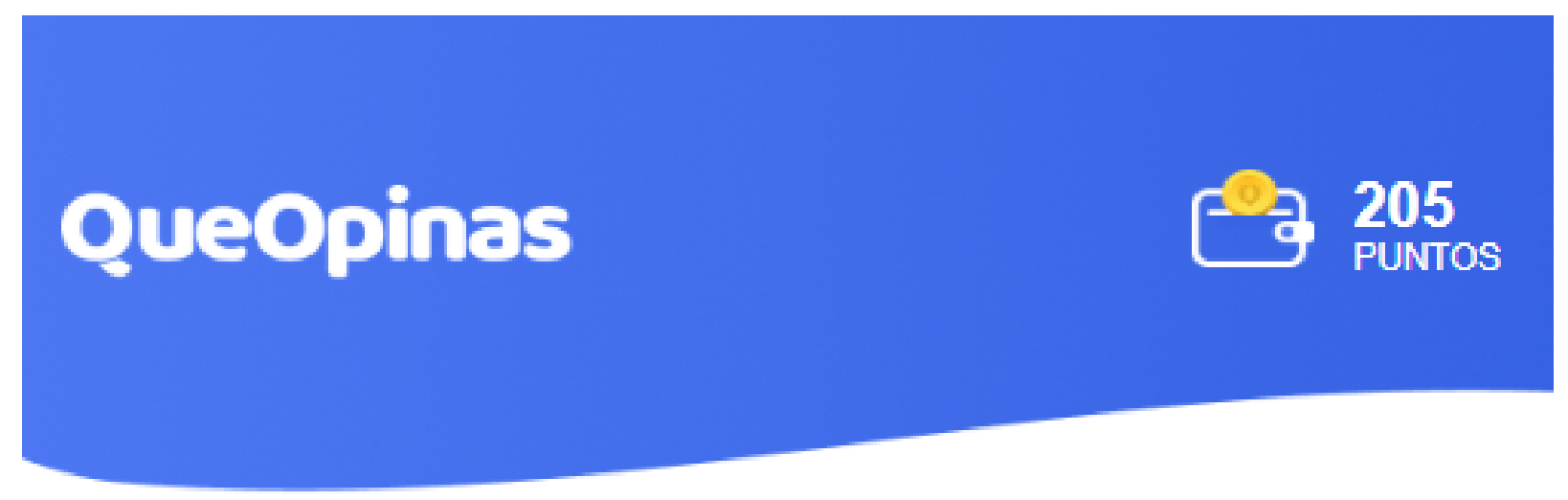

¡Hola!

¡HEMOS ENCONTRADO UNA ENCUESTA PARA TI!

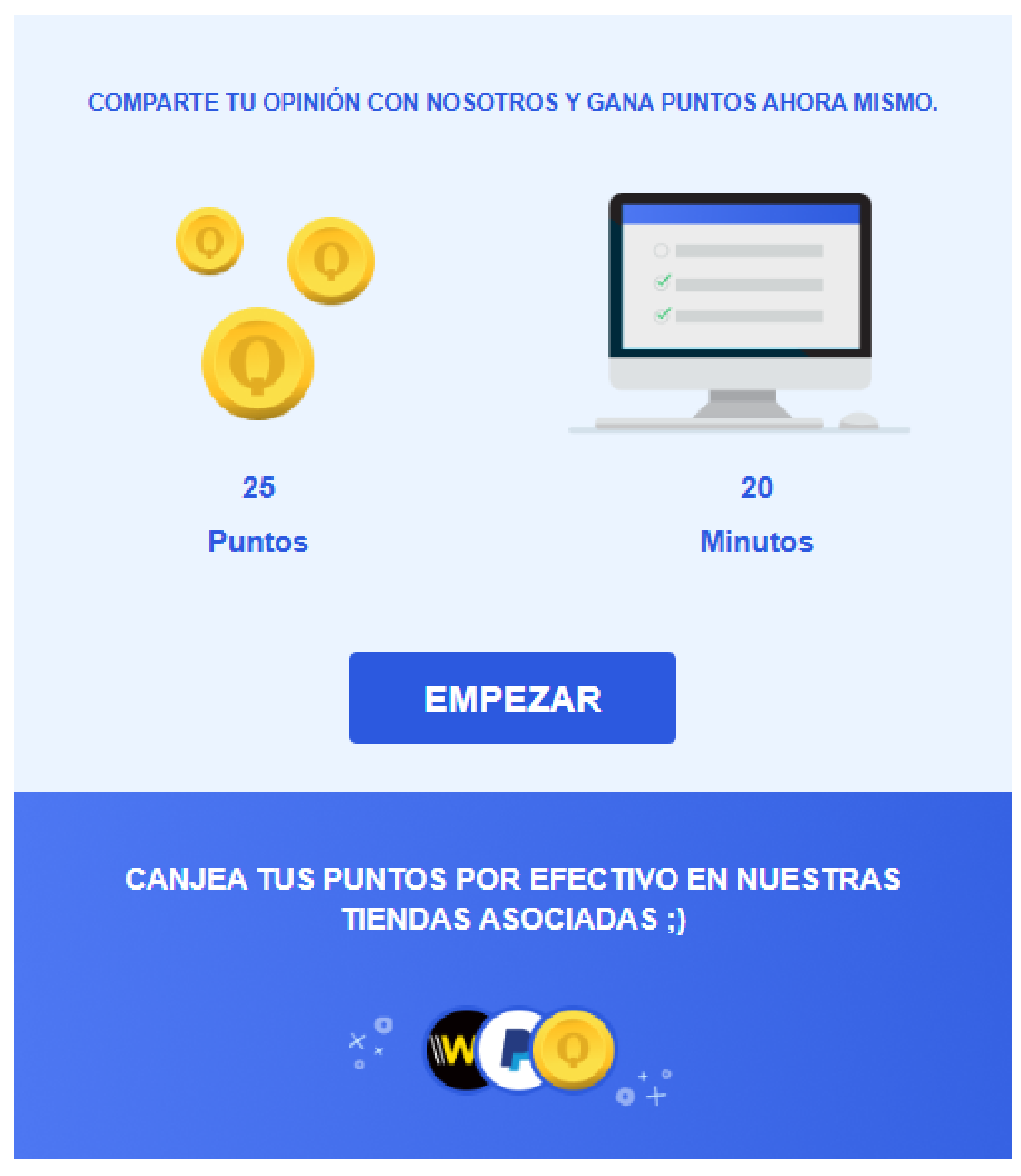

## F.2 Participant instructions

Below, we provide the exact wording (translated from Spanish) of the initial instructions and consent form given to participants at the start of the experiment (Appendix F.2.1), and the instructions given to participants right before the task (Appendix F.2.2) and follow-up questions (Appendix F.2.3).

### F.2.1 Initial instructions and consent form

We invite you to participate in this survey, conducted by university professors, to understand how people perform tasks in the workplace.

First, we will ask you some general questions, and then we will ask you to complete a task that simulates typical workplace situations. In addition to the payment for completing the survey, **you can earn up to an additional 25,000 pesos** in Carrefour supermarket gift cards, depending on **how you perform on the task. The better your performance on the task, the higher your reward.** If you do very well, you will earn 25,000 pesos; if you do very poorly, you will receive 0 pesos in additional rewards.

To participate, **we need you to**:

- **use a computer** (it cannot be done on a cellphone),
- **not consult external sources or ask for help**, and
- **complete the survey in one sitting**, without leaving this page (it will take about 20 minutes).

**The platform will verify that these conditions are met**.

Your responses are confidential and anonymous.

This survey is part of a study approved by the research ethics committees of CEDLAS (Center for Distributive, Labor and Social Studies) and the Nottingham School of Economics. If you have any questions, please contact us at: encuestas@cedlas.org.

Would you like to participate now, and do you meet the conditions? If you click "yes," **you will have one hour to complete the survey** (although **we estimate it will take about 20 minutes**).

### F.2.2 Instructions before the task

In the remainder of the survey, we will ask you to complete a task similar to a real work situation and then answer a few brief questions about the task. Please **take the task seriously, as if it were your job**: we will evaluate your responses as a boss would, considering both their content and presentation.

Remember that **you can earn up to 25,000 pesos in Carrefour gift cards depending on your performance on the task and on the subsequent questions**.

You are not allowed to consult external sources or ask for help from other people—we are interested in seeing how you solve the problem on your own. *For treatment group only:* However, **you will have access to a virtual assistant** based on ChatGPT (artificial intelligence), available on the same page as the task. Whether or not you use it is your choice: the reward depends solely on your answers, with or without the assistant.

On the next page, we provide a tutorial on how to use the virtual assistant.

### F.2.3 Instructions before the follow-up questions

Next, we will ask you a few short questions about the task you have just completed. **Please remember that you can earn up to 25,000 pesos in Carrefour gift cards, depending on your performance.**

## F.3 Tutorial on how to use the AI assistant

The virtual assistant with which you will interact uses artificial intelligence, and knows how to summarize, analyze or edit texts, interpret graphs or tables, and more. To use it, you just have to chat with the assistant as if you were asking someone for help: type your question or request in the chat box and press "Send." You will receive a reply and can continue talking about other aspects, ask for clarifications, etc.

You can use the virtual assistant in different ways to complete the task. For example:

- Share your answer with it, and ask it to edit or simplify it, or to correct it
- Ask it specific questions, or request suggestions on how to solve the task
- Provide it with the entire instructions, and ask it to solve it. Afterwards, you can ask more questions or request clarifications if necessary

You may freely use the assistant's responses —it is there to help you, we will not consider that you are "copying." You may even copy and paste part or all of what it writes.

In the remainder of this page, we present a short practice task (which will not be graded) to guide you on how to use the virtual assistant.

**Practice task**

This practice task is a simplified version of the task you will do on the next page. It includes an instruction and additional information needed to solve it. Below, you will see the virtual assistant, and finally a response box where you must write your solution to the task. **Important:** the virtual assistant **has access to the additional information, but not to the instructions.**

**Instructions:** Write a short paragraph with the name of the largest country on the map and the names of the countries that surround it.

*Additional information:*

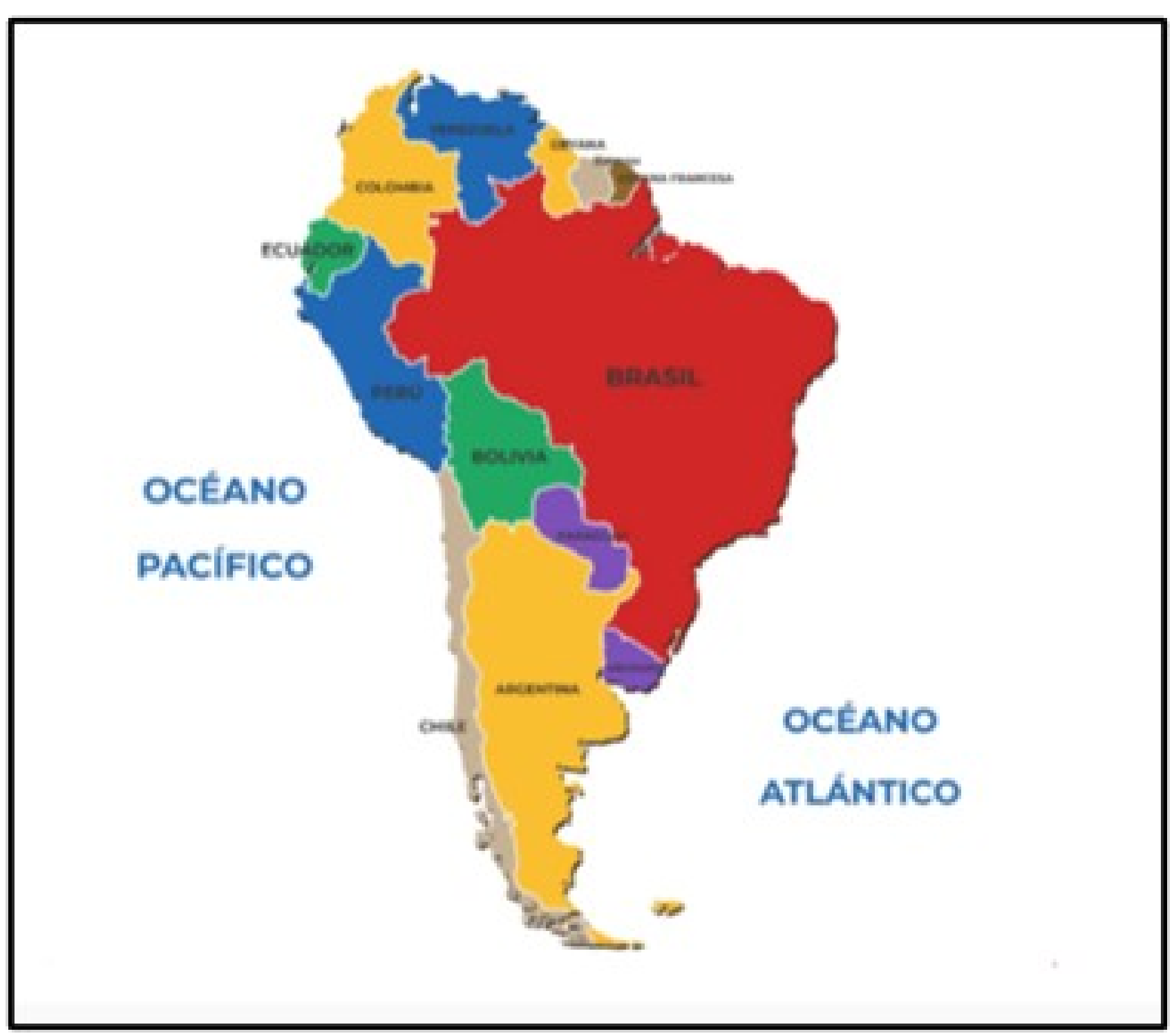

Below is the virtual assistant. We want to teach you how to use it. An important part of that is asking the right questions.

Please **follow the instructions below:**

(1) Type to the virtual assistant: *"Solve the task"*

You will see that the assistant describes the information on the map, but does not give you the answer, because it does not know the instructions.

(2) Now ask: *"What is the name of the largest country on the map, and what countries surround it?"*

(3) Finally, ask: *"Now write this for me in a short paragraph"*

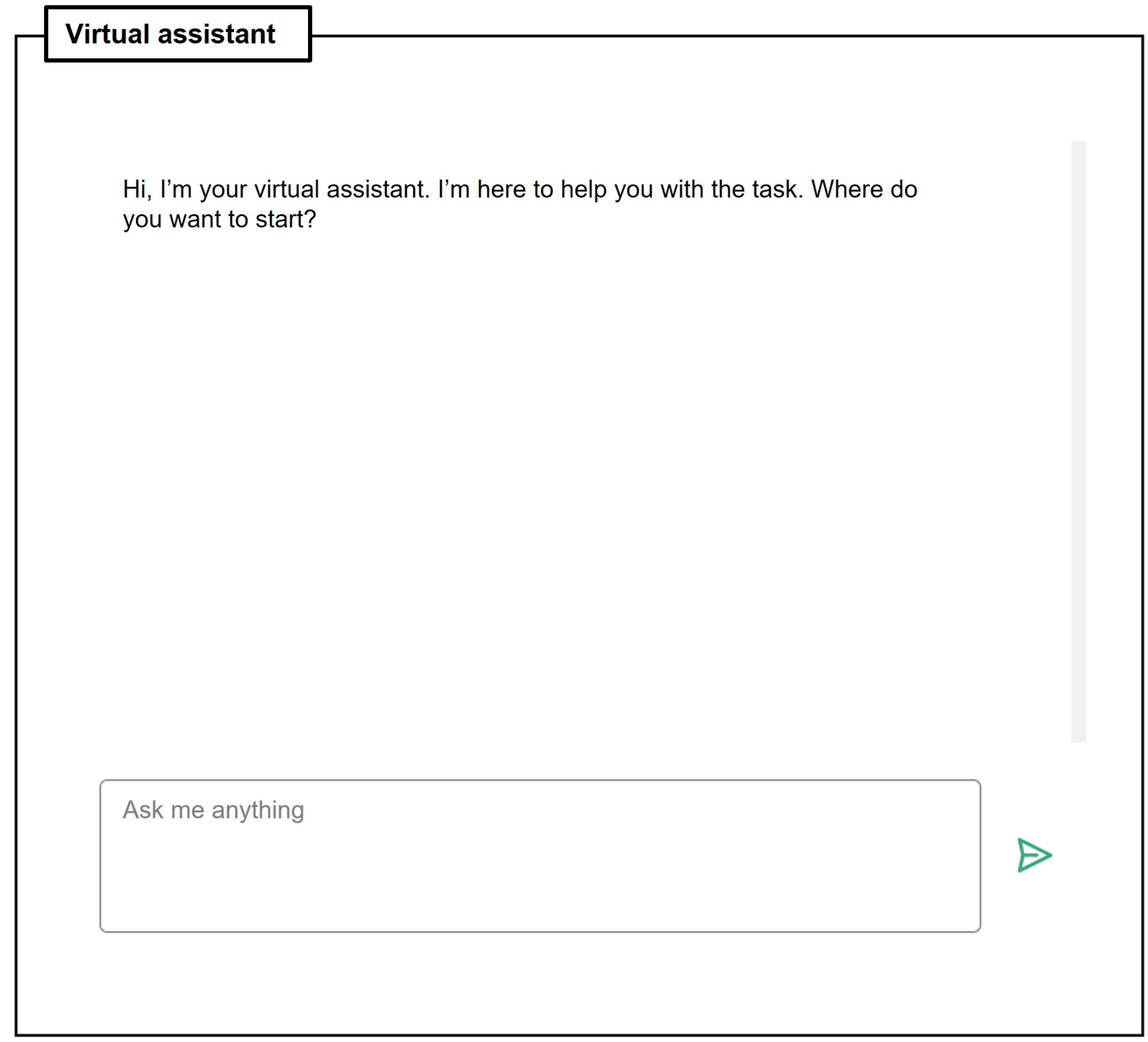

**Below is the answer box for the practice task. Write your answer there.** If you want, you can directly copy or past the assistant's response, following these steps:

1. Select the text of the assistant that you want to copy (by left-clicking and dragging the mouse over it). Once the text is selected, right-click and choose the "Copy" option.
2. Then, place the mouse over the answer box, right-click again, and select "Paste"

## F.4 Task and follow-up questions - Food delivery service

### F.4.1 Task

You work in the management of a premium burger delivery restaurant that has two branches (North and South). Each branch has its own employees, who sell burgers exclusively through a delivery platform (similar to PedidosYa). Yesterday you received the following email from your boss:

*From: Ramiro González* <`rgonzalez@gmail.com`>

***Subject:*** *problem with online reviews*

---

*Hello, how are you?*

*I am writing because I have been reviewing the online opinions of our branches, and I am a bit concerned about the North branch: in recent weeks we have been receiving low ratings due to excessive delivery delays. Could you help me understand what is happening?*

*I attach a report prepared by my assistant to help us better understand the situation.* ***Based on the data in this report, I would like your view on the following points:***

- *Could you check whether the problem originates from the delivery platform?*
- *Could it be that the North branch has more orders than the South branch, or that there has been a larger increase in the number of orders in this branch? Or are there no differences in orders between branches?*
- *Is there anything else in the report that you find relevant for understanding what is happening?*

*Please substantiate your answers well, since I plan to share this diagnosis with other team members.*

*Finally, according to your analysis,* ***what would you suggest we do to solve this problem?***

*Best regards,*

*Ramiro*

Attached report

**Summary of the problem:** In recent weeks, the North branch has been receiving low ratings due to excessive delivery delays.

1) Daily number of orders in the last three weeks in the North and South branches:

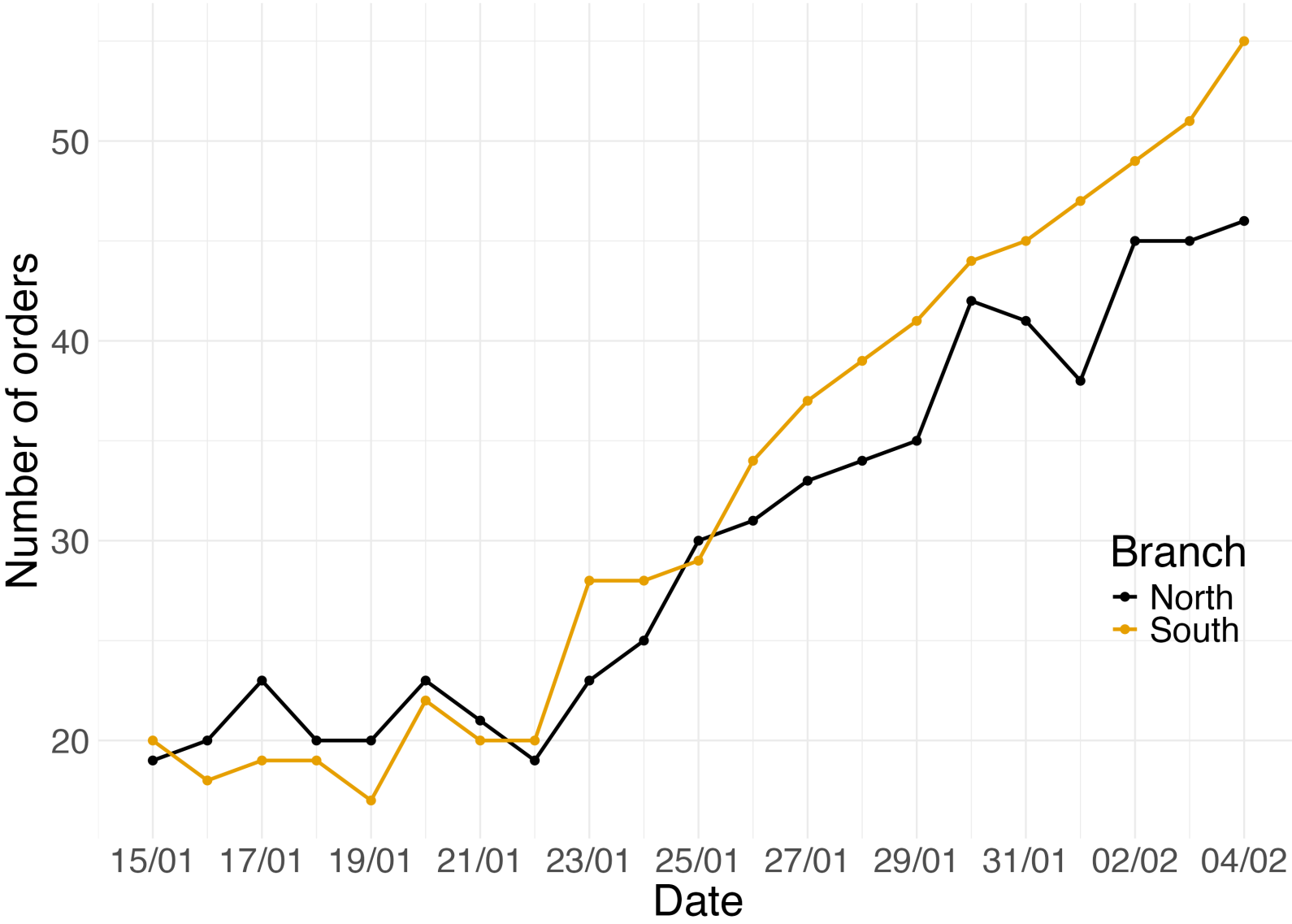

2) Information on preparation and delivery times for each branch in the last two weeks (provided by the delivery platform):

**South Branch**

- Average order acceptance time: 9 minutes
- Average preparation time at the restaurant: 18 minutes
- Average delivery time (from handover to the courier until delivery to the customer): 13 minutes

**North Branch**

- Average order acceptance time: 10 minutes
- Average preparation time in the restaurant: 29 minutes
- Average delivery time (from handover to the courier until delivery to the customer): 15 minutes

3) Number of employees and their average tenure in each branch:

| | North Branch | | South Branch | |
|---|---|---|---|---|
| | Number of Employees | Average Tenure (months) | Number of Employees | Average Tenure (months) |
| Grill staff | 2 | 0.5 | 2 | 4.0 |
| Assembly staff | 4 | 12.0 | 3 | 5.0 |
| Cleaning staff | 6 | 15.0 | 7 | 7.0 |
| **Total** | 12 | 11.6 | 12 | 6.0 |

*For control group only:* **Write an email below with your response to Ramiro.**

*For treatment group only:* **At the end of the page you will find the space to write your response to Ramiro. Below, you have the virtual assistant, in case you choose to use it. Remember that the virtual assistant only knows the information in the attached report.**

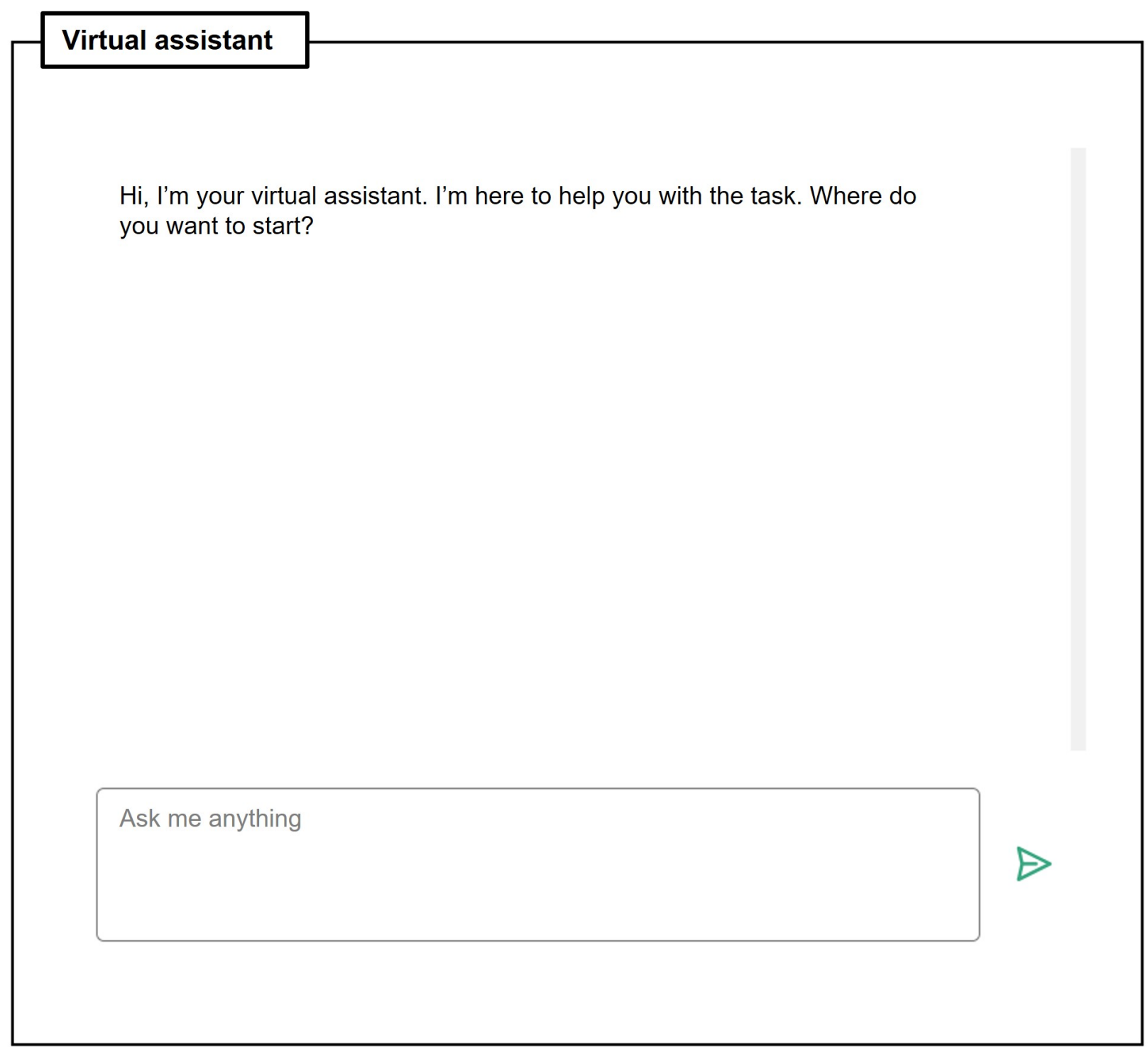

**Write an email below with your response to Ramiro.**

### F.4.2 Follow-up questions

One week after sending your email, you see Ramiro at a meeting. He says:

*"How are you? I've been sick these past few days and haven't had the chance to read what you sent me yet. I'm just about to go into a meeting where this topic will be discussed. Could you briefly tell me why the North branch is getting low ratings on the delivery platform?"*

Write your response to Ramiro below.

---

*Page break*

---

What did the graph show about the evolution of the number of orders in the two branches?

- Throughout the entire period, the North branch had more orders than the South branch
- In both branches, the number of orders remained constant throughout the entire period
- From a certain point onward, the number of orders increased in both branches
- The number of orders in the South branch remained constant, while those in the North branch increased

Which of the following statements is correct?

- There are more employees in the North branch than in the South branch
- There are fewer employees in the North branch than in the South branch
- There are more grill employees in the North branch than in the South branch
- The number of grill employees is the same in both branches

## F.5 Task and follow-up questions - Home appliances store

### F.5.1 Task

You work in an Argentina-based online business selling imported household appliances. In January 2025, the owner of the business sends you the following email:

*From: Luis González* <`lgonzalez@gmail.com`>

***Subject:*** *problem with sales of portable heaters and fans*

---

*Good morning,*

*I hope you're doing well. I'm writing because I was reviewing the sales of our products and noticed something unusual with the portable heaters and fans. Historically, our fan sales increase in summer and our heater sales in winter. However, this did not happen in 2024. Could you help me understand what's going on?*

*I attach a report prepared by my assistant to help us better understand the situation.* ***Based on the data in this report, I would like your view on the following points:***

- *To understand whether this problem is specific to these two products, could you check whether in 2024 air conditioner sales were different than usual?*
- *My assistant mentioned that last year there were restrictions on the import of certain electronic products. Could you check whether this might have something to do with it?*
- *Is there anything else in the report that you find relevant for understanding what is happening?*

*Please substantiate your answers well, since I plan to share this diagnosis with other team members.*

*Finally, according to your analysis,* ***what would you suggest we do to solve this problem?***

*I look forward to your reply.*

*Best regards,*

*Luis*

## Attached report

**Summary of the problem:** Historically, our portable fan sales increase in summer and our portable heater sales in winter. However, this did not happen in 2024.

1) Evolution of sales of air conditioners, portable heaters and portable fans during the last two years:

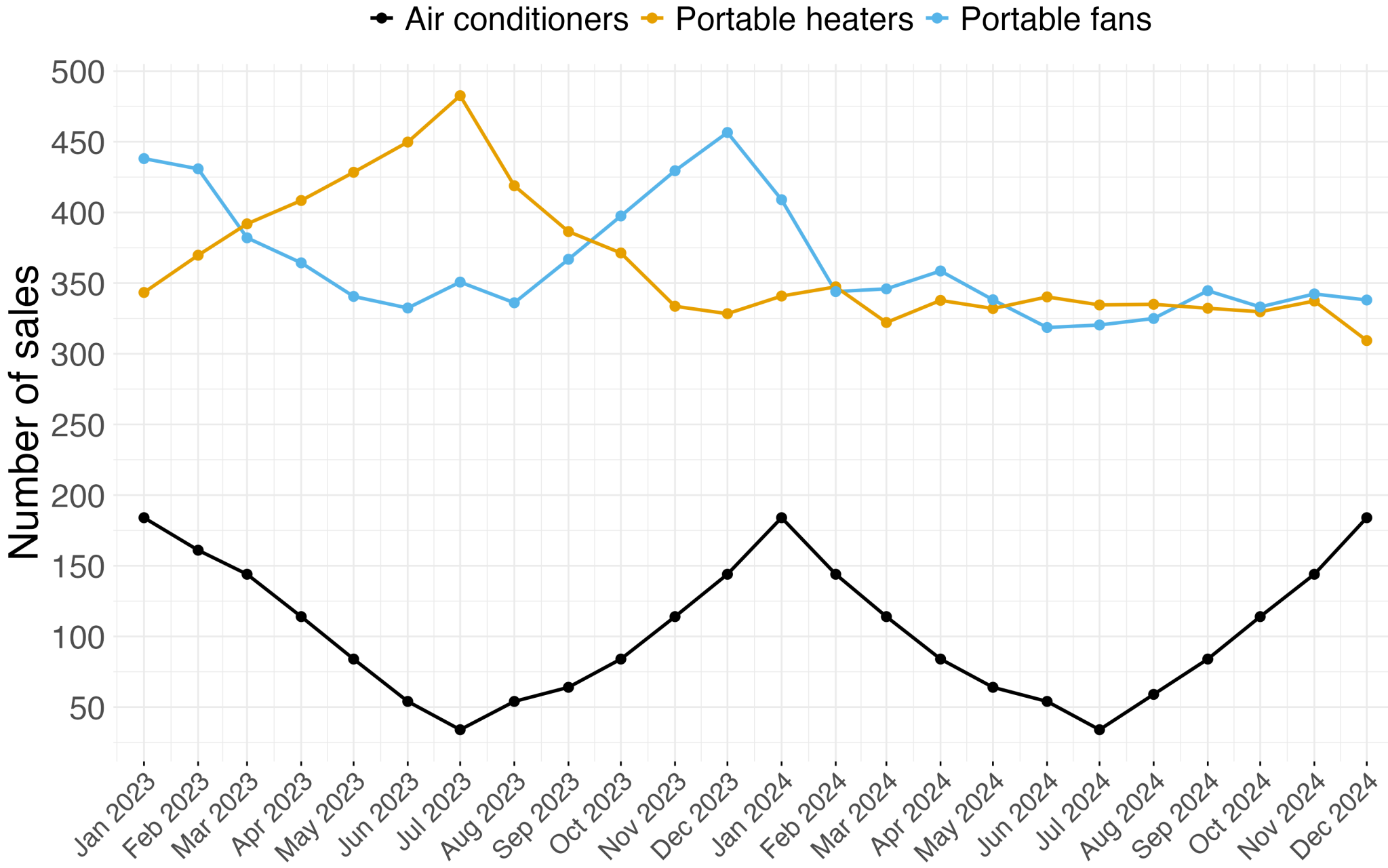

2) The table below indicates, for each product, the number of units ordered from the importer and delivered by the importer for each semester in 2023 and 2024.

| Period | Air conditioner | | Portable heaters | | Portable fans | |
|---|---|---|---|---|---|---|
| | Units ordered | Units delivered | Units ordered | Units delivered | Units ordered | Units delivered |
| Jan-Jun 2023 | 742 | 741 | 2,172 | 2,172 | 2,145 | 2,145 |
| Jul-Dec 2023 | 494 | 494 | 2,125 | 2,127 | 2,175 | 2,175 |
| Jan-Jun 2024 | 687 | 644 | 2,182 | 2,183 | 2,130 | 2,131 |
| Jul-Dec 2024 | 688 | 619 | 2,125 | 2,126 | 2,150 | 2,150 |

3) Each quarter we send an email to our customer base advertising certain products. Below are the products included in this email over the last two years:

- Jan-Mar 2023: Portable fan, television, air conditioner
- Apr-Jun 2023: Refrigerator, washing machine, microwave, portable heater
- Jul-Sep 2023: Portable heater, space heater, dishwasher
- Oct-Dec 2023: Computer, television, air conditioner, portable fan
- Jan-Mar 2024: Electric kettle, air conditioner, portable heater
- Apr-Jun 2024: Television, computer, portable fan, microwave
- Jul-Sep 2024: Dishwasher, washing machine, electric oven
- Oct-Dec 2024: Air conditioner, television, space heater

*For control group only:* **Write an email below with your response to Luis.**

*For treatment group only:* **At the end of the page you will find the space to write your response to Luis. Below, you have the virtual assistant, in case you choose to use it. Remember that the virtual assistant only knows the information in the attached report.**

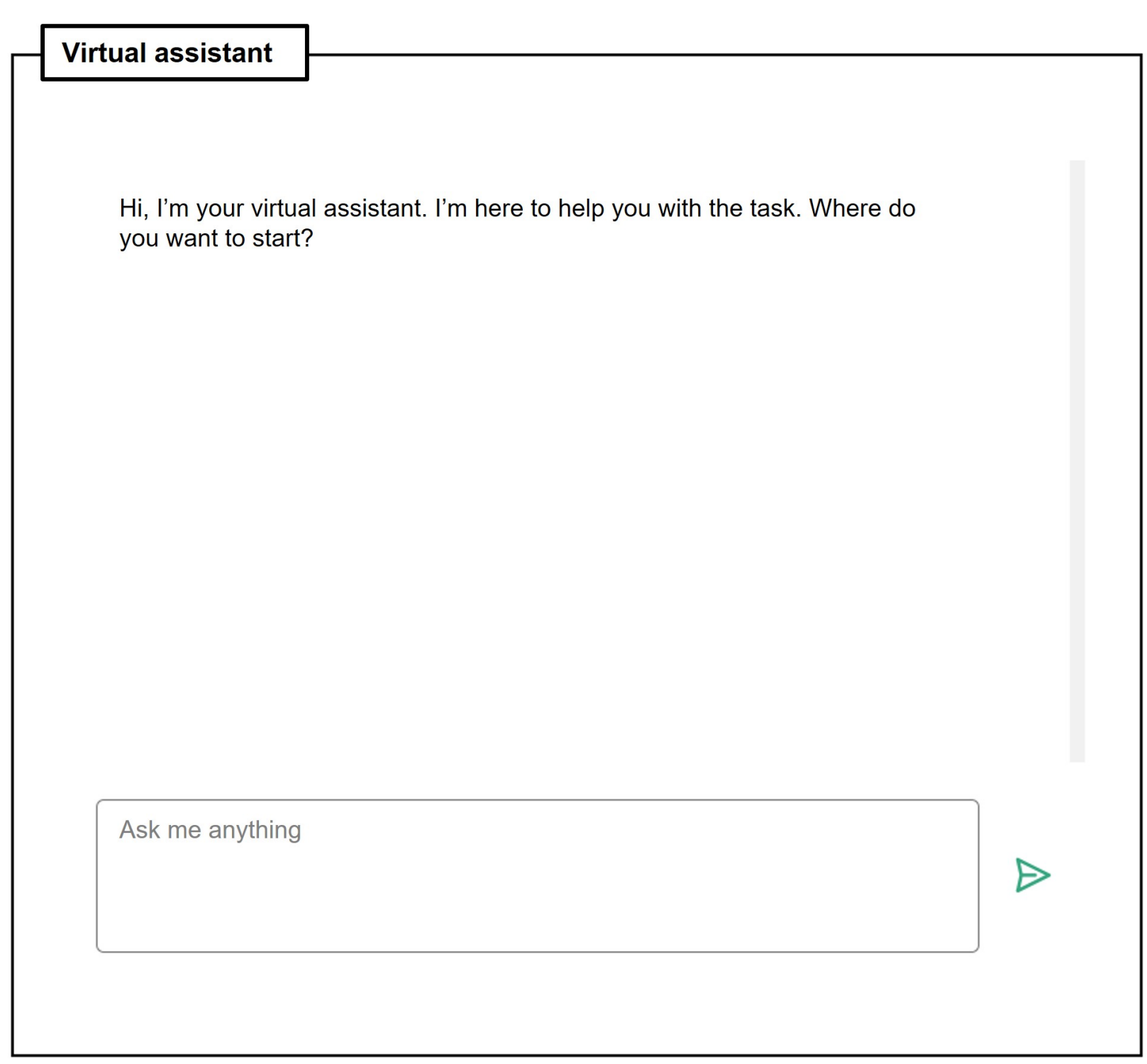

**Write an email below with your response to Luis.**

### F.5.2 Follow-up questions

One week after sending your email, you see Luis at the store. He says:

> *"How are you? I've been sick these past few days and haven't had the chance to read what you sent me yet. I'm just about to go into a meeting where this topic will be discussed. Could you briefly tell me why we are not selling more heaters in winter and portable fans in summer?"*

Write your response to Luis below.

---

*Page break*

---

What did the graph show about the evolution of portable fan sales during 2024?

- They followed the same pattern as air conditioner sales
- They remained constant throughout the year
- They increased at the end of the year
- They declined throughout the year

Which of the following statements is true regarding the email that the company sends to its customer base?

- The products advertised in a period may not match those of the same period in the previous year
- The company sends this email every month
- Portable heaters are never advertised in the email
- Air conditioners are always advertised in the email

## F.6 Task and follow-up questions - Cafeteria

### F.6.1 Task

You work in the management of "Sabores Express", a business hired to run the cafeteria at the North and South sites of Infotech, a major company. Yesterday you received the following email from your boss:

*From: Adriana González <agonzalez@gmail.com>*

***Subject:** problem with sales at the North site*

*Hello, how are you?*

*I'm writing because I've been looking at the sales from the past few weeks and I can't figure out what's going on in our cafeteria at the North site. Even though Infotech has almost the same number of employees at both sites, we're selling fewer lunches at the North site than at the South. Could you help me understand what's happening?*

*I attach a report prepared by my team to help us better understand the situation.* ***Based on the data in this report, I would like your view on the following points:***

- *Are customers at the North site choosing other lunch alternatives outside the cafeteria to a greater extent?*
- *Have sales at the North site decreased in general or only at a specific moment?*
- *Is there anything else in the report that you find relevant for understanding what is happening?*

*Please substantiate your answers well, since I plan to share this diagnosis with other team members.*

*Finally, according to your analysis,* ***what would you suggest we do to solve this problem?***

*Best regards,*

*Adriana*

## Attached report

**Summary of the problem:** Although Infotech has almost the same number of employees at both sites, fewer lunches are sold at the North than at the South site.

1) Daily number of lunches sold in the cafeteria at each site during the last three weeks:

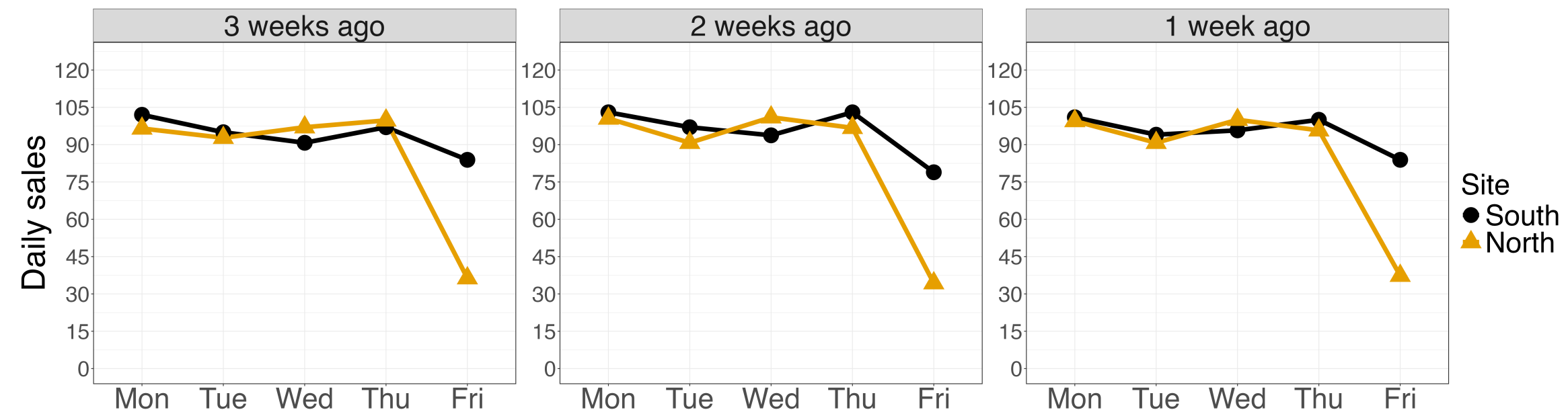

2) Last week, a survey was conducted among Infotech employees present at the office. They were asked where they got their lunch that day. The results are as follows:

| Site | Monday | Tuesday | Wednesday | Thursday | Friday |
|---|---|---|---|---|---|
| **South** | Cafeteria: 80%<br>Other place: 20% | Cafeteria: 76%<br>Other place: 24% | Cafeteria: 75%<br>Other place: 25% | Cafeteria: 80%<br>Other place: 20% | Cafeteria: 78%<br>Other place: 22% |
| **North** | Cafeteria: 78%<br>Other place: 22% | Cafeteria: 75%<br>Other place: 25% | Cafeteria: 80%<br>Other place: 20% | Cafeteria: 79%<br>Other place: 21% | Cafeteria: 78%<br>Other place: 22% |

3) Infotech has employees in two different areas: administration and sales. Administrative staff can work from home on Fridays, while sales staff are required to be on site from Monday to Friday. The table below shows the characteristics of employees at each site according to their area of work:

| | South site | | North site | |
|---|---|---|---|---|
| **Area** | **Number of Employees** | **Average Tenure (years)** | **Number of Employees** | **Average Tenure (years)** |
| Administration | 20 | 2.0 | 81 | 3.0 |
| Sales | 105 | 3.0 | 44 | 2.0 |
| **Total** | **125** | **2.8** | **125** | **2.6** |

*For control group only:* **Write an email below with your response to Adriana.**

*For treatment group only:* **At the end of the page you will find the space to write your response to Adriana. Below, you have the virtual assistant, in case you choose to use it. Remember that the virtual assistant only knows the information in the attached report.**

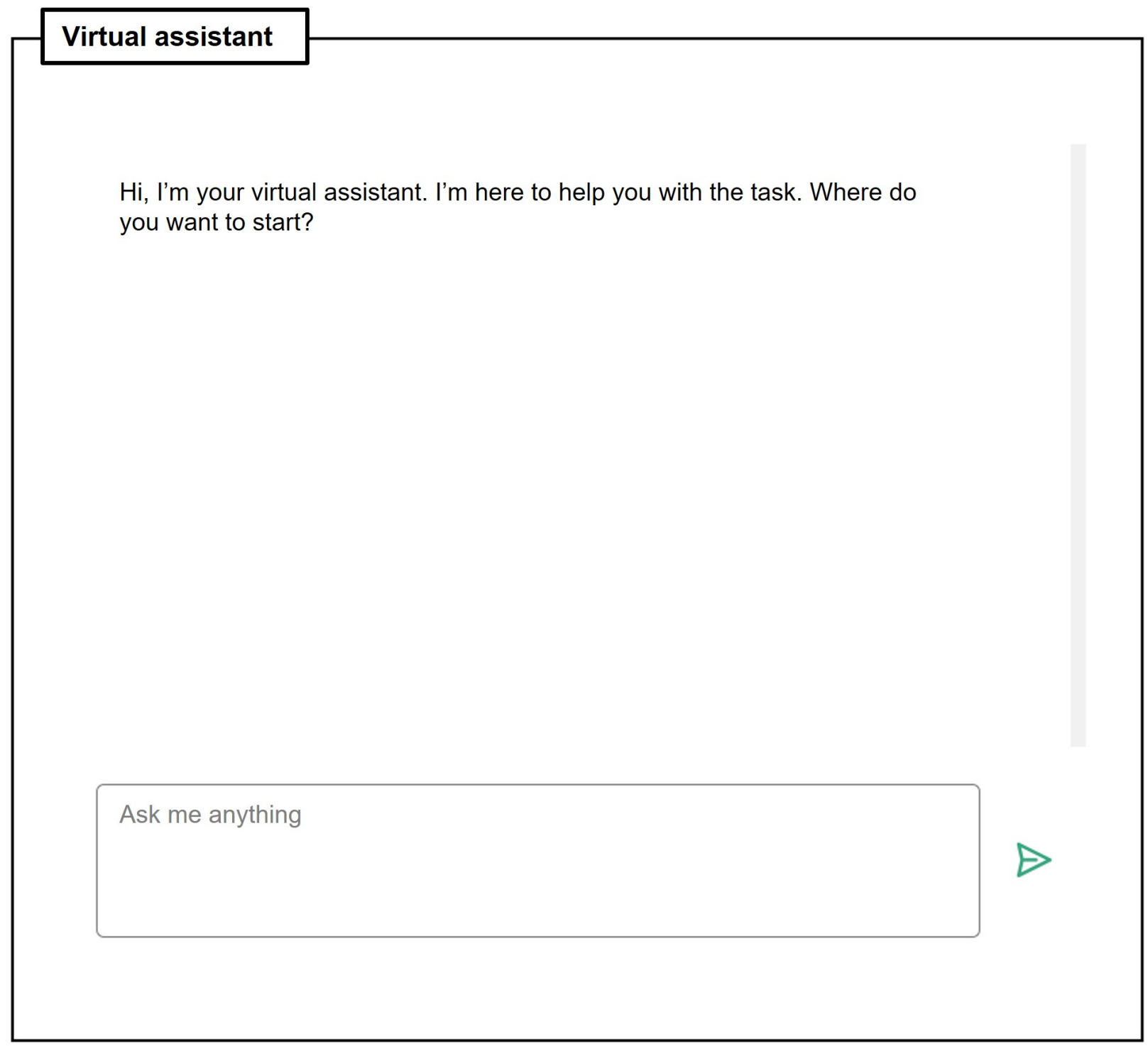

**Write an email below with your response to Adriana.**

### F.6.2 Follow-up questions

One week after sending your email, you see Adriana at a meeting. She says:

> *"How are you? I've been sick these past few days and haven't had the chance to read what you sent me yet. I'm just about to go into a meeting where this topic will be discussed. Could you briefly tell me why we are selling fewer lunches at the North site?"*

Write your response to Adriana below.

---

*Page break*

---

What did the graph show about daily lunch sales at both sites?

- Sales at both sites increase on Tuesdays compared to Mondays
- At both sites, Monday sales are higher than the previous week's Friday sales
- On Fridays, sales drop only at the North site
- Every day, the North site sells fewer lunches than the South site

Which of the following statements about Infotech employees is true?

- Only at the North site are Administration staff allowed to work remotely one day per week
- Infotech has Administration and Cleaning staff
- Both sites have the same total number of employees
- There are more employees at the South site than at the North

## F.7 Prompt given to the AI assistant

This appendix documents the exact prompts provided to the AI assistant during the experiment. It includes the task-specific prompts for each variation of the experimental task, as well as the prompt given to the assistant during the tutorial prompt used to familiarize treated participants with the assistant's functionality.

### F.7.1 Food delivery service task

You are a virtual assistant integrated into a website where people in Argentina must solve a task based on a given problem. As part of this task, users receive a report with the necessary information to solve the problem. Below you can find this report. Your job is to assist users in interpreting this information. Answer all user questions.

The first thing you will receive in the conversation is an image. Treat this image as part of the attached report and not as a message from the user. In other words, treat it the same way as you would treat tables or other information, naturally, and without mentioning that the image was sent as a message. Use it to complement your answers.

Provide clear and helpful responses.

---

**Report**

**Summary of the problem:** In recent weeks, the North branch has been receiving low ratings due to excessive delivery delays.

1) Here goes the image you will receive in the next message. Remember that this message is not from the user, but rather complements the information provided here. The chart title is *"Daily number of orders in the last three weeks in the North and South branches"*.

2) Information on preparation and delivery times for each branch in the last two weeks (provided by the delivery platform):

**South Branch**

- Average order acceptance time: 9 minutes
- Average preparation time at the restaurant: 18 minutes
- Average delivery time (from handover to courier until delivery to customer): 13 minutes

**North Branch**

- Average order acceptance time: 10 minutes
- Average preparation time in the restaurant: 29 minutes
- Average delivery time (from handover to the courier until delivery to the customer): 15 minutes

3) Number of employees and their average tenure in each branch:

| | North Branch | | South Branch | |
|---|---|---|---|---|
| **Category** | **No. of employees** | **Average tenure (months)** | **No. of employees** | **Average tenure (months)** |
| Grill staff | 2 | 0.5 | 2 | 4 |
| Assembly staff | 4 | 12 | 3 | 5 |
| Cleaning staff | 6 | 15 | 7 | 7 |
| **Total** | 12 | 11.6 | 12 | 6 |

---

Assist users with their requests, but if a user sends a message that makes no sense (such as "." or "bla bla" or "No" or "3"), do not provide the full solution to the problem. In that case, you can reply with something like: *"How can I help you?"*.

### F.7.2 Home appliances store task

You are a virtual assistant integrated into a website where people in Argentina must solve a task based on a given problem. As part of this task, users receive a report with the information needed to solve the problem. Below you can find this report. Your job is to assist users in interpreting this information. Answer all user questions.

The first thing you will receive in the conversation is an image. Treat this image as part of the attached report and not as a message from the user. In other words, treat it in the same way as you would treat tables or any other piece of information, naturally, and without mentioning that the image was sent as a message. Use it to complement your answers.

Provide clear and helpful responses.

---

**Report**

**Summary of the problem:** Historically, our portable fan sales increase in summer and our portable heater sales in winter. However, this did not happen in 2024.

1) Here goes the image you will receive in the next message. Remember that this message is not from the user, but rather complements the information provided here. The chart title is *"Evolution of sales of air conditioners, portable heaters and portable fans during the last two years:"*.

2) The table below indicates, for each product, the number of units ordered from the importer and the number delivered by the importer for each semester in 2023 and 2024:

| Period | Air conditioner | | Portable heaters | | Portable fans | |
|---|---|---|---|---|---|---|
| | Units ordered | Units delivered | Units ordered | Units delivered | Units ordered | Units delivered |
| Jan – Jun 2023 | 742 | 741 | 2172 | 2172 | 2145 | 2145 |
| Jul – Dec 2023 | 494 | 494 | 2125 | 2127 | 2175 | 2175 |
| Jan – Jun 2024 | 687 | 644 | 2182 | 2183 | 2130 | 2131 |
| Jul – Dec 2024 | 688 | 619 | 2125 | 2126 | 2150 | 2150 |

3) Each quarter we send an email to our customer base advertising certain products. Below are the products included in this email over the last two years:

- Jan-Mar 2023: Portable fan, television, air conditioner
- Apr-Jun 2023: Refrigerator, washing machine, microwave, portable heater

- Jul-Sep 2023: Portable heater, space heater, dishwasher
- Oct-Dec 2023: Computer, television, air conditioner, portable fan
- Jan-Mar 2024: Electric kettle, air conditioner, portable heater
- Apr-Jun 2024: Television, computer, portable fan, microwave
- Jul-Sep 2024: Dishwasher, washing machine, electric oven
- Oct-Dec 2024: Air conditioner, television, space heater

---

Assist users with their requests, but if a user sends a message that makes no sense (such as "." or "bla bla" or "No" or "3"), do not provide the full solution to the problem. In that case, you can reply with something like: *"How can I help you?"*.

#### F.7.3 Cafeteria task

You are a virtual assistant integrated into a website where people in Argentina must solve a task based on a given problem. As part of this task, users receive a report with the information needed to solve the problem. Below you can find this report. Your job is to assist users in interpreting this information. Answer all user questions.

The first thing you will receive in the conversation is an image. Treat this image as part of the attached report and not as a message from the user. In other words, treat it in the same way as you would treat tables or any other piece of information, naturally, and without mentioning that the image was sent as a message. Use it to complement your answers.

Provide clear and helpful responses.

---

**Report**

**Problem summary:** Although Infotech has almost the same number of employees at both sites, fewer lunches are sold at the North branch than at the South branch.

1) Here goes the image you will receive in the next message. Remember that this message is not from the user, but rather complements the information provided here. The chart title is *"Daily number of lunches sold in the cafeteria at each site during the last three weeks"*.

2) Last week, a survey was conducted among Infotech employees present at the office. They were asked where they got their lunch that day. The results are as follows:

| Site | Monday | Tuesday | Wednesday | Thursday | Friday |
|---|---|---|---|---|---|
| **South** | Cafeteria: 80%<br>Other: 20% | Cafeteria: 76%<br>Other: 24% | Cafeteria: 75%<br>Other: 25% | Cafeteria: 80%<br>Other: 20% | Cafeteria: 78%<br>Other: 22% |
| **North** | Cafeteria: 78%<br>Other: 22% | Cafeteria: 75%<br>Other: 25% | Cafeteria: 80%<br>Other: 20% | Cafeteria: 79%<br>Other: 21% | Cafeteria: 78%<br>Other: 22% |

3) Infotech has employees in two different areas: administration and sales. Administration staff can work from home on Fridays, while sales staff are required to be on site from Monday to Friday. The table below shows the characteristics of employees at each site according to their area of work:

---

Assist users with their requests, but if a user sends a message that makes no sense (such

| Work area | South Branch | | North Branch | |
|---|---|---|---|---|
| | **No. of employees** | **Average tenure (years)** | **No. of employees** | **Average tenure (years)** |
| Administration | 20 | 2 | 81 | 3 |
| Sales | 105 | 3 | 44 | 2 |
| **Total** | **125** | **2.8** | **125** | **2.6** |

as "." or "bla bla" or "No" or "3"), do not provide the full solution to the problem. In that case, you can reply with something like: *How can I help you?*".

#### F.7.4 Tutorial

You are a virtual assistant integrated into a website where people in Argentina must solve a task based on a given problem. As part of this task, users receive a report with the information needed to solve the problem. Below you can find this report. Your job is to assist users in interpreting this information. Answer all user questions.

The first thing you will receive in the conversation is an image. Treat this image as part of the attached report and not as a message from the user. In other words, treat it in the same way as you would treat tables or any other piece of information, naturally, and without mentioning that the image was sent as a message. Use it to complement your answers.

Provide clear and helpful responses.

---

**Report**

1) Here goes the image you will receive in the next message. Remember that this message is not from the user, but rather complements the information provided here. The chart title is *"Political map of South America"*.

---

Assist users with their requests, but if a user sends a message that makes no sense (such as "." or "bla bla" or "No" or "3"), do not provide the full solution to the problem. In that case, you can reply with something like: *"How can I help you?"*.